\documentclass[twocolumn]{aastex631}
\usepackage{CJK}
\usepackage{amsmath}
\usepackage{soul}
\usepackage{scalerel}
\usepackage{dashrule}
\definecolor{linkcolor}{cmyk}{1,1,0,0}
\newcommand\NB[1][0.3]{N\kern-#1emB}
\newcommand{\indep}{\perp \!\!\! \perp}
\newcommand{\dep}{\not\!\perp\!\!\!\perp}

\def\Pa{\mathbf{Pa}}
\def\X{\mathbf{X}}
\def\Y{\mathbf{Y}}
\def\T{\mathrm{T}}
\def\mh{m^{h}}

\def\mch{m_{c}^{h}}
\def\am{\alpha_{\mu}}
\def\aw{\alpha_{w}}

\received{September 16, 2024}
\revised{December 10, 2024}
\accepted{December 10, 2024}

\submitjournal{\apj}

\begin{document}
\begin{CJK*}{UTF8}{gbsn}

\title{Causal Discovery in Astrophysics: Unraveling Supermassive Black Hole and Galaxy Coevolution}
\shorttitle{Causal Discovery in Astrophysics}
\shortauthors{Jin et al.}

\correspondingauthor{Zehao Jin (金泽灏)}
\email{zj448@nyu.edu}

\author[0009-0000-2506-6645]{Zehao Jin (金泽灏)}
\affiliation{New York University Abu Dhabi, P.O. Box 129188, Abu Dhabi, United Arab Emirates}
\affiliation{Center for Astrophysics and Space Science (CASS), New York University Abu Dhabi, P.O. Box 129188, Abu Dhabi, UAE}
\affiliation{Montr\'{e}al Institute for Astrophysical Data Analysis and Machine Learning (Ciela), Montr\'{e}al, Canada}

\author[0000-0003-3784-5245]{Mario Pasquato}
\altaffiliation{These authors contributed equally to this work.}
\affiliation{Montr\'{e}al Institute for Astrophysical Data Analysis and Machine Learning (Ciela), Montr\'{e}al, Canada}
\affiliation{Montr\'{e}al Institute for Learning Algorithms (Mila), Quebec Artificial Intelligence Institute,
6666 Rue Saint-Urbain, Montr\'{e}al, Canada}
\affiliation{D\'{e}partement de Physique, Universit\'{e} de Montr\'{e}al, 1375 Avenue Th\'{e}r\`{e}se-Lavoie-Roux, Montr\'{e}al, Canada}
\affiliation{Dipartimento di Fisica e Astronomia, Universit\`{a} di Padova, Vicolo dell'Osservatorio 5, Padova, Italy}
\affiliation{Istituto di Astrofisica Spaziale e Fisica Cosmica (INAF IASF-MI), Via Alfonso Corti 12, I-20133, Milan, Italy}

\author[0000-0002-4306-5950]{Benjamin L.\ Davis}
\altaffiliation{These authors contributed equally to this work.}
\affiliation{New York University Abu Dhabi, P.O. Box 129188, Abu Dhabi, United Arab Emirates}
\affiliation{Center for Astrophysics and Space Science (CASS), New York University Abu Dhabi, P.O. Box 129188, Abu Dhabi, UAE}

\author[0009-0005-1943-3484]{Tristan Deleu}
\affiliation{Montr\'{e}al Institute for Learning Algorithms (Mila), Quebec Artificial Intelligence Institute, 6666 Rue Saint-Urbain, Montr\'{e}al, Canada}
\affiliation{D\'{e}partement d'Informatique et de Recherche Op\'{e}rationnelle, Universit\'{e} de Montr\'{e}al, 2920 Chemin de la Tour, Montr\'{e}al, Canada}

\author[0000-0003-2341-9755]{Yu Luo (罗煜)}
\affiliation{Department of Physics, School of Physics and Electronics, Hunan Normal University, Changsha 410081, China}
\affiliation{Purple Mountain Observatory, 10 Yuan Hua Road, Nanjing 210034, China}

\author[0000-0002-9879-1749]{Changhyun Cho}
\affiliation{New York University Abu Dhabi, P.O. Box 129188, Abu Dhabi, United Arab Emirates}
\affiliation{Center for Astrophysics and Space Science (CASS), New York University Abu Dhabi, P.O. Box 129188, Abu Dhabi, UAE}

\author[0000-0002-4728-8473]{Pablo Lemos}
\affiliation{Montr\'{e}al Institute for Astrophysical Data Analysis and Machine Learning (Ciela), Montr\'{e}al, Canada}
\affiliation{Montr\'{e}al Institute for Learning Algorithms (Mila), Quebec Artificial Intelligence Institute,
6666 Rue Saint-Urbain, Montr\'{e}al, Canada}
\affiliation{D\'{e}partement de Physique, Universit\'{e} de Montr\'{e}al, 1375 Avenue Th\'{e}r\`{e}se-Lavoie-Roux, Montr\'{e}al, Canada}

\author[0000-0003-3544-3939]{Laurence Perreault-Levasseur}
\affiliation{Montr\'{e}al Institute for Astrophysical Data Analysis and Machine Learning (Ciela), Montr\'{e}al, Canada}
\affiliation{Montr\'{e}al Institute for Learning Algorithms (Mila), Quebec Artificial Intelligence Institute,
6666 Rue Saint-Urbain, Montr\'{e}al, Canada}
\affiliation{D\'{e}partement de Physique, Universit\'{e} de Montr\'{e}al, 1375 Avenue Th\'{e}r\`{e}se-Lavoie-Roux, Montr\'{e}al, Canada}
\affiliation{Center for Computational Astrophysics, Flatiron Institute, New York, NY, United States of America}

\author[0000-0002-9322-3515]{Yoshua Bengio}
\affiliation{Montr\'{e}al Institute for Learning Algorithms (Mila), Quebec Artificial Intelligence Institute, 6666 Rue Saint-Urbain, Montr\'{e}al, Canada}
\affiliation{D\'{e}partement d'Informatique et de Recherche Op\'{e}rationnelle, Universit\'{e} de Montr\'{e}al, 2920 Chemin de la Tour, Montr\'{e}al, Canada}
\affiliation{Canadian Institute for Advanced Research Artificial Intelligence Chair}
\affiliation{Canadian Institute for Advanced Research Senior Fellow}

\author[0000-0002-5458-4254]{Xi Kang (康熙)}
\affiliation{Institute for Astronomy, Zhejiang University, Hangzhou 310027, China}
\affiliation{Purple Mountain Observatory, 10 Yuan Hua Road, Nanjing 210034, China}

\author[0000-0002-8171-6507]{Andrea Valerio Macci\`{o}}
\affiliation{New York University Abu Dhabi, P.O. Box 129188, Abu Dhabi, United Arab Emirates}
\affiliation{Center for Astrophysics and Space Science (CASS), New York University Abu Dhabi, P.O. Box 129188, Abu Dhabi, UAE}
\affiliation{Max-Planck-Institut f\"{u}r Astronomie, K\"{o}nigstuhl 17, 69117 Heidelberg, Germany}

\author[0000-0002-8669-5733]{Yashar Hezaveh}
\affiliation{Montr\'{e}al Institute for Astrophysical Data Analysis and Machine Learning (Ciela), Montr\'{e}al, Canada}
\affiliation{Montr\'{e}al Institute for Learning Algorithms (Mila), Quebec Artificial Intelligence Institute,
6666 Rue Saint-Urbain, Montr\'{e}al, Canada}
\affiliation{D\'{e}partement de Physique, Universit\'{e} de Montr\'{e}al, 1375 Avenue Th\'{e}r\`{e}se-Lavoie-Roux, Montr\'{e}al, Canada}
\affiliation{Center for Computational Astrophysics, Flatiron Institute, New York, NY, United States of America}

\begin{abstract}
Correlation \emph{does not} imply causation, but patterns of statistical association between variables can be exploited to infer a causal structure (even with purely observational data) with the burgeoning field of causal discovery.
As a purely observational science, astrophysics has much to gain by exploiting these new methods.
The supermassive black hole (SMBH)--galaxy interaction has long been constrained by observed scaling relations, that is low-scatter correlations between variables such as SMBH mass and the central velocity dispersion of stars in a host galaxy's bulge.
This study, using advanced causal discovery techniques and an up-to-date dataset, reveals a causal link between galaxy properties and dynamically-measured SMBH masses.
We apply a score-based Bayesian framework to compute the exact conditional probabilities of every causal structure that could possibly describe our galaxy sample.
With the exact posterior distribution, we determine the most likely causal structures and notice a probable causal reversal when separating galaxies by morphology.
In elliptical galaxies, bulge properties (built from major mergers) tend to influence SMBH growth, while in spiral galaxies, SMBHs are seen to affect host galaxy properties, potentially through feedback in gas-rich environments.
For spiral galaxies, SMBHs progressively quench star formation, whereas in elliptical galaxies, quenching is complete, and the causal connection has reversed.
Our findings support theoretical models of hierarchical assembly of galaxies and active galactic nuclei feedback regulating galaxy evolution.
Our study suggests the potentiality for further exploration of causal links in astrophysical and cosmological scaling relations, as well as any other observational science.
\end{abstract}

\keywords{
\href{http://astrothesaurus.org/uat/1882}{Astrostatistics (1882)};
\href{http://astrothesaurus.org/uat/159}{Black hole physics (159)};
\href{http://astrothesaurus.org/uat/573}{Galaxies (573)};
\href{http://astrothesaurus.org/uat/594}{Galaxy evolution (594)};
\href{http://astrothesaurus.org/uat/595}{Galaxy formation (595)};
\href{http://astrothesaurus.org/uat/612}{Galaxy physics (612)};
\href{http://astrothesaurus.org/uat/615}{Galaxy properties (615)};
\href{http://astrothesaurus.org/uat/2031}{Scaling relations (2031)};
\href{http://astrothesaurus.org/uat/1663}{Supermassive black holes (1663)}
}

\section{Introduction} \label{sec:intro}
\end{CJK*}

Purely observational sciences have long relied on correlations between variables to assess the validity of theoretical models.
However, observed correlations between two variables do not provide information about the direction of causality, making it impossible to discriminate between different causal mechanisms that predict the same correlational trends through traditional methods.
While interventions (such as randomized controlled trials) are commonly used to identify causal factors in laboratory settings, this is rarely possible in observational fields such as astrophysics.
Causal inference overcomes this limitation by exploiting the fact that different causal models produce distinct joint distributions of correlated variables with additional observables, allowing us to discriminate between these models and shed light on the direction of causality.
With this, it becomes possible to investigate causal relationships in a purely data-driven manner.

The existence of correlations between the mass of central supermassive black holes (SMBHs) and the properties of their host galaxies has long been observationally established \citep{Magorrian:1998,Ferrarese:2000,Gebhardt:2000} and reproduced by specific prescriptions in numerical simulations \citep{Soliman:2023}.
More recently, \cite{Natarajan_2023nat,Natarajan_2023} presented QUOTAS, a novel research platform for the data-driven investigation of SMBH populations, combining simulations, observations, and machine learning to explore the co-evolution of SMBHs and their hosts.
High-redshift, luminous quasar populations at $z\geq3$ alongside simulated data of the same epochs are assembled and co-located to demonstrate the connection between SMBH host galaxies and their parent dark matter halos.
However, unveiling the causal structure underpinning these correlations has remained an open problem: does galaxy evolution influence the growth of SMBHs by regulating accretion, or do SMBHs shape their host galaxies' properties via feedback \citep{Silk:1998,DiMatteo:2005,DiMatteo:2008,Hopkins:2006,Hopkins:2007,Hopkins:2008b,Hopkins:2008,Schaye:2010,Gaspari:2013,Kormendy:2013,Delvecchio:2014,Heckman:2014,Aird:2015,Peca:2023,Pouliasis:2024}?
With the advent of the \textit{James Webb Space Telescope}, this debate has been reinvigorated by the detection of massive high-redshift quasars \citep{Larson:2023}.

The few attempts at identifying causal relations in the astrophysical literature focus on two variables at a time or on estimating causal coefficients given a causal structure (causal inference).
\citet{2019RNAAS...3..179P} used a regression discontinuity approach \citep{imbens2008regression} to measure the causal effect of galactic disk-shocking \citep{1972ApJ...176L..51O} on open star cluster properties.
A similar approach was followed by \citet{2021ApJ...923...20P} to measure the causal effect of a supernova explosion in the Vela OB2 complex on star formation.

\citet{2019MNRAS.487.2491E} used a matching strategy to measure the causal effect of galaxy mergers on active galactic nuclei (AGNs) activity.
Matching is a popular way of controlling for confounds in quasi-experimental data, where assignment to treatment is not determined at random \citep{greenwood1945experimental}. 
A precursor to matching in the astrophysical literature is the study of ``second-parameter pairs'' in globular clusters \citep{catelan2001horizontal}: globular clusters were matched based on metallicity and other properties, looking for the reason one member of the pair displayed a hot horizontal branch and its match would not.

\citet{Bluck:2022} utilized a Random Forest classifier to extract causal insights from observations to find the most predictive parameters associated with the quenching of star formation. 
\citet{2022A&A...666A...9G} applied a denoising technique based on causal principles, half-sibling regression \citep{2016PNAS..113.7391S}, to exoplanet imaging. 
In physics, outside of the context of astronomy and cosmology, causal techniques have found direct application in geophysics \citep{runge2019inferring} and climate science \citep{2020WCD.....1..519D}, and have functioned as a basis for theoretical development in quantum foundations \citep{pirsa_PIRSA:23030069, PhysRevX.7.031021, 2013PhRvA..88e2130L, 2015NJPh...17c3002W} and thermodynamics \citep{2007arXiv0708.3411J, 2008JSMTE..04..001A}.
Our work builds upon a preliminary pilot study to explore causal connections in galaxy--SMBH systems \citep{Pasquato:2023,Pasquato:2024}.

In this paper, we present a first-of-its-kind causal study of galaxies and their SMBHs, ultimately finding a compelling data-driven result.
In the text, we provide a primer on causal inference/discovery (\S\ref{sec:primer}), a detailed accounting of our data (\S\ref{sec:data}), our exact posterior methodology (\S\ref{sec:methodology}), the causal structures identified (\S\ref{sec:compendium}), a deeper look into the distribution of identified causal structures (\S\ref{sec:analysis}), our findings as they pertain to galaxy evolution and AGNs' feedback (\S\ref{sec:findings}), an experiment with semi-analytical models (\S\ref{sec:SAMs}), a cross-check with alternative causal discovery methods (\S\ref{sec:PC}), a series of discussions on limitations of the data and the method (\S\ref{sec:limitations}) including unobserved confounders (\S\ref{sec:confounders}), observational errors (\S\ref{sec:Obs_errors}), outliers (\S\ref{sec:outliers}), cyclicity (\S\ref{sec:cyclicity}), and different priors (\S\ref{sec:prior}).
Finally, we show possible extension of our methods into more complex scenarios (\S\ref{sec:DAG-GFN}), and conclude by offering some insights and future directions (\S\ref{sec:conclusions}).
All uncertainties are quoted at 1\,$\sigma\equiv68.3\%$ confidence intervals.

\vspace{4mm}
\section{Causal Inference and Discovery}\label{sec:primer}

The seminal book \textit{Causality} \citep{pearl2009causality} introduced operational definitions for the presence of several types of causal relations between different variables.\footnote{We refer the interested reader to \citet{pearl2016} and two online courses found at \url{https://www.bradyneal.com/causal-inference-course} and \url{https://www.bilibili.com/video/BV1sJ41177sg} for further information about causal inference.}
While these build on empirically observable statistical dependencies between pairs of variables, they leverage the presence of additional variables to break the symmetry inherent in such associations.
For instance, if variables $X$ and $Y$ are dependent when conditioning on any set of other variables $S$ (that is, they are persistently associated) and there exists a third variable $Z$ such that (conditional on some $S$) $X$ and $Z$ are independent while $Z$ and $Y$ are dependent (there is something else, independent of $X$, with which $Y$ is associated), then $X$ is dubbed a \emph{potential cause of} $Y$.
In addition to \emph{potential cause}, \textit{Causality} \citep{pearl2009causality} also defines the notions of \emph{genuine cause} and \emph{spurious association}.

In the following subsections, we gradually introduce readers to casual structures and the methods to infer them.
We start with the three fundamental causal structures (\S\ref{sec:structures}) and discussing how larger, more complicated causal structures can be built from the three fundamental building blocks (\S\ref{sec:composite}).
In \S\ref{sec:causaldiscovery}, we present two common approaches for discovering causal structures from purely-observational data: \emph{constraint-based methods} (\S\ref{sec:constraint-based}) and \emph{score-based methods} (\S\ref{sec:score-based}).
Finally, we provide a list of terminology that is used in the field of causal inference, in \S\ref{sec:terminology} and its Table~\ref{tab:terminology}.

\subsection{Basic Causal Structures}\label{sec:structures}

\begin{figure*}
  \centering
  \includegraphics[width=0.745\linewidth]{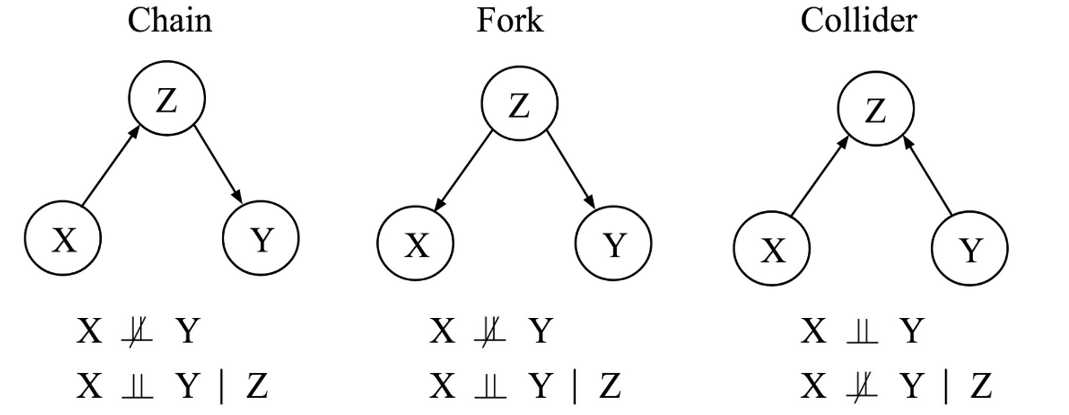}
  \includegraphics[width=0.245\linewidth]{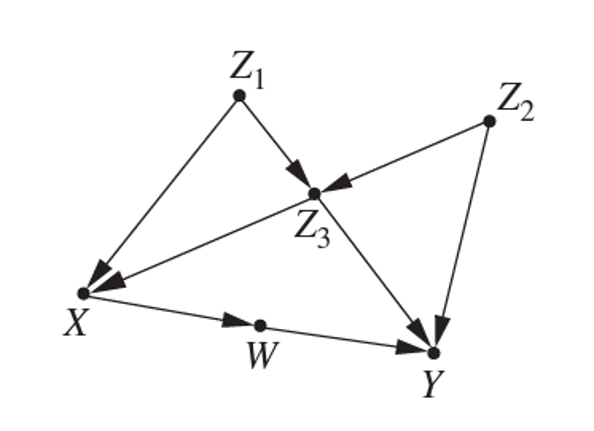}
  \caption{
  Three basic causal models, their (conditional) independencies, and a more complex graph on the right. In a complex graph, variables are potentially connected through multiple paths made of chains, forks, and colliders.
  Following the \emph{d-separation} rules introduced in \S\ref{sec:composite} and Table~\ref{tab:d-separation}, one should find $Z_2 \dep X$, $Z_2 \indep X \mid (Z_3,Z_1)$, and $Z_2 \dep X \mid (Z_3,Z_1,Y)$.}
\label{fig:chain_fork_collider}
\end{figure*}

Causal structures are often represented by Directed Acyclic Graphs (DAGs), which are made of nodes and edges.
A directed edge between nodes suggests the direction of causality, i.e., $A\rightarrow B$ means $A$ causes $B$.
There are three basic causal structures: \emph{chains}, \emph{forks}, and \emph{colliders} (Figure~\ref{fig:chain_fork_collider}).

\begin{itemize}

\begin{table}[]
\centering
\begin{tabular}{ccc}
\hline
         & uncondition & condition \\ \hline
Chain    & unblock     & block     \\ \hline
Fork     & unblock     & block     \\ \hline
Collider & block       & unblock   \\ \hline
\end{tabular}
\caption
{In a DAG with multiple nodes, variables can potentially be connected through multiple paths, with several chains, forks, or colliders.
Two variables are defined to be \emph{d-separated}, i.e., (conditionally) independent, if every path between them is blocked.
A path is blocked or unblocked based on the rules outlined in the above table.
A more detailed description and a case study can be found in \S\ref{sec:composite}.}
\label{tab:d-separation}
\end{table}

\item In the case of a \emph{chain}, $X$ causes ($\rightarrow$) $Z$, and $Z$ causes ($\rightarrow$) $Y$.
In a chain model, $X$ and $Y$ are \emph{not} independent ($X\dep Y$) without conditioning on $Z$.
For example, consider three standing dominoes in order $X$, $Z$, and $Y$.
The falling of $X$ will cause $Z$ to fall, which in turn will cause $Y$ to fall.
However, when we condition on $Z$, the other two variables, $X$ and $Y$, will be independent ($X\indep Y \mid Z$).
In other words, if we let domino $Z$ fall, the subsequent domino $Y$ will fall regardless of whether the prior domino $X$ fell or not.

\item In the case of a \emph{fork}, a single variable $Z$, called a confounder, causally influences two other variables $X$ and $Y$.
For instance, consider the influence of rainy weather ($Z$) on both umbrella sales ($X$) and the number of people jogging outside ($Y$).
On rainy days, more umbrellas are likely to be sold, and less people will go out for a jog.
In a fork model, without conditioning on the confounder $Z$, the other variables, $X$ and $Y$, will be dependent on each other ($X\dep Y$).
If one were to analyze umbrella sales and jogging activity without considering the weather, they will find them to be dependent.
However, once we condition on the confounder $Z$ and compare days with the same weather condition, umbrella sales and jogging activity should be independent of each other ($X\indep Y \mid Z$).

\item A \emph{collider} refers to the case that two variables, $X$ and $Y$, independently cause a third variable $Z$.
Consider the tossing of two fair coins $X$ and $Y$, and a bell $Z$ that rings whenever both coins lands on heads (this example is still valid when $Z$ is a bell that rings whenever at least one of the coins lands on heads, see \citet{pearl2016} for a detailed Bayesian proof).
Without revealing if the bell rings or not, the head/tail states of two coins are independent to each other ($X\indep Y$)---simply as how coin tosses naturally works. 
However, if we condition on the bell $Z$ not ringing, knowing one of the coins landed on heads immediately informs us that the other coin landed on tails ($X\dep Y \mid Z$), otherwise the bell would have rung.
\end{itemize}

These three causal models each encode (conditional) independencies as discussed above and summarized in Figure~\ref{fig:chain_fork_collider}.
Notice that chains and forks share the same (conditional) independencies, while colliders have a different set of (conditional) independencies.
Chains and forks are then considered as the same Markov Equivalence Class (MEC), while colliders belong to a different MEC.
Note that these examples operate under the Markov assumption: $X \indep_\textup{\scriptsize{Graph}} Y \mid Z \Rightarrow X \indep_\textup{\scriptsize{Data}} Y \mid Z$, meaning that the (conditional) independencies encoded in a causal graph should appear in its data.\footnote{Here, we are referring to the global Markov assumption, which is implied by the local Markov assumption.
The local Markov assumption states that given its parents in a
DAG, a node $X$ is independent of all its non-descendants.}

\subsection{Composite Causal Structures}\label{sec:composite}

In cases with more than three variables, such as in the right side of Figure~\ref{fig:chain_fork_collider}, variables can potentially be connected through multiple paths, with several chains, forks, or colliders.
Following the (conditional) independencies encoded by chains, forks, and colliders, a path is \emph{blocked} when conditioning on the middle variable of a chain or a fork and \emph{unblocked} when not conditioning on the middle variable of a chain or a fork.
Furthermore, a path is \emph{blocked} when not conditioning on the middle variable of a collider and \emph{unblocked} when conditioning on the middle variable of a collider. 
Two variables are defined to be \emph{d-separated} if every path between them is blocked, and \emph{d-separated} variables are (conditional) independent.\footnote{With the exception of intransitive cases \citep{pearl2016}.} The above rules are summarized in Table~\ref{tab:d-separation}.

In the Figure~\ref{fig:chain_fork_collider}, one will find $Z_2 \dep X$ without any conditioning, since there is an unblocked chain path $Z_2 \rightarrow Z_3 \rightarrow X$.
One should also find $Z_2 \indep X \mid (Z_3,Z_1)$.
Conditioning on $Z_3$ blocks the $Z_2 \rightarrow Z_3 \rightarrow X$ chain.
Over the $Z_2 \rightarrow Z_3 \leftarrow Z_1 \rightarrow X$ path, although conditioning on $Z_3$ unblocks the $Z_2 \rightarrow Z_3 \leftarrow Z_1$ collider, the conditioning on $Z_1$ blocks the $Z_3 \leftarrow Z_1 \rightarrow X$ fork, making this path blocked.
The remaining $Z_2 \rightarrow Y \leftarrow W \leftarrow X$ path is blocked by the collider $Z_2 \rightarrow Y \leftarrow W$ without conditioning on $Y$.
Similarly, $Z_2 \dep X \mid (Z_3,Z_1,Y)$.

\subsection{Causal Discovery from Observational Data}\label{sec:causaldiscovery}

Causal discovery is most easily achieved through interventions.
However, in observational fields such as astrophysics, interventions are rarely possible \citep{2018P&SS..157..104C}.
Therefore, the field of causal discovery aims to reveal the causal relations between variables from purely observational data without interventions using alternative strategies.
As the universe is one of the most challenging entities for humankind to intervene with, there is perhaps no field of science that stands to gain more from the study of causation than astronomy \citep{Clark:2012}.

\subsubsection{Constraint-based Methods}\label{sec:constraint-based}

One of the most straightforward approaches to discovering causal structures from observational data is conducting conditional independence tests among variables since different MECs encode distinct conditional independencies.\footnote{Here, we adopt the faithfulness assumption, or the converse of the Markov assumption: $X \indep_\textup{Data} Y \mid Z \Rightarrow X \indep_\textup{Graph} Y \mid Z$.}
These approaches are generally referred to as \emph{constraint-based methods}.

A commonly used constraint-based method is the Peter-Clark \citep{spirtes2000causation}, PC, algorithm. 
The PC algorithm consists of three steps:
\begin{enumerate}
    \item Start with a fully-connected, undirected graph among all variables, and remove edges based on conditional independence tests to arrive at a graph skeleton.

    \item Identify colliders with conditional independence tests and orient them.

    \item Orient edges that are incident on colliders such that no new colliders will be constructed.
\end{enumerate}
Note that in addition to the Markov assumption and faithfulness assumption, the PC algorithm further assumes causal sufficiency (i.e., no unobserved confounders) and acyclicity.

Another constraint-based method, the Fast Causal Inference \citep{SpirtesManuscript-SPIAAA}, FCI, algorithm relaxes the assumption of causal sufficiency, allowing unobserved confounders.
The FCI algorithm is based on the same independence testing procedure as the PC algorithm but differs at the stage of labeling and orienting edges.

It has also been proven that PC and FCI algorithms are sound and complete under cyclic settings \citep{Mooij_2020}.
However, these methods only provide a point estimate for the true MEC without quantifying their uncertainties; this becomes particularly problematic when the number of data points is small and the reliability of conditional independence tests degrades.

\subsubsection{Score-based Methods}\label{sec:score-based}

Instead of finding a single causal structure, we can adopt a Bayesian perspective and define a posterior over all possible DAGs, $P(G \mid D)$.
To do this, \emph{score-based methods} assign a numerical score to every DAG given the data.
There are several possible ways to define such a score, such as the Bayesian Information Criterion (BIC) score \citep{Chickering:2002}, generalized score \citep{Huang:2018}, and the Bayesian Gaussian equivalent (BGe) score \citep{Geiger:1994,Geiger:2002,Kuipers:2014}.\footnote{The scores listed above are for continuous data.
The Bayesian Dirichlet equivalent (BDe) score \citep{Heckerman:1995} is one of the scores for discrete or categorical data.}

In an exact posterior approach, one evaluates the chosen score for every possible DAG, i.e., $P(G \mid D)$. 
However, the cost of an exact search grows super-exponentially as the number of variables (nodes) increases. 
For example, the number of possible DAGs for three variables is only 25 but exceeds $4\times10^{18}$ for ten variables\footnote{See the online encyclopedia of integer sequences \citep{OEIS} for the number of DAGs corresponding to $n$ nodes.}, making an exact posterior search quickly computationally intractable.
For the current study, the total number of DAGs for seven variables is 1,138,779,265, which is near the limit of computational feasibility.

As a result, sampling algorithms have been developed to approximate the exact posterior distribution without going over all DAGs. 
Such approximation is often done with Markov Chain Monte Carlo (MCMC) methods, such as the \texttt{MC3} algorithm \citep{Madigan:1995} and \texttt{Gadget} \citep{Viinikka:2020}. 
More recently, \citet{deleu2022daggflownet} developed \texttt{DAG-GFN} as an alternative to MCMC, and showed that \texttt{DAG-GFN} compares favorably against other methods based on MCMC.
Here, we also show that \texttt{DAG-GFN} does recover the exact posterior distribution fairly well under the SMBH--galaxy context in \S\ref{sec:DAG-GFN}.
Some benchmark studies on causal discovery algorithms can be found in several references \citep{emezue2023benchmarking,Vowels2022,Menegozzo2022,runge2019inferring,ahmed2020causalworld,Kalainathan2020,Scutari2014}.

\subsection{List of Terminology}\label{sec:terminology}
See Table~\ref{tab:terminology} for a list of causal inference terminology and their explanations.

\begin{table*}
\centering
\caption{Causal Inference Terminology}
\label{tab:terminology}
\begin{tabular}{lp{11cm}}
\hline
\textbf{Term} & \textbf{Definition} \\
\hline
\textbf{DAG} & A Directed Acyclic Graph (DAG) is a graphical representation of causal relationships between variables, consists of nodes and edges, with each edge directed from one node to another (``directed"), such that following those directions will never form a closed loop (``acyclic").
Directed edges suggest the direction of causality, i.e., $A\rightarrow B$ means $A$ causes $B$.\\ \hline
\textbf{MEC} & A Markov Equivalence Class (MEC) is a set of DAGs that encode the same conditional independence relationships among variables, making them observationally indistinguishable based on data alone.
See Figure~\ref{fig:MEC+DAG} for examples of MECs and their corresponding DAGs.\\ \hline
\textbf{PDAG} & A Partially Directed Acyclic Graph (PDAG) is a graph that contains both directed and undirected edges to represent a MEC.
An undirected edge $A\ \text{---}\ B$ suggests both directions are possible (either $A\rightarrow B$ or $A\leftarrow B$), with the exception when new MEC/conditional independencies are introduced by creating new colliders.
For example, $A\ \text{---}\ B \text{---}\ C$ corresponds to $A\rightarrow C\rightarrow B$, $A\leftarrow C\leftarrow B$, and $A\leftarrow C\rightarrow B$, but \textbf{not} $A\rightarrow C\leftarrow B$.
Examples of PDAGs can be found in the left column of Figure~\ref{fig:MEC+DAG}.\\ \hline
\textbf{PAG} & A Partial Ancestral Graph (PAG) is a graphical model that represents equivalence classes of causal structures when unmeasured confounders or selection biases are present, capturing possible ancestral relationships among variables.
There are two additional edge types: $A\longleftrightarrow B$ corresponding to a confounding relation (i.e., a third variable causes both A and B).
Empty circles ($\circ$) representing uncertainty regarding the ending symbol of the edge (i.e., $A\ \circ\hspace{-1.2mm}\rightarrow B$ may correspond to either $A\rightarrow B$ or to $A\longleftrightarrow B$, but rules out $B\rightarrow A$).
Examples of PAGs can be found in the bottom row of Figure~\ref{fig:PCFCI}.\\ \hline
\textbf{Child} & A variable that is directly influenced by another variable (its causal parent) in a causal relationship.
For example, if $A\rightarrow B\rightarrow C$, then $B$ is the child of $A$, and $C$ is the child of $B$.\\ \hline
\textbf{Parent} & A variable that directly influences another variable (its causal child) in a causal relationship.
For example, if $A\rightarrow B\rightarrow C$, then $A$ is the parent of $B$, and $B$ is the parent of $C$.\\ \hline
\textbf{Ancestor} & A variable that influences another variable either directly or indirectly through one or more causal paths.
For example, if $A\rightarrow B\rightarrow C$, then both $A$ and $B$ are ancestors of $C$.\\ \hline
\textbf{Descendant} & A variable that is influenced by another variable either directly or indirectly through one or more causal paths.
For example, if $A\rightarrow B\rightarrow C$, then both $B$ and $C$ are descendants of $A$.\\ \hline
\textbf{PC, FCI} & Peter-Clark \citep{spirtes2000causation}, PC, algorithm and Fast Causal Inference \citep{SpirtesManuscript-SPIAAA}, FCI, algorithm are two time-tested causal discovery algorithms. 
They are described in detail in \S\ref{sec:constraint-based}.\\ \hline
\textbf{BGe score} & The Bayesian Gaussian equivalent (BGe) score \citep{Geiger:1994,Geiger:2002,Kuipers:2014} gives the exact posterior probabilities of every DAG given the data, $P(G \mid D)$.
The BGe score is explained in detail in \S\ref{sec:score-based} and \S\ref{sec:bge}.\\ \hline
\end{tabular}
\end{table*}

\section{Data}\label{sec:data}

Our data is composed of an up-to-date sample of most known galaxies with dynamically-measured SMBH masses.\footnote{
For our purposes, we consider dynamical (i.e., direct) measurements from published studies that utilize resolved (sphere-of-influence) measurements of stellar dynamics, gas dynamics, maser emission, proper motion, or direct imaging.
}
These include 145 nearby galaxies with a median luminosity distance of 21.5\,Mpc that host SMBHs with directly resolved spheres of influence.
From this parent sample, 101 out of 145 galaxies have all of the seven desired variables of interest for our study: dynamically-measured black hole mass ($M_\bullet$), central stellar velocity dispersion ($\sigma_0$), effective (half-light) radius of the spheroid\footnote{Throughout this paper, we use the terms ``bulge'' and ``spheroid'' interchangeably to refer to the spheroid component of spiral and lenticular galaxies or the entirety of pure elliptical galaxies.} ($R_\mathrm{e}$), the average projected density within $R_\mathrm{e}$ ($\langle{\Sigma_\mathrm{e}}\rangle$), total stellar mass ($M^*$), color ($W2-W3$), and specific star formation rate (sSFR). 

We include the three parameters that define the fundamental plane of elliptical galaxies \citep{Djorgovski:1987}, i.e., central stellar velocity dispersion ($\sigma_0$), effective (half-light) radius of the spheroid\footnote{Here, we use the terms ``bulge'' and ``spheroid'' interchangeably to refer to the spheroid component of spiral and lenticular galaxies or the entirety of pure elliptical galaxies.
} ($R_\mathrm{e}$), and the average projected density within $R_\mathrm{e}$ ($\langle{\Sigma_\mathrm{e}}\rangle$).
$M_\bullet$ values are compiled by a series of progressive studies on black hole mass scaling relations \citep{Graham:2013,Scott:2013,Savorgnan:2013,Savorgnan:2016,Sahu:2019,Sahu:2019b,Sahu:2020,Graham:2023,Davis:2017,Davis:2018,Davis:2019,Davis:2019b,Davis:2023,Davis:2024}.
$\sigma_0$ values are collected from several works \citep{Davis:2017,Davis:2019b,Sahu:2019b}, which are obtained primarily from the HyperLeda database \citep{Makarov:2014} and homogenized to produce an estimate of the mean velocity dispersion within an aperture of 595\,pc.
$R_\mathrm{e}$ and $\langle{\Sigma_\mathrm{e}}\rangle$ measurements were produced via the multi-component decompositions of surface brightness light profiles (primarily of $3.6\,\mu\mathrm{m}$ \textit{Spitzer} Space Telescope imaging) from succeeding works \citep{Savorgnan:2016a,Davis:2019,Sahu:2019,Graham:2023c}.\footnote{
$R_\mathrm{e}$ (and $\langle{\Sigma_\mathrm{e}}\rangle$) is calculated from the equivalent (i.e., geometric-mean) axis surface brightness profile of each galaxy.
Radii computed along the equivalent axis of quasi-elliptical isophotes are equal to $\sqrt{ab}$, where $a$ is the semi-major axis and $b$ is the semi-minor axis of an isophote, and thus produce a circle with the \emph{equivalent} area as the quasi-elliptical isophote.
}

This choice of parameters also allows us to explore the well-known $M_\bullet$--$\sigma_0$ relation \citep{Ferrarese:2000,Gebhardt:2000}.
Indeed, one impetus for this study of SMBH--galaxy causality was the significant difference in intrinsic scatter ($\epsilon$) observed in elliptical galaxies ($\epsilon=0.31$\,dex) vs.\ spiral galaxies ($\epsilon=0.67$\,dex) as determined by \citet{Sahu:2019b}.
This implies that the $M_\bullet$--$\sigma_0$ relation is $\approx$2.3 times less accurate for predicting SMBH masses in spiral galaxies as opposed to elliptical galaxies.
As shown in this work, this difference in the scatter of the relationship between morphological types foreshadows their inherent dichotomy in causal structures.

The remaining parameters we explore concern properties related to the star-formation rate (SFR) in galaxies.
For this, we consider data from the Wide-field Infrared Survey Explorer \citep{Wright:2010}, WISE, to provide the color, total stellar masses ($M^*$), and SFRs for our galaxies.
These WISE values are all compiled from \citet{Graham:2024}: $M^*$ is derived from the prescriptions of \citet{Jarrett:2023} for W1 ($3.4\,\mu\mathrm{m}$) photometry and colors from WISE; and SFRs accounting for activity over the past 100\,Myr is ascertained via the WISE total integrated fluxes as per the calibrations of \citet{Cluver:2017,Cluver:2024}.
For WISE colors, we considered both W1$-$W2 ($3.4\,\mu\mathrm{m}-4.6\,\mu\mathrm{m}$) and W2$-$W3 ($4.6\,\mu\mathrm{m}-12.1\,\mu\mathrm{m}$) colors, but ultimately elected to conduct our analyses with only the latter color, which exhibits a greater range of diversity across morphological classes of galaxies.
Rather than absolute SFR, we convert to specific star-formation rate (sSFR) by normalizing each SFR by the stellar mass of each galaxy (i.e., sSFR $\equiv$ SFR/$M^*$).

\begin{figure*}[h]
  \centering
  \includegraphics[width=\linewidth]{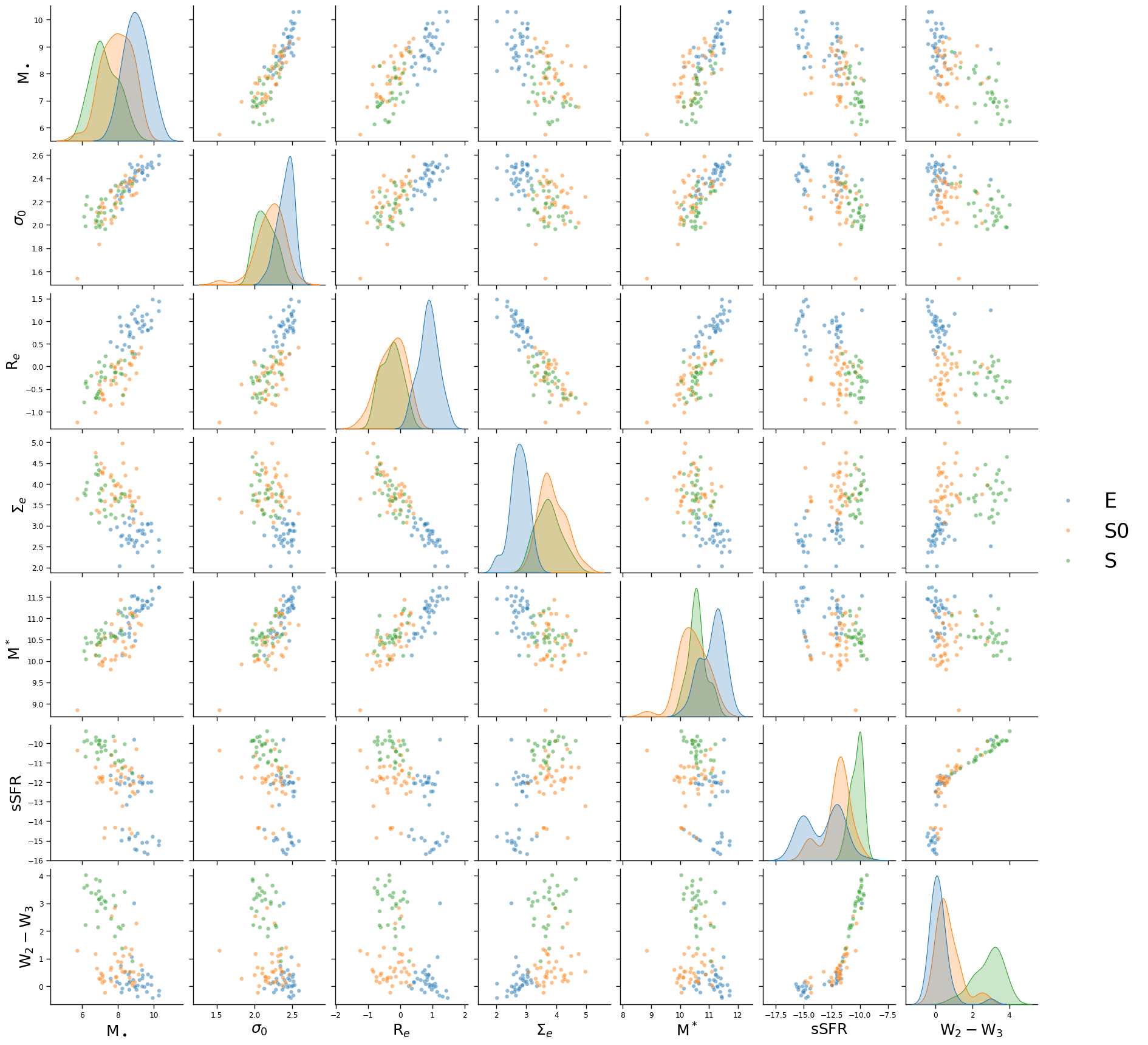}
  \caption{
  A pairplot of the data listed in Table~\ref{tab:sample}.
  }
\label{fig:pairplot}
\end{figure*}

We split the sample of galaxies into three morphological classes:
\begin{itemize}
    \item highly-evolved, massive, gas-poor \underline{elliptical} (E) galaxies, which have been exposed to the full range of feedback and merging processes throughout their long histories spanning large fractions of the age of the Universe,
    \item \underline{spiral} (S) galaxies, at the opposite end of galaxy morphological classification schemes \citep{Jeans:1928,Hubble:1936,Graham:2019}, which are unlikely to have encountered any major mergers and still retain a large fraction of their gas, and
    \item \underline{lenticular} (S0) galaxies, which represent a bridging population between E and S types.
\end{itemize}
Altogether, this gives us a sample of 35 elliptical, 38 lenticular, and 28 spiral galaxies for a total of 101 galaxies, each with six physical measurements of the host galaxy plus a dynamically-measured SMBH mass (see Table~\ref{tab:sample} and its pairplot in Figure~\ref{fig:pairplot}).
Although this division makes our already small sample into even smaller subsets, this exploration of morphologically-distinct causality is the overarching goal of our study.
All morphologies have been determined by the multi-component decompositions of surface brightness light profiles \citep{Savorgnan:2016a,Davis:2019,Sahu:2019,Graham:2023c}.
Our general classification scheme defines elliptical galaxies as spheroids (with or without embedded disk components), lenticular galaxies as spheroids with extended disk components (without spiral structure), and spiral galaxies as disk galaxies (with classical bulges, pseudobulges, or no bulges) with spiral structure.
For our purposes in this study, we have not considered barred morphologies as a distinct classification element.

\startlongtable
\begin{deluxetable*}{lrrrrrrr}
\tablecolumns{8}
\tablecaption{Sample of 101 Galaxies with Dynamical SMBH Mass Measurements}\label{tab:sample}
\tablehead{
\colhead{Galaxy} & \colhead{$\log({M}_\bullet)$} & \colhead{$\log(\sigma_0)$} & \colhead{$\log(R_\mathrm{e})$} & \colhead{$\log(\langle{\Sigma_\mathrm{e}}\rangle$)} & \colhead{W2$-$W3} & \colhead{$\log({M}^*)$} & \colhead{$\log(\mathrm{sSFR})$} \\
\colhead{} & \colhead{[M$_\sun$]} & \colhead{[km\,s$^{-1}$]} & \colhead{[kpc]} & \colhead{[M$_\sun$\,pc$^{-2}$]} & \colhead{[mag]} & \colhead{[M$_\sun$]} & \colhead{[yr$^{-1}$]} \\
\colhead{(1)} & \colhead{(2)} & \colhead{(3)} & \colhead{(4)} & \colhead{(5)} & \colhead{(6)} & \colhead{(7)} & \colhead{(8)}
}
\startdata
\multicolumn8c{35 Elliptical Galaxies}
\\ \hline
IC~1459  & $9.38\pm0.18$  & $2.47\pm0.01$ & $0.89\pm0.09$ & $2.89\pm0.13$ & $0.39\pm0.06$  & $11.24\pm0.08$ & $-12.00\pm0.10$ \\
IC~4296  & $9.10\pm0.09$  & $2.52\pm0.01$ & $0.96\pm0.31$ & $2.87\pm0.09$ & $0.02\pm0.08$  & $11.47\pm0.08$ & $-12.47\pm0.11$ \\
NGC~821  & $7.59\pm0.19$  & $2.30\pm0.01$ & $0.54\pm0.01$ & $2.91\pm0.09$ & $0.27\pm0.13$  & $10.64\pm0.08$ & $-11.65\pm0.11$ \\
NGC~1275 & $8.90\pm0.24$  & $2.39\pm0.02$ & $1.24\pm0.31$ & $2.51\pm0.13$ & $3.00\pm0.04$  & $11.52\pm0.09$ & $-9.81\pm0.11$  \\
NGC~1399 & $8.67\pm0.06$  & $2.52\pm0.01$ & $0.76\pm0.09$ & $3.01\pm0.09$ & $0.17\pm0.07$  & $11.23\pm0.08$ & $-12.72\pm0.13$ \\
NGC~1407 & $9.65\pm0.06$  & $2.42\pm0.01$ & $0.80\pm0.31$ & $3.03\pm0.12$ & $0.07\pm0.09$  & $11.39\pm0.08$ & $-15.69\pm0.44$ \\
NGC~1600 & $10.28\pm0.04$ & $2.52\pm0.01$ & $1.22\pm0.09$ & $2.66\pm0.07$ & $-0.34\pm0.10$ & $11.71\pm0.09$ & $-15.23\pm0.44$ \\
NGC~3091 & $9.61\pm0.02$  & $2.49\pm0.01$ & $1.15\pm0.09$ & $2.61\pm0.17$ & $-0.21\pm0.11$ & $11.39\pm0.09$ & $-15.09\pm0.44$ \\
NGC~3377 & $8.24\pm0.23$  & $2.13\pm0.01$ & $0.36\pm0.01$ & $2.77\pm0.07$ & $-0.09\pm0.08$ & $10.13\pm0.08$ & $-14.43\pm0.44$ \\
NGC~3379 & $8.63\pm0.11$  & $2.31\pm0.00$ & $0.43\pm0.31$ & $3.19\pm0.14$ & $0.12\pm0.05$  & $10.64\pm0.08$ & $-14.94\pm0.44$ \\
NGC~3414 & $8.38\pm0.05$  & $2.38\pm0.01$ & $0.47\pm0.09$ & $3.07\pm0.15$ & $0.50\pm0.07$  & $10.63\pm0.08$ & $-12.05\pm0.10$ \\
NGC~3585 & $8.49\pm0.14$  & $2.33\pm0.01$ & $0.90\pm0.31$ & $2.67\pm0.10$ & $0.06\pm0.09$  & $11.03\pm0.09$ & $-13.16\pm0.28$ \\
NGC~3607 & $8.17\pm0.17$  & $2.35\pm0.01$ & $0.90\pm0.31$ & $2.74\pm0.13$ & $0.73\pm0.06$  & $11.13\pm0.08$ & $-11.73\pm0.09$ \\
NGC~3608 & $8.63\pm0.10$  & $2.29\pm0.01$ & $0.66\pm0.31$ & $2.75\pm0.09$ & $-0.13\pm0.10$ & $10.56\pm0.09$ & $-14.86\pm0.44$ \\
NGC~3842 & $9.94\pm0.12$  & $2.49\pm0.01$ & $1.48\pm0.09$ & $2.03\pm0.08$ & $-0.43\pm0.07$ & $11.45\pm0.09$ & $-14.80\pm0.44$ \\
NGC~3923 & $9.47\pm0.13$  & $2.39\pm0.01$ & $0.92\pm0.09$ & $2.80\pm0.13$ & $-0.03\pm0.08$ & $11.30\pm0.08$ & $-15.60\pm0.44$ \\
NGC~4261 & $9.21\pm0.08$  & $2.47\pm0.01$ & $0.84\pm0.31$ & $2.89\pm0.10$ & $0.22\pm0.08$  & $11.17\pm0.08$ & $-11.99\pm0.10$ \\
NGC~4291 & $8.97\pm0.14$  & $2.47\pm0.01$ & $0.27\pm0.31$ & $3.35\pm0.14$ & $0.00\pm0.09$  & $10.47\pm0.08$ & $-14.77\pm0.44$ \\
NGC~4374 & $8.95\pm0.04$  & $2.44\pm0.00$ & $1.04\pm0.31$ & $2.59\pm0.08$ & $-0.04\pm0.07$ & $11.14\pm0.08$ & $-15.44\pm0.44$ \\
NGC~4472 & $9.36\pm0.03$  & $2.45\pm0.00$ & $1.01\pm0.09$ & $2.81\pm0.08$ & $0.28\pm0.12$  & $11.41\pm0.08$ & $-12.08\pm0.10$ \\
NGC~4473 & $7.95\pm0.22$  & $2.25\pm0.01$ & $0.43\pm0.31$ & $2.96\pm0.08$ & $0.15\pm0.08$  & $10.53\pm0.08$ & $-12.05\pm0.11$ \\
NGC~4486 & $9.85\pm0.02$  & $2.51\pm0.01$ & $0.85\pm0.31$ & $3.05\pm0.08$ & $0.33\pm0.05$  & $11.31\pm0.08$ & $-12.02\pm0.10$ \\
NGC~4552 & $8.67\pm0.04$  & $2.40\pm0.01$ & $0.71\pm0.31$ & $2.68\pm0.09$ & $0.50\pm0.10$  & $10.77\pm0.09$ & $-12.12\pm0.11$ \\
NGC~4621 & $8.59\pm0.04$  & $2.36\pm0.01$ & $0.88\pm0.09$ & $2.58\pm0.10$ & $0.42\pm0.13$  & $10.89\pm0.08$ & $-11.82\pm0.10$ \\
NGC~4649 & $9.67\pm0.10$  & $2.52\pm0.01$ & $0.80\pm0.09$ & $3.04\pm0.09$ & $0.42\pm0.10$  & $11.24\pm0.08$ & $-12.02\pm0.09$ \\
NGC~4697 & $8.10\pm0.02$  & $2.22\pm0.00$ & $1.09\pm0.40$ & $2.03\pm0.08$ & $0.09\pm0.06$  & $10.65\pm0.08$ & $-12.11\pm0.11$ \\
NGC~4889 & $10.29\pm0.33$ & $2.59\pm0.01$ & $1.43\pm0.09$ & $2.38\pm0.09$ & $-0.17\pm0.09$ & $11.72\pm0.09$ & $-15.02\pm0.44$ \\
NGC~5077 & $8.85\pm0.23$  & $2.40\pm0.01$ & $0.64\pm0.09$ & $3.16\pm0.17$ & $0.22\pm0.07$  & $11.02\pm0.08$ & $-11.94\pm0.10$ \\
NGC~5419 & $9.86\pm0.14$  & $2.54\pm0.01$ & $1.01\pm0.01$ & $2.87\pm0.09$ & $0.04\pm0.12$  & $11.64\pm0.08$ & $-12.47\pm0.12$ \\
NGC~5576 & $8.20\pm0.10$  & $2.26\pm0.01$ & $0.76\pm0.09$ & $2.52\pm0.09$ & $-0.23\pm0.05$ & $10.70\pm0.08$ & $-15.00\pm0.44$ \\
NGC~5846 & $9.04\pm0.04$  & $2.38\pm0.01$ & $0.98\pm0.31$ & $2.64\pm0.10$ & $-0.13\pm0.08$ & $11.18\pm0.09$ & $-15.48\pm0.44$ \\
NGC~6251 & $8.77\pm0.14$  & $2.49\pm0.03$ & $1.16\pm0.09$ & $2.66\pm0.09$ & $1.05\pm0.04$  & $11.51\pm0.08$ & $-11.49\pm0.09$ \\
NGC~7052 & $9.35\pm0.02$  & $2.45\pm0.02$ & $0.77\pm0.09$ & $3.04\pm0.07$ & $0.58\pm0.05$  & $11.22\pm0.08$ & $-11.69\pm0.10$ \\
NGC~7619 & $9.35\pm0.10$  & $2.50\pm0.01$ & $1.11\pm0.31$ & $2.52\pm0.07$ & $-0.01\pm0.12$ & $11.29\pm0.08$ & $-15.11\pm0.44$ \\
NGC~7768 & $9.10\pm0.15$  & $2.46\pm0.02$ & $1.32\pm0.31$ & $2.36\pm0.09$ & $-0.38\pm0.05$ & $11.44\pm0.09$ & $-14.63\pm0.44$ \\
\hline
\multicolumn8c{38 Lenticular Galaxies}
\\ \hline
NGC~404  & $5.74\pm0.10$ & $1.54\pm0.04$ & $-1.24\pm0.31$ & $3.64\pm0.12$ & $1.28\pm0.05$  & $8.85\pm0.09$  & $-10.36\pm0.16$ \\
NGC~524  & $8.68\pm0.10$ & $2.37\pm0.01$ &  $0.04\pm0.31$ & $3.83\pm0.07$ & $0.52\pm0.06$  & $11.10\pm0.08$ & $-12.16\pm0.10$ \\
NGC~1023 & $7.62\pm0.04$ & $2.29\pm0.01$ & $-0.41\pm0.09$ & $4.21\pm0.09$ & $0.18\pm0.07$  & $10.61\pm0.08$ & $-12.35\pm0.11$ \\
NGC~1194 & $7.82\pm0.04$ & $2.17\pm0.07$ & $-0.04\pm0.40$ & $3.96\pm0.09$ & $2.83\pm0.04$  & $10.46\pm0.08$ & $-9.87\pm0.09$  \\
NGC~1316 & $8.16\pm0.22$ & $2.35\pm0.01$ &  $0.14\pm0.31$ & $3.94\pm0.30$ & $0.65\pm0.05$  & $11.43\pm0.08$ & $-11.96\pm0.09$ \\
NGC~1332 & $9.16\pm0.06$ & $2.47\pm0.02$ &  $0.28\pm0.40$ & $3.68\pm0.10$ & $0.42\pm0.05$  & $10.88\pm0.08$ & $-11.70\pm0.11$ \\
NGC~1374 & $8.76\pm0.04$ & $2.25\pm0.01$ &  $0.03\pm0.31$ & $3.35\pm0.08$ & $0.12\pm0.07$  & $10.33\pm0.08$ & $-14.63\pm0.44$ \\
NGC~2549 & $7.14\pm0.23$ & $2.15\pm0.01$ & $-0.73\pm0.09$ & $4.25\pm0.13$ & $0.33\pm0.06$  & $9.97\pm0.08$  & $-11.97\pm0.15$ \\
NGC~2778 & $7.14\pm0.43$ & $2.19\pm0.01$ & $-0.63\pm0.31$ & $3.85\pm0.14$ & $0.12\pm0.05$  & $9.89\pm0.08$  & $-11.70\pm0.17$ \\
NGC~2787 & $7.60\pm0.05$ & $2.28\pm0.01$ & $-0.86\pm0.31$ & $4.16\pm0.16$ & $0.59\pm0.04$  & $9.80\pm0.08$  & $-11.89\pm0.14$ \\
NGC~3115 & $8.94\pm0.31$ & $2.42\pm0.01$ &  $0.19\pm0.09$ & $3.58\pm0.08$ & $0.14\pm0.12$  & $10.63\pm0.08$ & $-12.60\pm0.13$ \\
NGC~3245 & $8.30\pm0.11$ & $2.32\pm0.02$ & $-0.63\pm0.09$ & $4.50\pm0.10$ & $1.09\pm0.04$  & $10.45\pm0.08$ & $-11.22\pm0.09$ \\
NGC~3384 & $7.02\pm0.20$ & $2.16\pm0.01$ & $-0.52\pm0.09$ & $4.29\pm0.08$ & $0.24\pm0.05$  & $10.37\pm0.08$ & $-11.85\pm0.11$ \\
NGC~3489 & $6.76\pm0.06$ & $2.02\pm0.01$ & $-1.02\pm0.31$ & $4.74\pm0.09$ & $1.16\pm0.04$  & $10.14\pm0.08$ & $-11.24\pm0.09$ \\
NGC~3665 & $8.76\pm0.10$ & $2.33\pm0.02$ &  $0.33\pm0.31$ & $3.57\pm0.09$ & $1.33\pm0.04$  & $11.13\pm0.08$ & $-11.27\pm0.09$ \\
NGC~3998 & $8.42\pm0.18$ & $2.42\pm0.02$ & $-0.51\pm0.40$ & $4.20\pm0.10$ & $1.39\pm0.05$  & $10.30\pm0.08$ & $-11.18\pm0.09$ \\
NGC~4026 & $8.26\pm0.11$ & $2.24\pm0.01$ & $-0.83\pm0.40$ & $4.97\pm0.13$ & $0.52\pm0.05$  & $10.18\pm0.08$ & $-13.23\pm0.39$ \\
NGC~4339 & $7.63\pm0.12$ & $2.05\pm0.01$ & $-0.31\pm0.31$ & $3.48\pm0.10$ & $0.67\pm0.14$  & $10.02\pm0.08$ & $-11.16\pm0.12$ \\
NGC~4342 & $8.65\pm0.18$ & $2.38\pm0.01$ & $-0.29\pm0.31$ & $3.68\pm0.07$ & $0.31\pm0.04$  & $10.10\pm0.08$ & $-14.40\pm0.44$ \\
NGC~4350 & $8.87\pm0.14$ & $2.26\pm0.01$ &  $0.20\pm0.31$ & $3.08\pm0.07$ & $0.51\pm0.06$  & $10.35\pm0.08$ & $-12.05\pm0.10$ \\
NGC~4371 & $6.83\pm0.07$ & $2.11\pm0.01$ & $-0.15\pm0.31$ & $3.37\pm0.19$ & $0.64\pm0.09$  & $10.38\pm0.08$ & $-12.02\pm0.11$ \\
NGC~4429 & $8.18\pm0.03$ & $2.24\pm0.01$ & $-0.05\pm0.31$ & $3.76\pm0.08$ & $0.86\pm0.05$  & $10.75\pm0.08$ & $-11.59\pm0.09$ \\
NGC~4434 & $7.84\pm0.05$ & $2.07\pm0.01$ & $-0.25\pm0.31$ & $3.59\pm0.07$ & $0.00\pm0.06$  & $10.03\pm0.08$ & $-14.33\pm0.44$ \\
NGC~4459 & $7.83\pm0.08$ & $2.24\pm0.01$ & $-0.01\pm0.31$ & $3.69\pm0.11$ & $1.17\pm0.10$  & $10.56\pm0.08$ & $-11.12\pm0.09$ \\
NGC~4526 & $8.66\pm0.01$ & $2.35\pm0.02$ &  $0.06\pm0.31$ & $3.73\pm0.10$ & $1.14\pm0.05$  & $10.84\pm0.08$ & $-11.38\pm0.09$ \\
NGC~4564 & $7.90\pm0.12$ & $2.19\pm0.01$ & $-0.38\pm0.09$ & $3.96\pm0.09$ & $0.31\pm0.05$  & $10.12\pm0.08$ & $-12.54\pm0.13$ \\
NGC~4578 & $7.28\pm0.08$ & $2.05\pm0.02$ & $-0.31\pm0.31$ & $3.58\pm0.07$ & $-0.24\pm0.10$ & $10.05\pm0.08$ & $-14.35\pm0.44$ \\
NGC~4596 & $7.90\pm0.20$ & $2.15\pm0.01$ & $-0.13\pm0.09$ & $3.64\pm0.07$ & $0.34\pm0.08$  & $10.54\pm0.08$ & $-11.84\pm0.09$ \\
NGC~4742 & $7.13\pm0.15$ & $2.01\pm0.01$ & $-0.61\pm0.31$ & $4.28\pm0.09$ & $0.40\pm0.05$  & $9.99\pm0.09$  & $-11.79\pm0.15$ \\
NGC~4762 & $7.24\pm0.05$ & $2.15\pm0.01$ & $-0.74\pm0.31$ & $4.38\pm0.07$ & $0.18\pm0.07$  & $10.56\pm0.08$ & $-14.86\pm0.44$ \\
NGC~5018 & $8.00\pm0.08$ & $2.33\pm0.01$ &  $0.05\pm0.31$ & $4.00\pm0.09$ & $0.89\pm0.07$  & $11.10\pm0.08$ & $-11.40\pm0.09$ \\
NGC~5128 & $7.65\pm0.13$ & $2.01\pm0.03$ &  $0.04\pm0.40$ & $3.75\pm0.07$ & $2.53\pm0.03$  & $10.86\pm0.08$ & $-10.61\pm0.09$ \\
NGC~5252 & $9.03\pm0.35$ & $2.27\pm0.06$ & $-0.15\pm0.31$ & $4.37\pm0.09$ & $2.27\pm0.04$  & $11.05\pm0.08$ & $-10.30\pm0.09$ \\
NGC~5813 & $8.83\pm0.04$ & $2.37\pm0.01$ &  $0.32\pm0.31$ & $3.44\pm0.10$ & $0.03\pm0.12$  & $11.10\pm0.08$ & $-12.49\pm0.12$ \\
NGC~5845 & $8.41\pm0.16$ & $2.36\pm0.01$ & $-0.20\pm0.31$ & $3.70\pm0.11$ & $0.53\pm0.04$  & $10.14\pm0.08$ & $-11.75\pm0.11$ \\
NGC~6861 & $9.30\pm0.22$ & $2.59\pm0.02$ &  $0.41\pm0.31$ & $3.29\pm0.14$ & $0.76\pm0.05$  & $10.84\pm0.08$ & $-11.81\pm0.09$ \\
NGC~7332 & $7.06\pm0.20$ & $2.11\pm0.01$ & $-0.59\pm0.40$ & $4.49\pm0.10$ & $0.46\pm0.05$  & $10.48\pm0.08$ & $-11.78\pm0.11$ \\
NGC~7457 & $6.96\pm0.26$ & $1.83\pm0.02$ & $-0.40\pm0.31$ & $3.32\pm0.10$ & $0.27\pm0.11$  & $9.92\pm0.08$  & $-11.74\pm0.14$ \\
\hline
\multicolumn8c{28 Spiral Galaxies}
\\ \hline
Circinus & $6.22\pm0.08$ & $2.17\pm0.05$ & $-0.34\pm0.03$ & $3.86\pm0.17$ & $4.02\pm0.03$ & $10.04\pm0.09$ & $-9.37\pm0.13$  \\
IC~2560  & $6.51\pm0.09$ & $2.14\pm0.01$ & $-0.21\pm0.03$ & $3.29\pm0.15$ & $3.38\pm0.04$ & $10.43\pm0.08$ & $-10.01\pm0.09$ \\
NGC~224  & $8.15\pm0.19$ & $2.19\pm0.01$ & $-0.16\pm0.00$ & $3.67\pm0.08$ & $2.08\pm0.04$ & $10.71\pm0.08$ & $-10.94\pm0.09$ \\
NGC~253  & $7.00\pm0.30$ & $1.98\pm0.08$ & $-0.33\pm0.01$ & $3.61\pm0.07$ & $3.81\pm0.04$ & $10.43\pm0.08$ & $-9.85\pm0.09$  \\
NGC~1097 & $8.38\pm0.09$ & $2.29\pm0.01$ & $0.13\pm0.07$  & $3.75\pm0.07$ & $3.41\pm0.04$ & $11.22\pm0.08$ & $-10.04\pm0.10$ \\
NGC~1300 & $7.86\pm0.31$ & $2.34\pm0.06$ & $-0.13\pm0.10$ & $3.19\pm0.09$ & $2.91\pm0.04$ & $10.56\pm0.08$ & $-10.41\pm0.09$ \\
NGC~1320 & $6.77\pm0.16$ & $2.04\pm0.04$ & $-0.70\pm0.07$ & $4.24\pm0.09$ & $3.34\pm0.04$ & $10.13\pm0.09$ & $-9.68\pm0.10$  \\
NGC~1398 & $8.03\pm0.08$ & $2.29\pm0.04$ & $0.09\pm0.04$  & $3.58\pm0.17$ & $2.14\pm0.04$ & $11.17\pm0.08$ & $-10.88\pm0.09$ \\
NGC~2273 & $6.95\pm0.06$ & $2.15\pm0.03$ & $-0.55\pm0.03$ & $3.77\pm0.15$ & $3.14\pm0.04$ & $10.43\pm0.08$ & $-10.02\pm0.09$ \\
NGC~2960 & $7.07\pm0.04$ & $2.22\pm0.04$ & $-0.13\pm0.05$ & $3.89\pm0.09$ & $2.98\pm0.04$ & $10.72\pm0.08$ & $-10.42\pm0.09$ \\
NGC~2974 & $8.23\pm0.05$ & $2.37\pm0.01$ & $-0.17\pm0.01$ & $3.76\pm0.12$ & $1.36\pm0.08$ & $10.61\pm0.08$ & $-11.28\pm0.09$ \\
NGC~3031 & $7.83\pm0.09$ & $2.18\pm0.01$ & $-0.14\pm0.01$ & $3.65\pm0.08$ & $1.80\pm0.03$ & $10.57\pm0.08$ & $-11.00\pm0.09$ \\
NGC~3079 & $6.28\pm0.30$ & $2.24\pm0.03$ & $-0.46\pm0.05$ & $4.04\pm0.17$ & $3.64\pm0.04$ & $10.41\pm0.08$ & $-9.84\pm0.09$  \\
NGC~3227 & $7.25\pm0.25$ & $2.10\pm0.02$ & $0.01\pm0.03$  & $3.38\pm0.13$ & $3.09\pm0.04$ & $10.65\pm0.08$ & $-10.11\pm0.09$ \\
NGC~3368 & $6.89\pm0.09$ & $2.07\pm0.01$ & $-0.60\pm0.02$ & $4.22\pm0.14$ & $2.13\pm0.04$ & $10.55\pm0.08$ & $-10.81\pm0.09$ \\
NGC~3627 & $6.94\pm0.05$ & $2.10\pm0.02$ & $-0.71\pm0.07$ & $4.33\pm0.17$ & $3.44\pm0.04$ & $10.63\pm0.08$ & $-10.06\pm0.09$ \\
NGC~4151 & $7.29\pm0.37$ & $1.96\pm0.05$ & $-0.25\pm0.02$ & $3.97\pm0.13$ & $2.82\pm0.04$ & $10.53\pm0.08$ & $-9.89\pm0.09$  \\
NGC~4258 & $7.61\pm0.01$ & $2.12\pm0.02$ & $-0.01\pm0.06$ & $3.27\pm0.07$ & $2.44\pm0.04$ & $10.56\pm0.08$ & $-10.59\pm0.09$ \\
NGC~4303 & $6.78\pm0.04$ & $1.98\pm0.04$ & $-0.70\pm0.02$ & $4.65\pm0.07$ & $3.87\pm0.04$ & $10.71\pm0.09$ & $-9.86\pm0.10$  \\
NGC~4388 & $6.95\pm0.09$ & $2.00\pm0.04$ & $0.09\pm0.02$  & $3.07\pm0.19$ & $3.15\pm0.04$ & $10.12\pm0.08$ & $-9.85\pm0.09$  \\
NGC~4501 & $7.31\pm0.08$ & $2.22\pm0.02$ & $0.22\pm0.04$  & $3.22\pm0.07$ & $3.05\pm0.04$ & $10.89\pm0.08$ & $-10.35\pm0.09$ \\
NGC~4594 & $8.81\pm0.03$ & $2.35\pm0.01$ & $0.28\pm0.02$  & $3.48\pm0.08$ & $0.90\pm0.05$ & $11.06\pm0.08$ & $-11.55\pm0.09$ \\
NGC~4699 & $8.28\pm0.05$ & $2.28\pm0.02$ & $-0.64\pm0.06$ & $3.25\pm0.18$ & $2.20\pm0.04$ & $11.06\pm0.08$ & $-10.86\pm0.09$ \\
NGC~4736 & $6.83\pm0.10$ & $2.03\pm0.01$ & $-0.64\pm0.01$ & $4.46\pm0.10$ & $2.71\pm0.04$ & $10.38\pm0.08$ & $-10.46\pm0.09$ \\
NGC~4826 & $6.20\pm0.11$ & $1.99\pm0.02$ & $-0.38\pm0.01$ & $3.73\pm0.11$ & $2.21\pm0.04$ & $10.54\pm0.08$ & $-10.68\pm0.09$ \\
NGC~4945 & $6.12\pm0.30$ & $2.07\pm0.06$ & $-0.80\pm0.14$ & $3.78\pm0.07$ & $3.56\pm0.03$ & $10.23\pm0.08$ & $-9.91\pm0.09$  \\
NGC~7582 & $7.74\pm0.18$ & $2.07\pm0.02$ & $-0.32\pm0.11$ & $4.08\pm0.17$ & $3.29\pm0.04$ & $10.59\pm0.08$ & $-9.64\pm0.09$  \\
UGC~3789 & $7.07\pm0.04$ & $2.03\pm0.05$ & $-0.24\pm0.01$ & $3.59\pm0.13$ & $3.22\pm0.04$ & $10.51\pm0.08$ & $-10.19\pm0.09$ \\
\enddata
\tablecomments{
\textbf{Column~(1):} galaxy name.
\textbf{Column~(2):} black hole mass.
\textbf{Column~(3):} central stellar velocity dispersion.
\textbf{Column~(4):} equivalent-axis, effective (half-light) radius of the spheroid component.
\textbf{Column~(5):} average projected density within $R_\mathrm{e}$.
\textbf{Column~(6):} WISE W2$-$W3 color.
\textbf{Column~(7):} galaxy stellar mass.
\textbf{Column~(8):} specific star-formation rate, i.e., $\log(\mathrm{sSFR})\equiv\log(\mathrm{SFR})-\log(M^*)$.
}
\end{deluxetable*}

As can be seen from the shape of the data in Figure~\ref{fig:pairplot}, the data is predominantly characterized by linear relations and appears normally distributed in their logarithmic form, which underpins the general assumption for calculating BGe scores.\footnote{The BGe score has also been shown to perform well in real-world scenarios where the data is not strictly linear and Gaussian, for example, in benchmark studies \citep{deleu2022daggflownet,emezue2023benchmarking} using real flow cytometry data \citep{Sachs:2005}, as well as simulated noisy data following an Erd\H{o}s-R\'{e}nyi model \citep{Erdos:1960}.}

\subsection{Notes on Sample Selection}

Our sample is initially derived from a total sample of 145 galaxies that host directly-measured SMBHs, 103 of which (with WISE luminosities) are listed in \citet{Graham:2024}.
We reduced the sample further down to 101 galaxies by excluding the galaxies NGC~4395 and NGC~6926 because they are bulge-less galaxies and hence lack a \citet{Sersic:1963} bulge component (including $R_\mathrm{e}$ values).
Readers are invited to read \citet[][\S2]{Graham:2024} for a detailed description of the galaxies, SMBHs, stellar masses, SFRs, etc.
Extended accounts of the provenances behind the directly-measured SMBH masses and photometric multi-component decompositions are given in \citet{Sahu:2020} and \citet{Graham:2023}, respectively.

In the first instance, our sample is selected from the available pool of directly-measured SMBHs in the literature, which numbers at only 145 galaxies at last check.
Because SMBHs generally scale along with the stellar mass of their host galaxies \citep{Davis:2018}, this leads to an unavoidable selection bias where only the closest and/or most massive galaxies have dynamically-measurable SMBHs (i.e., no galaxy in our sample is further than NGC~7768, at a luminosity distance of 108.2\,Mpc, or a redshift of $z = 0.02424$ assuming the \citealt{Planck:2020} cosmological parameters).
This is more noticeable when considering the spiral galaxies; our sample is primarily composed of massive spiral galaxies (i.e., earlier/redder types) and is devoid of irregular galaxies.
The latest type galaxy in our sample is NGC~3079, which is an SBc or $T=6.4\pm1.1$ Hubble type galaxy \citep{Makarov:2014}.
Recently, \citet{Winkel:2024} have shown from their study of direct black hole mass measurements that active and quiescent galaxies follow the same black hole mass scaling relations, which strengthens the merits of applying local relations to high-$z$ AGNs.

Our sample contains one dwarf galaxy, which is the dwarf lenticular galaxy NGC~404 with $M^*=(7.08\pm1.47)\times10^8\,\mathrm{M}_\sun$.
Not surprisingly, it also hosts the least massive black hole in our sample, $M_\bullet=(5.50\pm1.27)\times10^5\,\mathrm{M}_\sun$, which is near the intermediate-mass black hole (IMBH) range ($10^2\leq \mathrm{M}_{\sun}<10^5$).
Notably, we lack any dwarf elliptical galaxies.
If it were possible to include them, they would likely warrant being segregated from other ellipticals due to their distinctive diminutive masses and higher sSFRs.
This is something we plan on testing in a similar causal study using simulated galaxies.
For now, the lack of more dwarf galaxies or any IMBHs is not detrimental to our study; \citet{Limberg:2024} recently demonstrated that dwarf galaxies and IMBHs appear to follow the $M_\bullet$--$\sigma_0$ and $M_\bullet$--$M^*$ relations extrapolated from local massive galaxies.

The second criteria for inclusion is the availability of mid-infrared luminosities from WISE.
This provides us with a robust tracer of dust-obscured star formation activity.
It is helpful for us to have consistency regarding the source of imaging when calculating stellar masses.
Indeed, \citet{Sahu:2023} revealed that inconsistencies between stellar mass derivations across mixed samples led to the misled claim \citep{Shankar:2016} that galaxies with directly-measured SMBHs represented a biased sample relative to the general population.
Although, mid-infrared sample selection is known for its propensity to miss low-mass dust-poor S0 galaxies with low sSFRs \citep{Eales:2018,Graham:2024}.
However, this effect will be somewhat mitigated in our sample because our galaxies are all nearby.
These galaxies hold some significance as being possible primordial galaxies that have avoided mergers \citep{Eggen:1962,Harikane:2023,Graham:2023e,Graham:2024}.
If so, that would negate the effect of any causal mechanism from hierarchical growth for these galaxies, which would be another prime opportunity for testing with galaxy simulations.

Despite the inherent limitations of restricting our data to a small sample, it is crucial for our study of causality to use only directly-measured black holes.
One of the primary focuses of our study is to determine the causal direction in the $M_\bullet$--$\sigma_0$ relation.
Because of this, we are restricted to using only directly-measured black holes, i.e., black holes masses that are not derived from the $M_\bullet$--$\sigma_0$ relation.
For example, the $M_\bullet$--$\sigma_0$ relation governs all black hole masses that are calculated using single-epoch spectra or reverberation mapping, because both of these temporal methods require calibrating their viral products with the $M_\bullet$--$\sigma_0$ relation \citep{Graham:2011}.
Therefore, any causal study we might attempt using indirect methods of measuring black holes masses, that are at their heart calibrated to the $M_\bullet$--$\sigma_0$ relation, will inevitably be biased by the artificial $\sigma_0\rightarrow M_\bullet$ causal direction.

\section{Exact Posterior Methodology}\label{sec:methodology}
\subsection{General Description}
To represent the causal structure of the dataset, we use Directed Acyclic Graphs (DAGs).
Each DAG encodes a set of conditional independencies, and DAGs that encode the same conditional independencies belong to the same Markov Equivalence Class (MEC).\footnote{See \S\ref{sec:primer} for a brief introduction to causal inference.}
This choice assumes that no cyclical dependencies between variables exist.
This is a reasonable assumption, given the clear differences in gas fractions and merger histories between the different morphological classes (see \S\ref{sec:cyclicity} for more details).
To achieve a purely data-driven study, we adopt a uniform prior, giving an equal prior probability, $P(G)$, to every one of the nearly $1.14\times10^9$ possible DAGs \citep{OEIS}.
We calculate the exact posterior probabilities of every DAG given the data, $P(G \mid D)$, using the Bayesian Gaussian equivalent (BGe) score \citep{Geiger:1994,Geiger:2002,Kuipers:2014}.
The BGe score gives the marginal likelihood by examining conditional independencies and ensures that DAGs belonging to the same MEC are scored equally.

\subsection{Theory}
\label{sec:bge}
The posterior probability of a graph given the data $P(G \mid D)$ is proportional to the posterior probability of the data given a graph $P(D \mid G)$ under a uniform prior, through Bayes' rule $P(G\mid D) \propto P(D\mid G)P(G)$.
Under the assumption of linear and Gaussian data, the BGe score gives the marginal likelihood that the distribution of the data sample $d=\{\mathbf{x}_{1},\ldots,\mathbf{x}_{N}\}$ of $N$ variables is faithful (i.e., the data satisfies only and all the conditional independencies encoded by the DAG) to a hypothetical DAG model $m^h$ as a product of local scores:
\begin{equation}
\label{bgescore} p\bigl(d\mid\mh\bigr) = \prod_{i=1}^{n}
\frac{p(d^{\Pa_i\cup\{X_i\}}\mid\mch
)}{p(d^{\Pa_i}\mid\mch)},
\end{equation}
where $\Pa_i$ is the parent variables of the vertex $i$, and $d^Y$ is the data restricted to the subset of data $Y$. 
The modularity (i.e., the full score is a product of local scores over all vertices $i$) of local scores ensures that all DAGs in the same MEC are scored equally, and simplifies the posterior calculation over a large amount of DAGs.
The local scores are further characterized by 
\begin{multline}
\label{pdYresult} %
p\bigl(d^{\Y}\mid\mch\bigr)
= \biggl(\frac
{\am}{N+\am}
\biggr)^{{l}/{2}} \\
\frac{\Gamma_{l} ({(N+\aw-n+l)}/{2}
)}{\pi^{
{lN}/{2}}\Gamma_{l} ({(\aw-n+l)}/{2} )} 
\frac{\vert T_{\Y\Y}
\vert^{{(\aw-n+l)}/{2}}}{\vert R_{\Y\Y} \vert^{{(N+\aw-n+l)}/{2}}}.
\end{multline}
The detailed explanation and full derivation of Equations~\ref{bgescore} and \ref{pdYresult} can be found in \citet{Kuipers:2014}. 
Empirically, many causal discovery methods based on the BGe score have been proven to successfully recover the ground truth causal structures in benchmark tests \citep{deleu2022daggflownet,emezue2023benchmarking}.

\subsection{Calculating Exact Posteriors}
The steps of calculating exact posteriors can be summarized in the following steps:
\begin{enumerate}
  \item Generate all possible DAGs for $N$ variables represented by $N\times N$ adjacency matrices $A$, with $A_{i,j}=1$ if there is an arrow from node $i$ to node $j$. 
  \item For every DAG, generate its transitive closure represented by an adjacency matrix.
  \item Calculate the posterior probability for every DAG given the data with the BGe score following Equation~\ref{bgescore} (the sum of the scores over all DAGs is equal to unity by construction).
  \item Perform a weighted average on all DAG adjacency matrices according to their posterior probabilities to get the matrix of edge marginals.
  \item Perform a weighted average on all transitive closure adjacency matrices according to their posterior probabilities to get the matrix of path marginals.
\end{enumerate}
For a given value of $N$, steps 1 and 2 only need to be done once (i.e., the possible DAGs for $N$ variables are unique), and only steps 3--5 need to be repeated for different datasets.

In this work, steps 2--5 are coded in a highly optimized and parallelized way on graphics processing units (GPUs), powered by a \texttt{Python} package \texttt{JAX} \citep{jax2018github}.
The calculation of transitive closure adopts Warshall's algorithm \citep{Warshall:1962}.
The MECs for analysis are generated with a \texttt{Python} package \texttt{causal-learn} \citep{causallearn}.
The visualization of causal graphs is made possible through \texttt{Python} packages \texttt{NetworkX} \citep{networkx} and \texttt{PyGraphviz}\footnote{\url{https://pygraphviz.github.io/}}.

\section{A Compendium of Causal Structures}\label{sec:compendium}
\begin{figure*}
\centering
\begin{tabular}{c|c|c}
  & Most probable MEC & Corresponding DAGs \\ \hline
E & \raisebox{-0.5\height}{\includegraphics[width=0.165\textwidth]{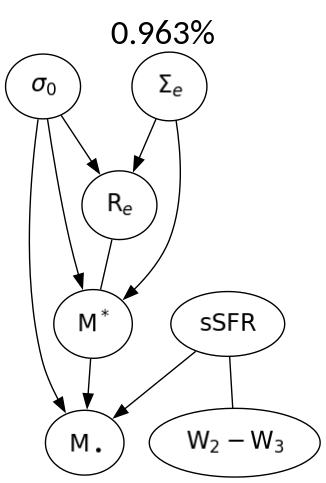}} &
\begin{tabular}{cc}
\includegraphics[width=0.15\textwidth]{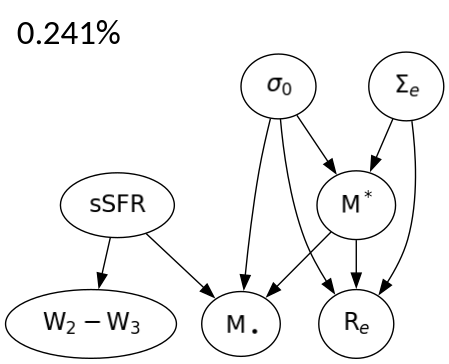} &
\includegraphics[width=0.15\textwidth]{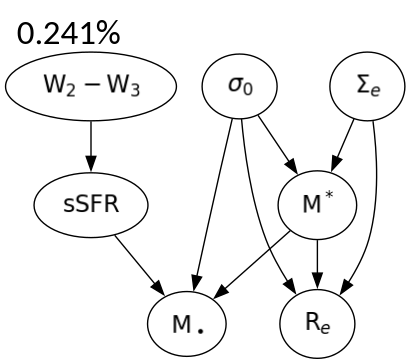} \\
\includegraphics[width=0.15\textwidth]{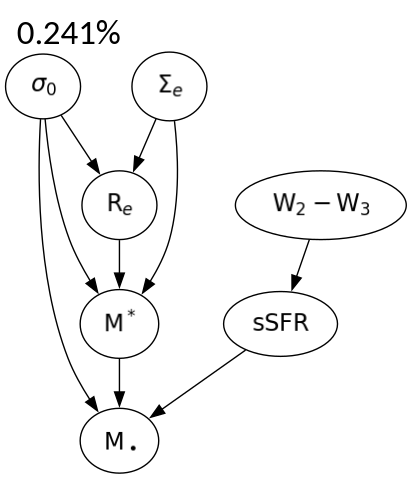} &
\includegraphics[width=0.115\textwidth]{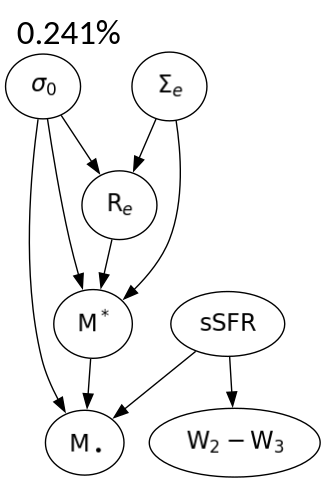} \\
\end{tabular} \\ \hline

S0 & \raisebox{-0.5\height}{\includegraphics[width=0.265\textwidth]{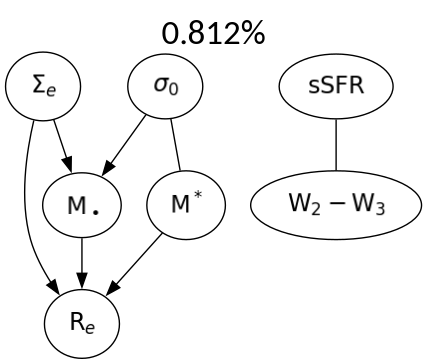}} &
\begin{tabular}{cc}
\includegraphics[width=0.165\textwidth]{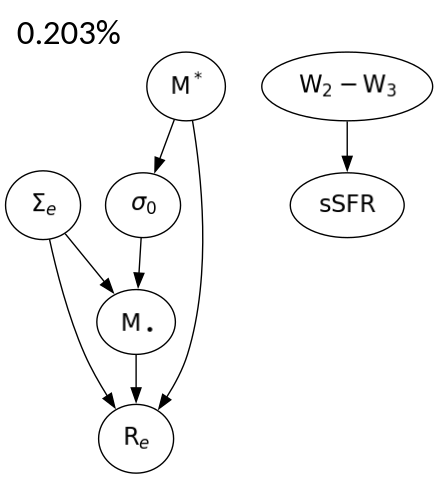} &
\includegraphics[width=0.165\textwidth]{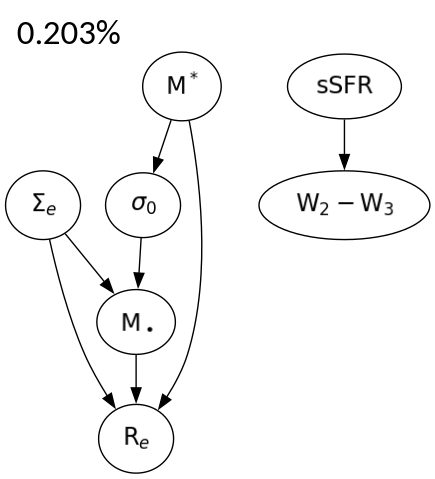} \\
\includegraphics[width=0.165\textwidth]{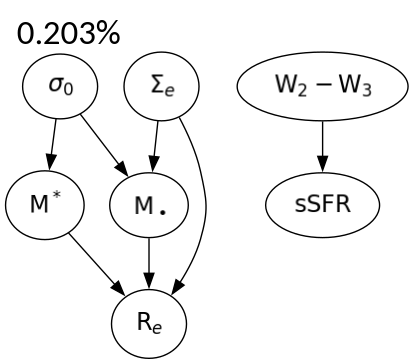} &
\includegraphics[width=0.165\textwidth]{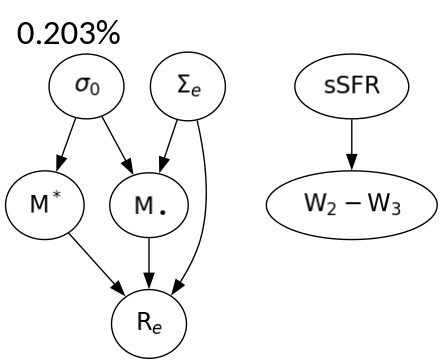} \\
\end{tabular} \\ \hline

S & \raisebox{-0.5\height}{\includegraphics[width=0.315\textwidth]{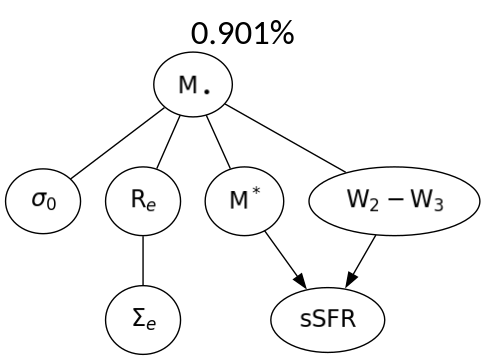}} &
\begin{tabular}{ccc}
\includegraphics[width=0.115\textwidth]{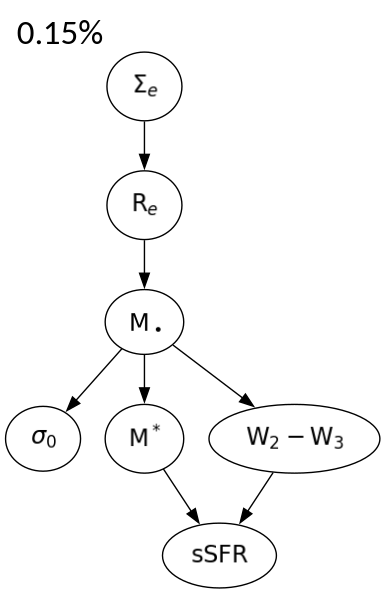} &
\includegraphics[width=0.115\textwidth]{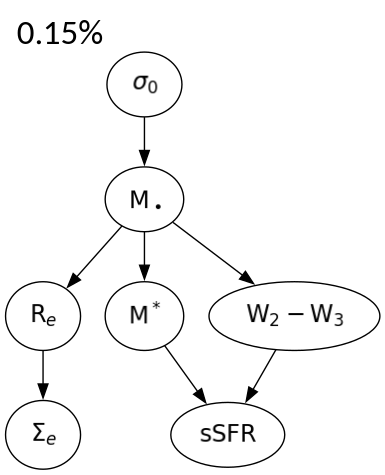} &
\includegraphics[width=0.115\textwidth]{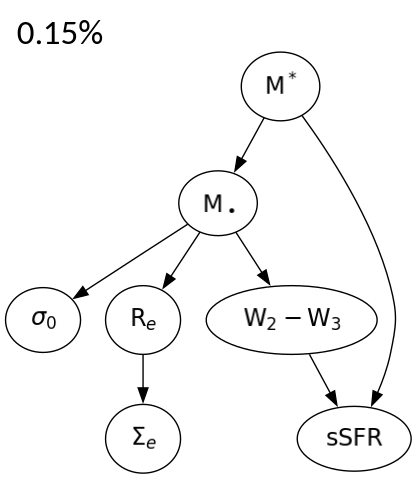} \\
\includegraphics[width=0.165\textwidth]{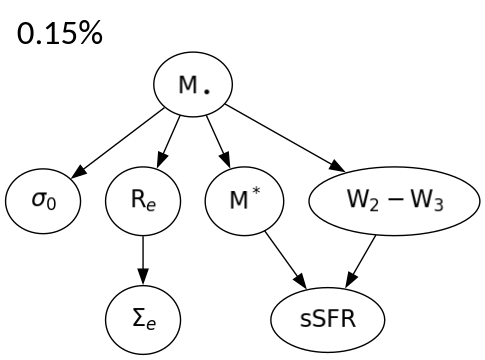} &
\includegraphics[width=0.115\textwidth]{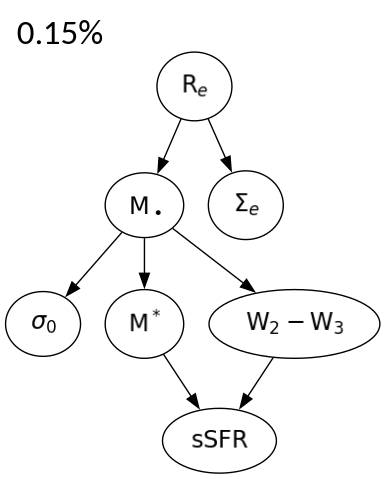} &
\includegraphics[width=0.115\textwidth]{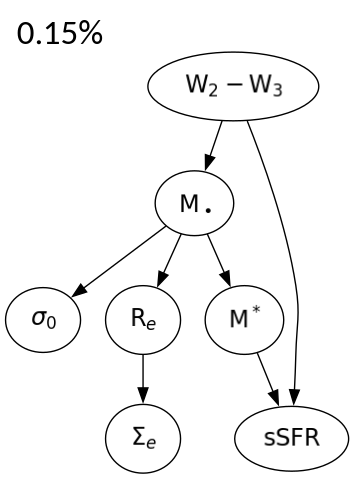} \\
\end{tabular}
\end{tabular}
\caption{
The most probable Markov Equivalence Class (MEC) for each morphology and their corresponding Directed Acyclic Graphs (DAGs) demonstrate a reversal of the causal position of $M_\bullet$.
MECs are represented as Partially Directed Acyclic Graphs (PDAGs).
Directed edges suggest the direction of causality.
The undirected edge $A$ --- $B$ suggests both directions are possible (either $A\rightarrow B$ or $A\leftarrow B$), as long as no new MEC/conditional independencies are introduced by creating new colliders (i.e., two nodes both pointing towards a third node, $A\rightarrow C\leftarrow B$).
In the ellipticals, $M_\bullet$ is strictly a child, while in spiral galaxies, $M_\bullet$ is \emph{always} connected with four galaxy properties through four undirected edges, suggesting either $M_\bullet$ is the parent of all of the four galaxy properties, or $M_\bullet$ is the parent of three of the galaxy properties, and the child of the remaining one (as shown in the corresponding DAGs), ruling out more than one galaxy property pointing towards $M_\bullet$, since this creates a new collider and breaks the encoded conditional independencies.
The percentage listed above each graph indicates the posterior probability of the graph, whereas the prior probability for each individual DAG is equal to the reciprocal of the total number of DAGs, approximately $8.78\times10^{-10}$ \citep{OEIS}.
The MEC probabilities are the sum of their corresponding DAGs.
}
\label{fig:MEC+DAG}
\end{figure*}

Among all possible causal structures, the most probable MEC and its corresponding DAGs for E, S0, and S galaxies are shown in Figure~\ref{fig:MEC+DAG}.
More detailed information about the DAGs, MECs, and their exact posterior distributions can be found in \S\ref{sec:methodology}.
We find that in the most probable MEC for elliptical galaxies, the SMBH mass is a causal child, i.e., an effect of galaxy properties, while in the most probable MEC for spirals, the SMBH mass is a parent of galaxy properties (with lenticulars being in the middle).

\begin{figure*}[]
  \centering
  \includegraphics[width=\linewidth]{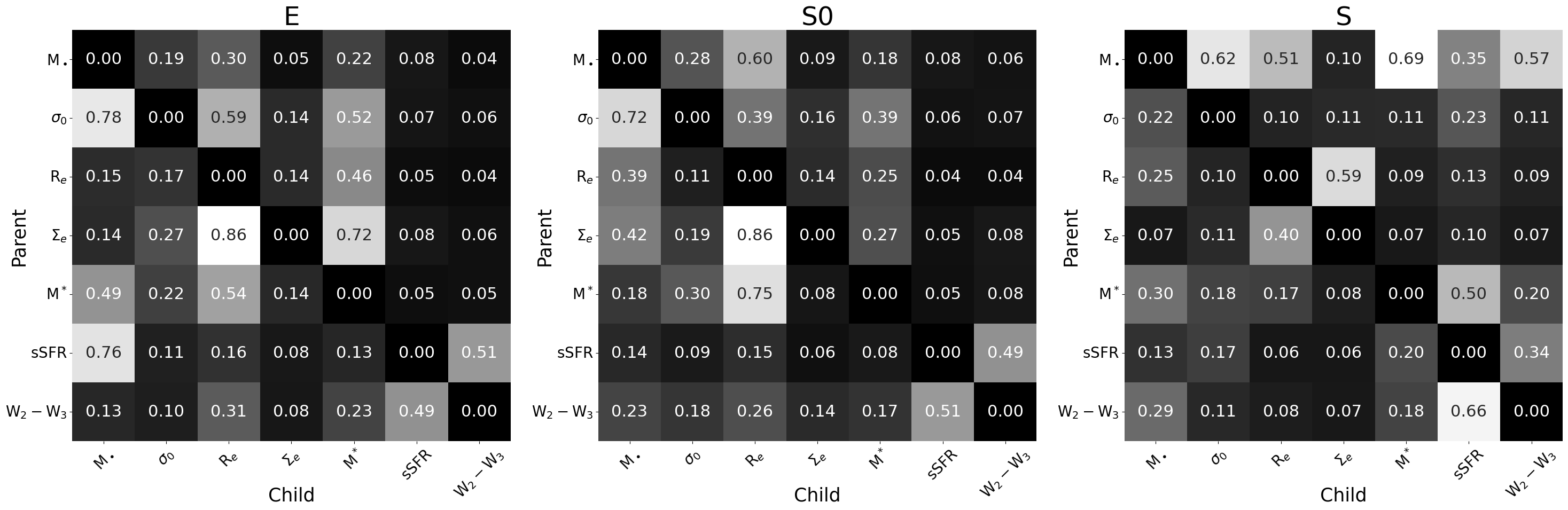}
  \includegraphics[width=\linewidth]{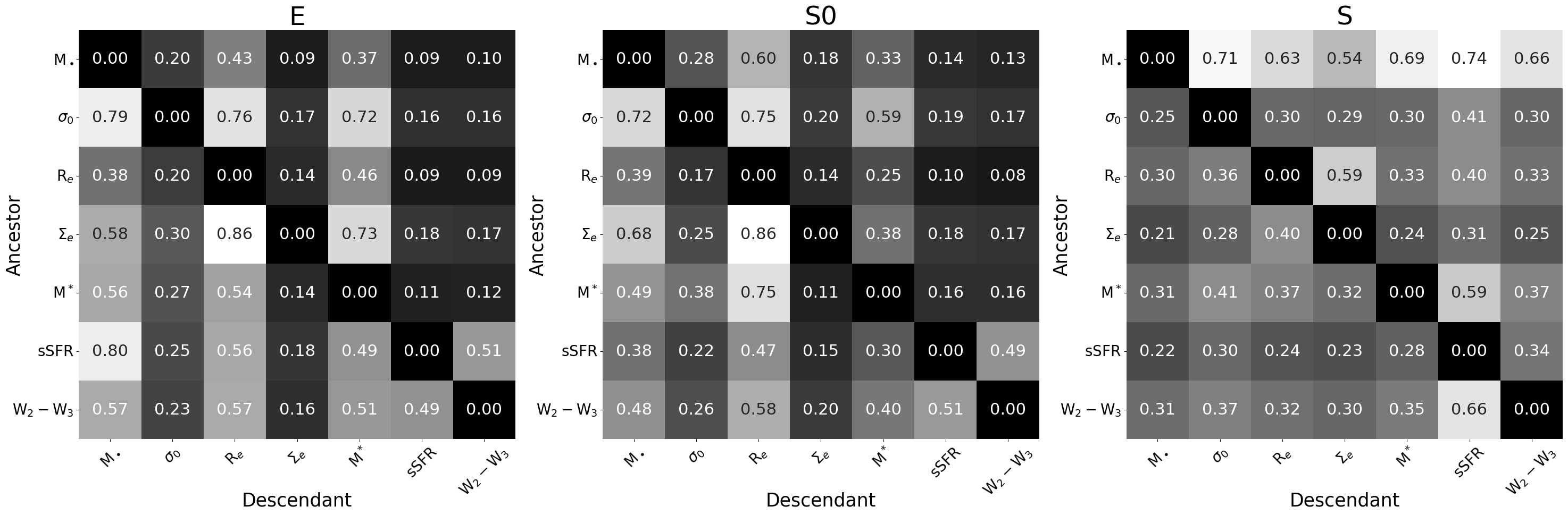}
  \caption{
  Exact posterior \emph{edge} marginals (\emph{top} matrices) and \emph{path} marginals (\emph{bottom} matrices) for \emph{elliptical} (\emph{left} matrices), \emph{lenticular} (\emph{middle} matrices), and \emph{spiral} (\emph{right} matrices) galaxies.
  Edge marginals give the probability of Parent $\rightarrow$ Child through directed edges summed over all DAGs and their probabilities, and path marginals give the probability of Ancestor $\rightarrow$ Descendant through both direct and indirect paths.
  Focusing on the first row and column of each matrix, it is generally evident the first row becomes progressively more light and and first column becomes progressively more dark as you look across the matrices from left to right, going from ellipticals to spirals, i.e., $M_\bullet$ becomes more of a causal parent/ancestor.
  }
\label{fig:exact_matrix}
\end{figure*}

The morphologically-dependent trend holds not only in the most probable graphs but is common over the entire posterior distribution.
This can be quantified using edge and path marginals.
Edge marginals are the posterior probability of a direct causal relation between two variables, marginalized over the causal structures of the other nodes.
Similarly, path marginals provide the probability of a causal connection between two variables through a potentially indirect path (e.g., through intermediate nodes).
These marginal causal structures can be represented in matrix form as shown in Figure~\ref{fig:exact_matrix}.
The first row ($M_\bullet\rightarrow$ galaxy) and column (galaxy $\rightarrow M_\bullet$) of each matrix contain information pertaining to the inferred causal relationship between SMBH masses and their host galaxy properties.

Among all possible DAGs, the percentage of graphs exhibiting a direct edge from $\sigma_0$ to $M_\bullet$ is $78\%$ in ellipticals, $72\%$ in lenticulars, and only $22\%$ in spirals.
The path marginals in the bottom row of matrices support a similar picture, as by considering all possible paths relating these two nodes, we find that $79\%$ of DAGs in ellipticals and $72\%$ in lenticulars have $\sigma_0$ as an ancestor of $M_\bullet$, whereas this is the case in only $25\%$ of DAGs in spirals.
For comparison, the null results (i.e., the posterior from a uniform prior without any data) for the edge marginals are $P(\mathrm{Parent})=29\%$, $P(\mathrm{Child})=29\%$, and $P(\mathrm{Disconnected})=42\%$; for the path marginals these probabilities are $P(\mathrm{Ancestor})=42\%$, $P(\mathrm{Descendant})=42\%$, and $P(\mathrm{Disconnected})=16\%$ (see \S\ref{sec:analysis}).

\section{Posterior Distribution Inspection}\label{sec:analysis}

\begin{figure*}
  \centering
  \includegraphics[width=0.22\linewidth]{exact_MEC_ell_top1.png}
  \hspace{0.02\linewidth}
  \includegraphics[width=0.22\linewidth]{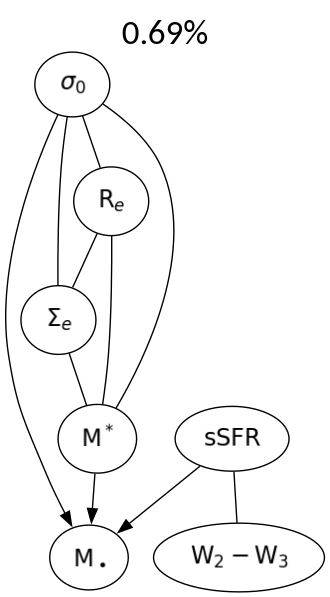}
  \hspace{0.02\linewidth}
  \includegraphics[width=0.22\linewidth]{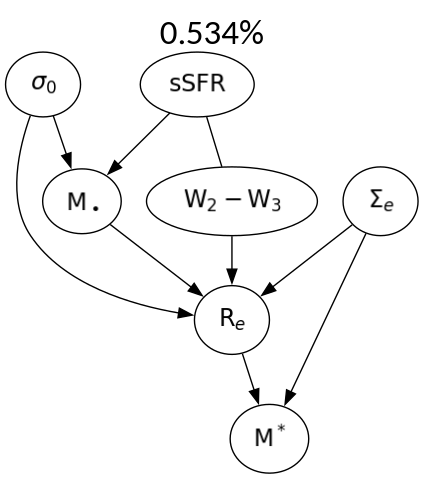}
  \hspace{0.02\linewidth}
  \includegraphics[width=0.22\linewidth]{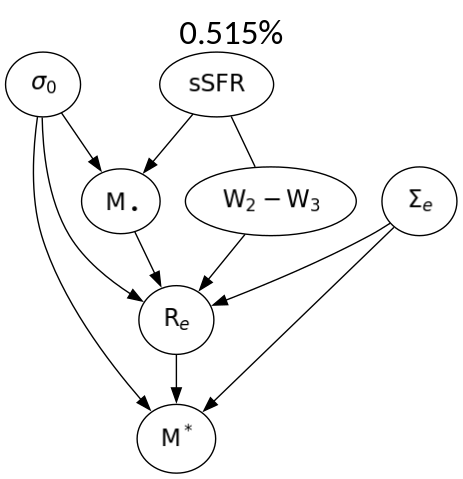}
  \noindent\rule{\textwidth}{0.5pt}
  \vskip 0.1in
  \includegraphics[width=0.22\linewidth]{exact_MEC_len_top1.png}
  \hspace{0.02\linewidth}
  \includegraphics[width=0.22\linewidth]{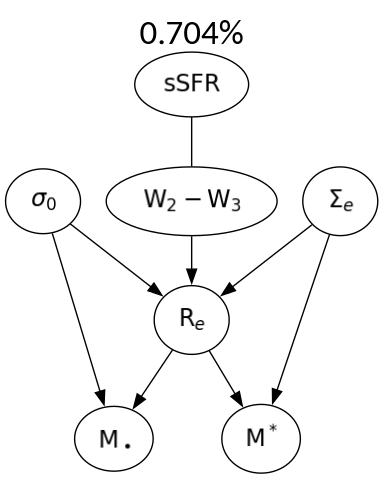}
  \hspace{0.02\linewidth}
  \includegraphics[width=0.22\linewidth]{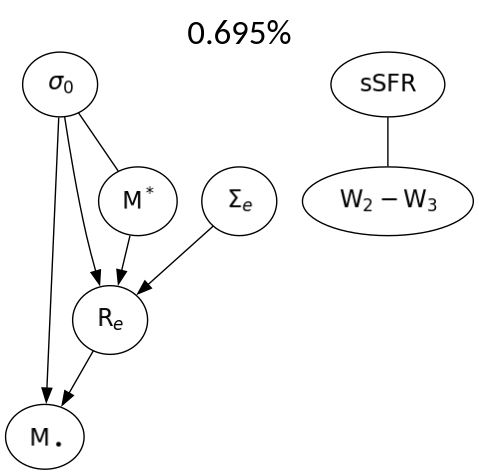}
  \hspace{0.02\linewidth}
  \includegraphics[width=0.22\linewidth]{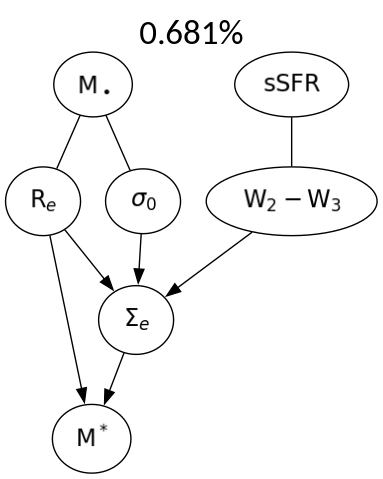}
  \noindent\rule{\textwidth}{0.5pt}
  \vskip 0.1in
  \includegraphics[width=0.22\linewidth]{exact_MEC_spr_top1.png}
  \hspace{0.02\linewidth}
  \includegraphics[width=0.22\linewidth]{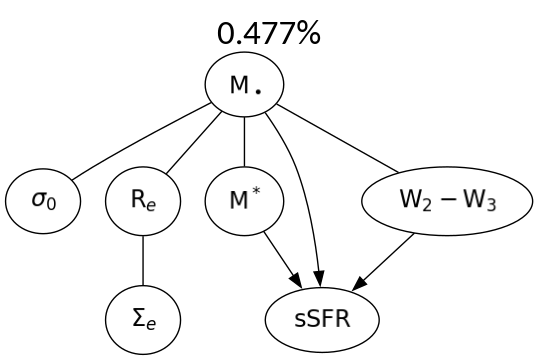}
  \hspace{0.02\linewidth}
  \includegraphics[width=0.22\linewidth]{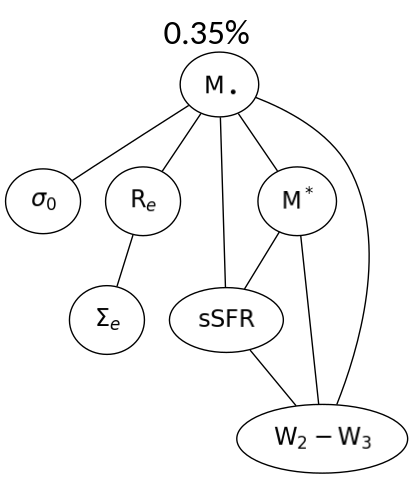}
  \hspace{0.02\linewidth}
  \includegraphics[width=0.22\linewidth]{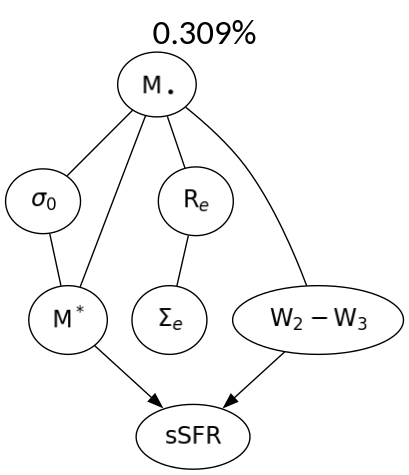}
  \caption{
  Exact posterior result for the top four Markov Equivalence Classes (MECs), represented as Partially Directed Acyclic Graphs (PDAGs)  for \emph{elliptical} (\emph{top} panel), \emph{lenticular} (\emph{middle} panel), and \emph{spiral} (\emph{bottom} panel) galaxies.
  The posterior probability is labeled on top of each MEC, and is calculated by the sum of all DAG posterior probabilities within that MEC.
  The top MECs within each morphology class share noticeable similarities. 
  In ellipticals, $M_\bullet$ generally sits at the bottom of the graph as a descendant of galaxy properties, and rises up as an ancestor of galaxy properties in spirals.
  }
\label{fig:exact_MECs}
\end{figure*}

\begin{figure*}
  \centering
  \includegraphics[width=0.125\linewidth]{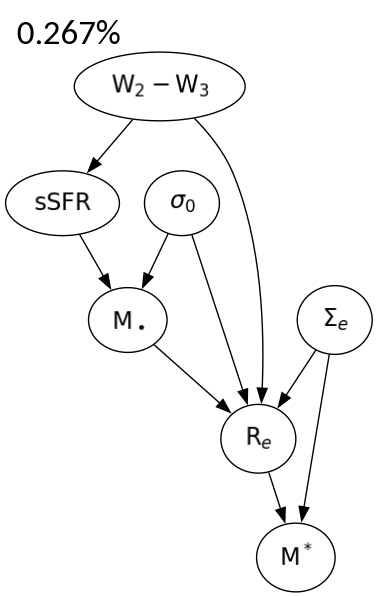}
  \hspace{0.045\linewidth}
  \includegraphics[width=0.125\linewidth]{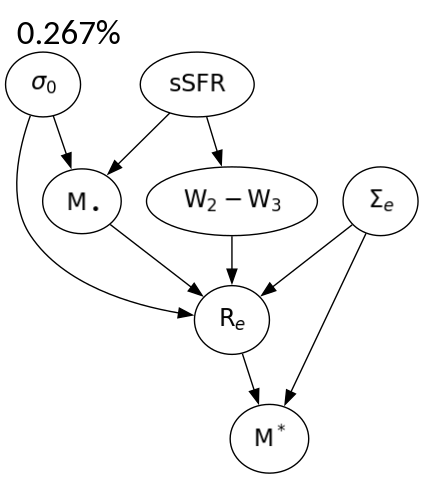}
  \hspace{0.045\linewidth}
  \includegraphics[width=0.125\linewidth]{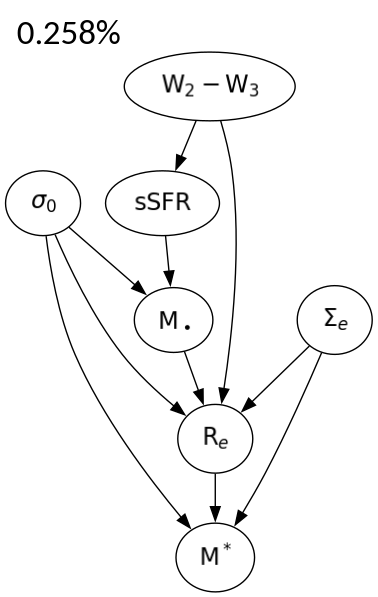}
  \hspace{0.045\linewidth}
  \includegraphics[width=0.125\linewidth]{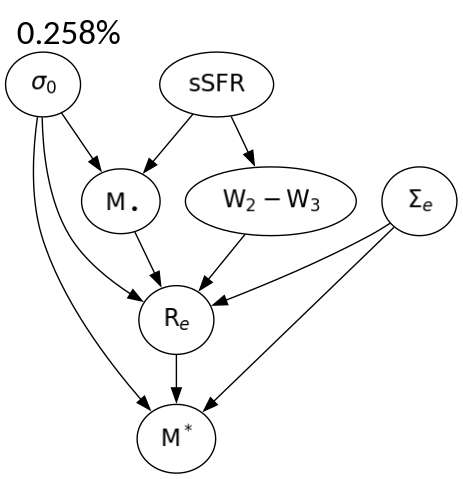}
  \hspace{0.045\linewidth}
  \includegraphics[width=0.125\linewidth]{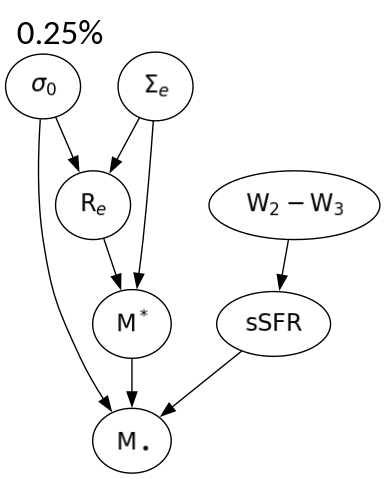}
  \vskip 0.1in
  \includegraphics[width=0.125\linewidth]{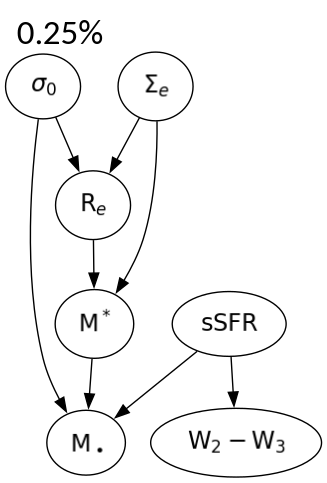}
  \hspace{0.045\linewidth}
  \includegraphics[width=0.125\linewidth]{ell_top7.png}
  \hspace{0.045\linewidth}
  \includegraphics[width=0.125\linewidth]{ell_top8.png}
  \hspace{0.045\linewidth}
  \includegraphics[width=0.125\linewidth]{ell_top9.png}
  \hspace{0.045\linewidth}
  \includegraphics[width=0.125\linewidth]{ell_top10.png}
  \noindent\rule{\textwidth}{0.5pt}
  \vskip 0.1in
  \includegraphics[width=0.125\linewidth]{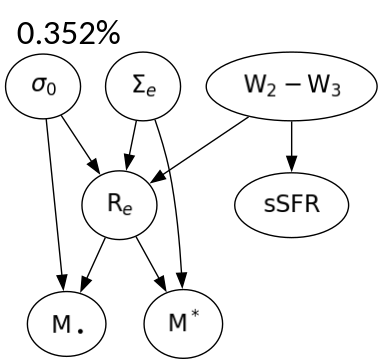}
  \hspace{0.045\linewidth}
  \includegraphics[width=0.125\linewidth]{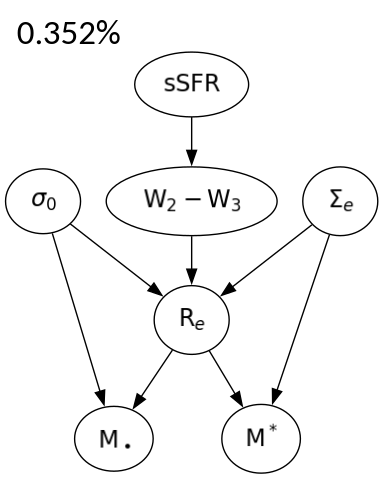}
  \hspace{0.045\linewidth}
  \includegraphics[width=0.125\linewidth]{len_top3.png}
  \hspace{0.045\linewidth}
  \includegraphics[width=0.125\linewidth]{len_top4.png}
  \hspace{0.045\linewidth}
  \includegraphics[width=0.125\linewidth]{len_top5.png}
  \vskip 0.1in
  \includegraphics[width=0.125\linewidth]{len_top6.png}
  \hspace{0.045\linewidth}
  \includegraphics[width=0.125\linewidth]{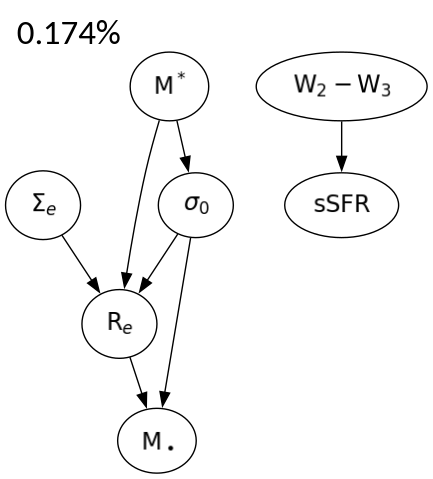}
  \hspace{0.045\linewidth}
  \includegraphics[width=0.125\linewidth]{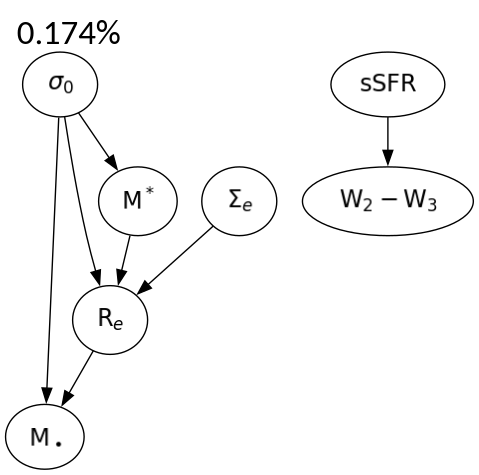}
  \hspace{0.045\linewidth}
  \includegraphics[width=0.125\linewidth]{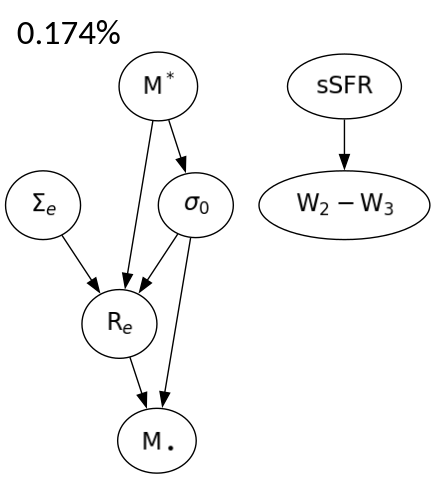}
  \hspace{0.045\linewidth}
  \includegraphics[width=0.125\linewidth]{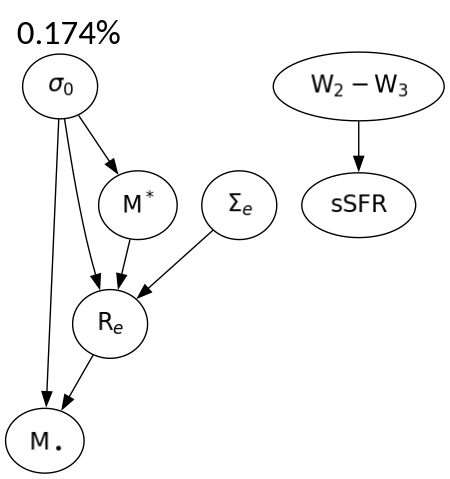}
  \noindent\rule{\textwidth}{0.5pt}
  \vskip 0.1in
  \includegraphics[width=0.125\linewidth]{spr_top1.png}
  \hspace{0.045\linewidth}
  \includegraphics[width=0.125\linewidth]{spr_top2.png}
  \hspace{0.045\linewidth}
  \includegraphics[width=0.125\linewidth]{spr_top3.png}
  \hspace{0.045\linewidth}
  \includegraphics[width=0.125\linewidth]{spr_top4.png}
 \hspace{0.045\linewidth}
  \includegraphics[width=0.125\linewidth]{spr_top5.png}
  \vskip 0.1in
  \includegraphics[width=0.125\linewidth]{spr_top6.png}
  \hspace{0.045\linewidth}
  \includegraphics[width=0.125\linewidth]{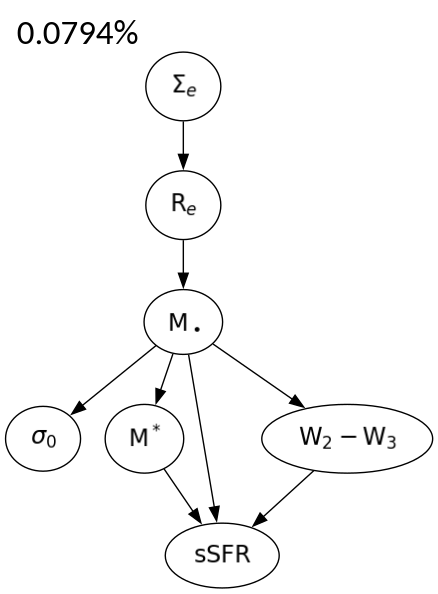}
  \hspace{0.045\linewidth}
  \includegraphics[width=0.125\linewidth]{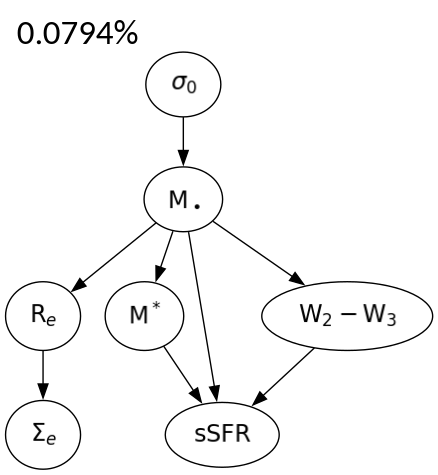}
  \hspace{0.045\linewidth}
  \includegraphics[width=0.125\linewidth]{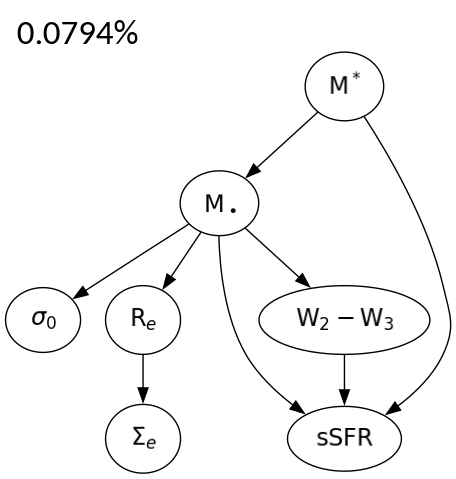}
  \hspace{0.045\linewidth}
  \includegraphics[width=0.125\linewidth]{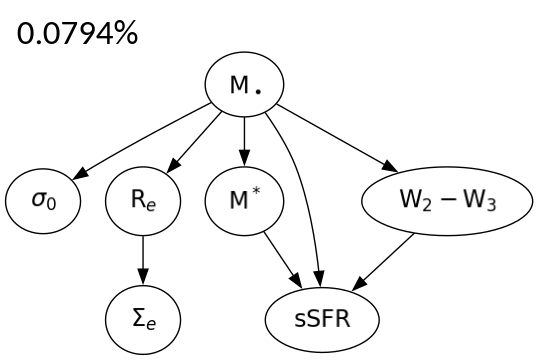}
  \caption{
  Exact posterior result for the top 10 most probable Directed Acyclic Graphs (DAGs) for \emph{elliptical} (\emph{top} panel), \emph{lenticular} (\emph{middle} panel), and \emph{spiral} (\emph{bottom} panel) galaxies.
  The percentage listed above each DAG indicates the posterior probability of the DAG, whereas the prior probability for every DAGs is equal to precisely $8.781333053161975\times10^{-10}$ \citep{OEIS}.
  The top DAGs within each morphology are similar to each other, and the general trend of $M_\bullet$ predominantly rising to the top of the DAGs (from a descendant to an ancestor of galaxy properties) when going from the top (ellipticals) to bottom (spirals) panel is again manifest.
  }
\label{fig:exact_DAGs}
\end{figure*}

In addition to Figure~\ref{fig:MEC+DAG} (which shows the MEC with the highest posterior probability along with its corresponding DAGs) and Figure~\ref{fig:exact_matrix} (which shows the edge and path marginals), here we take a deeper look at the posterior distribution.
Figure~\ref{fig:exact_MECs} shows the top four (in terms of posterior probabilities) MECs and Figure~\ref{fig:exact_DAGs} shows the top ten DAGs.
The top graphs within each morphology class are similar to each other, and most of them convey the idea that elliptical galaxy properties $\rightarrow$ SMBH mass, in spirals SMBH mass $\rightarrow$ galaxy properties, and lenticulars occupy the middle ground.
The paltry percentage of the total population for individual DAGs or MECs is not a rare and surprising phenomenon; due to the huge space of possible causal structures, the number of possible DAGs grows \emph{super-exponentially} as the number of variables increases.
The chance of drawing any DAG from a uniform distribution out of all possible DAGs is $8.781333053161975\times10^{-10}$, which is $\sim$$10^6$ times smaller than the typical proportion around $\sim$$10^{-3}$ we observed for the top DAGs (see Figure~\ref{fig:exact_DAGs}).

\begin{figure*}
  \centering
  \includegraphics[width=0.495\linewidth]{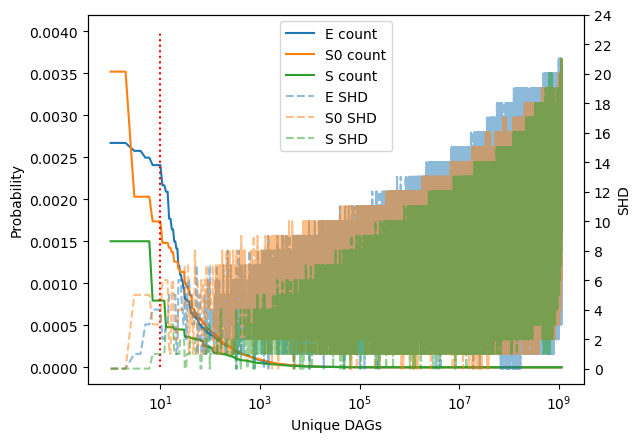}
  \includegraphics[width=0.495\linewidth]{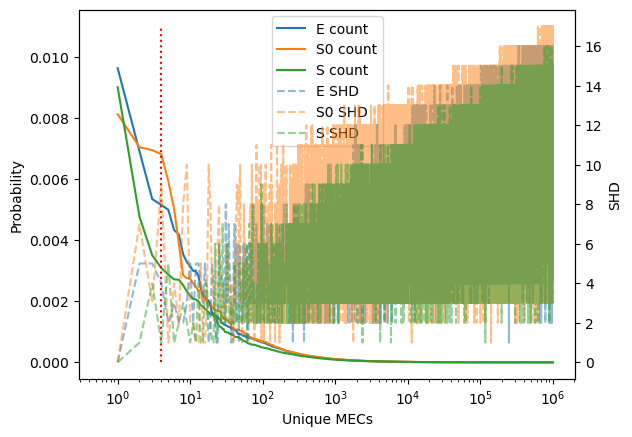}
  \caption{
  The exact posterior probability distribution and the structural Hamming distances (SHDs) to the most probable graph.
  There are in total $3.12510571\times10^8$ MECs in the case of seven nodes \citep{OEISmec}.
  Here, only the first $10^6$ MECs are plotted for simplicity.
The DAGs (left plot) and MECs (right plot) are ordered by their posterior probabilities from highest to lowest.
The solid lines and the left $y$-axes show the posterior probability of the DAGs/MECs.
The dashed lines and the right $y$-axes show the SHD, a measure of distance between graphs, from each DAG or MEC to the most probable DAG or MEC. 
The red dashed line marks the $10^{th}$ DAG and the $4^{th}$ MEC, which are shown in Figures~\ref{fig:exact_DAGs} and \ref{fig:exact_MECs}, respectively.
Together, these panels demonstrate that a few DAGs or MECs that are similar to each other (in terms of SHDs) have much higher posterior probabilities than the rest of the distinctive DAGs and MECs that reside in the very long tail of nearly zero probability.
}
\label{fig:exact_distribution}
\end{figure*}

To better understand the relative posterior probability distribution and quantify the difference between graphs, in Figure~\ref{fig:exact_distribution} we ordered DAGs and MECs by their posterior probabilities from highest to lowest.
The posterior probability is shown as solid lines and labeled on the left $y$-axis, and the rapidly dropping curves show that a few leading DAGs and MECs are relatively much more probable than the DAGs and MECs in the long tail (note that the $x$-axis is in a logarithmic scale). 
The dashed lines and the right $y$-axis show the \emph{structural Hamming distance} (SHD), a standard metric to evaluate the distance between graphs\footnote{$\textup{SHD}(G, G^*)$ counts the number of edges needed to add, delete, or revert, to move from one graph $G$ to another graph $G^*$.}, from each unique DAG or MEC to the most probable DAG or MEC. 
From the SHDs, the top few DAGs or MECs are more similar to each other with fewer edges away from each other, and less prominent DAGs or MECs are statistically more and more distinctive from the top ones.

The SHD increases the slowest in spirals, making the posterior distribution of spirals the most unimodal, while the SHD grows fastest in lenticulars, reinforcing the picture that lenticulars, as the middle ground between ellipticals and spirals, have more sub-modes of causal structures and no clear dominance of one particular causal direction between black hole mass and galaxy properties.
The probability distribution and SHD distribution together shows that despite the existence of some sub-modes, a general mode of causal structure is detected in ellipticals and spirals respectively, and this general mode can be visualized statistically via the edge marginals and path marginals in Figure~\ref{fig:exact_matrix}, as discussed in \S\ref{sec:compendium}.

\begin{figure*}
  \centering
  \includegraphics[width=0.49\linewidth]{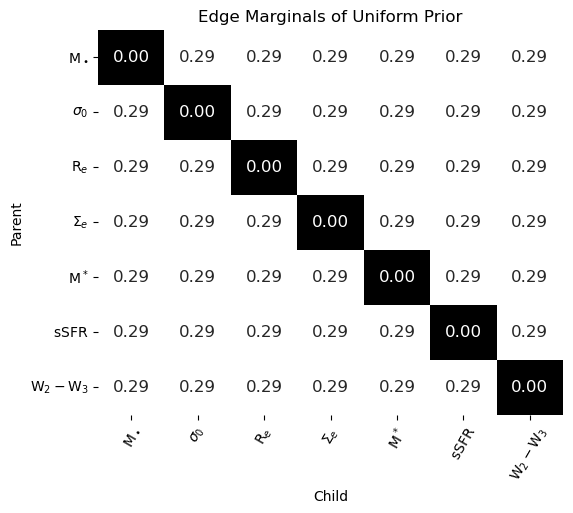}
  \includegraphics[width=0.49\linewidth]{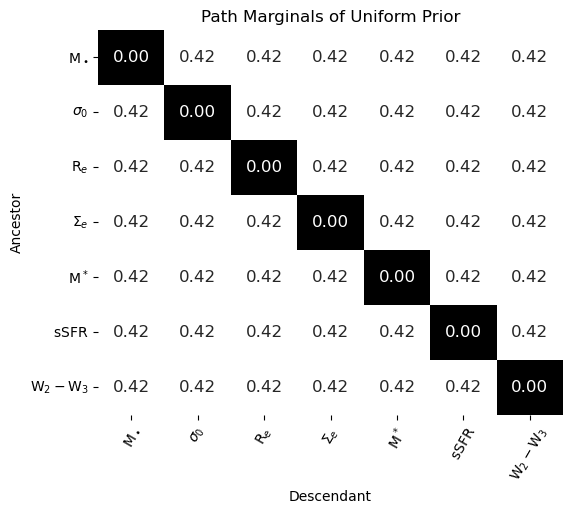}
  \caption{
  The \emph{edge} marginals (\emph{left} matrix) and \emph{path} marginals (\emph{right} matrix) for a uniform prior, i.e., all possible DAGs share the same probability.
  Note that under a uniform prior, the probability of having $A\rightarrow B$, $A\leftarrow B$, and $A$ disconnected to $B$ is not 1/3 each.
  The edge marginals and path marginals under the uniform prior will vary slightly as the number of nodes change, but the probability of having opposite directions of causality (i.e., $A\rightarrow B$ and $A\leftarrow B$) will remain equal.
  }
\label{fig:uniform}
\end{figure*}

Finally, we note that it is useful to remember for all interpretations of edge/path marginals that the null solution is nontrivial.
Indeed, there is not a 50--50 chance of one direction of causality or the other, but rather there are differing probabilities of one direction, its reverse, or no direction of causality that vary depending on whether it is an edge or path marginal, and it is not intuitive (see Figure~\ref{fig:uniform})

\section{Findings}\label{sec:findings}

\subsection{Causal Connections for Galaxy Evolution}\label{sec:causalGE}

We find that these results are consistent with theoretical models of galactic evolution.
Ellipticals are highly-evolved galaxies, being the result of a large number of galactic mergers. 
Modern hydrodynamical cosmological simulations such as EAGLE \citep{Crain:2015} or IllustrisTNG \citep{Marinacci:2018,Naiman:2018,Nelson:2018,Pillepich:2018,Springel:2018} show that elliptical galaxies with $\log(M^*/\mathrm{M}_\odot)\geq11$ are generally the end result of two or more major merger events, such that the typical present-day fraction of stars with \textit{ex situ} origins is greater than 50\% \citep{Cannarozzo:2023}\footnote{
Here, \citet{Cannarozzo:2023} follow previous work \citep{Rodriguez-Gomez:2015,Rodriguez-Gomez:2016} and define a major merger as a stellar mass ratio greater than 1/4 between the two progenitors of a given galaxy.}.
In even more general terms, the process of successive mergers will act to erase the preexisting causal connection from the SMBH to its host galaxy and establish new correlations via the central limit theorem \citet{Jahnke:2011}.

During a merger, the SMBHs at the center of each merging galaxy play no role in the large-scale dynamics; it is the galaxy properties (chiefly size and mass) that shape the galaxy mergers and their outcomes.
Central SMBHs are passively driven to the bottom of the post-galaxy-merger potential well by dynamical friction, eventually merging together \citep{Chandrasekhar:1943,Khan:2020}.
So it stands to reason to expect that in ellipticals, the distribution of SMBH masses is determined by that of galaxy properties and \emph{not} vice versa.

For spiral galaxies, this is not the case, since they experience at most a few relatively minor mergers.
Unlike elliptical galaxies, spirals are predominantly composed of \textit{in situ} stellar populations.
Causal relations between SMBH mass and galaxy properties may thus be set primordially in a secular co-evolution phase, and they are not erased by mergers.
As a result, spiral galaxies behave markedly differently compared to ellipticals.
Interestingly, lenticulars appear to lie in-between, as expected, based on the fact that lenticulars have undergone enough mergers to erase spiral structure while still maintaining an extended disk structure, but are not yet comparable to ellipticals in terms of mass and pressure support.\footnote{The co-evolution of lenticular galaxies and their black holes is also strongly influenced by the presence of dust \citep{Graham:2023d,2024MNRAS.531..230G,Graham:2024b}.}
Moreover, by extension of \citet{Cannarozzo:2023}'s results to all early-type (i.e., lenticular and elliptical) galaxies, all but the most massive lenticular galaxies should still maintain \textit{in situ} stellar fractions greater than 50\%.

The six galaxy variables studied here can be split into the three parameters defining the fundamental plane (FP) of elliptical galaxies and three parameters related to star formation.
The FP is a manifestation of dynamical equilibrium reached in the largely pressure-supported stellar dynamics of massive elliptical galaxies \citep{Mould:2020}.
Moreover, it is a consequence of the merger formation of these galaxies via dissipation and feedback that ultimately places them on the FP.
Although only 35/101 of our galaxies are ellipticals, the classical bulges of lenticular and spiral galaxies are also governed by the FP.
Indeed, it has been found that the bulges of type S0--Sbc galaxies tightly follow the same FP relation as ellipticals \citep{Jesus:2002}.

The matrices in Figure~\ref{fig:exact_matrix} also provide information about the causal nature of the observed FP relationship.
By looking at the path marginals for elliptical galaxies (bottom left), we find that $\langle{\Sigma_\mathrm{e}}\rangle$ is the ancestor (86\%) of $R_\mathrm{e}$ and that $\sigma_0$ is an ancestor (76\%) of $R_\mathrm{e}$.
This implies $\langle{\Sigma_\mathrm{e}}\rangle$ and $\sigma_0$ are both upstream of $R_\mathrm{e}$, confirming that the density and dynamics of stellar populations in an elliptical galaxy govern its size.
Furthermore, we find that there is nearly no chance that $M^*$ is disconnected from $R_\mathrm{e}$ (i.e., $54\%+46\%=100\%$, they are \emph{never} $d$-separated, thus \emph{always} correlated), indicating the existence of a size--mass relation due to the virial theorem (i.e., $M\sim\sigma^2R$).

\subsection{Causal Active Galactic Nuclei Feedback}\label{sec:causalAGN}

From Figure~\ref{fig:exact_matrix}, we find that, in spirals, $M_\bullet$ is the ancestor (74\%) of sSFR, in lenticulars, there is no dominant causal direction between the two parameters (38\% and 14\%), while in ellipticals, $M_\bullet$ becomes the descendant (80\%) of a galaxy's sSFR.
This can be interpreted as a direct consequence of the presence or absence of gas through AGNs feedback.
If there is a substantial gas reservoir (as in spirals), the SMBH is the ancestor since its feedback is responsible for shutting down star formation and hence stopping the growth of stellar mass.
With a dearth of gas, as in ellipticals, even large AGNs bursts will not affect the stellar mass, and thus the SMBH cannot be an ancestor of galaxy properties.
This is further supported by the fact that we find that $M_\bullet$ is the parent (69\%) of $M^*$ in spirals, but becomes the descendant (56\%) or child (49\%) of $M^*$ in elliptical galaxies.
However, it is true that in the absence of gas, mergers are the main pathway for SMBH growth, and this will also cause the SMBH to become a descendant or child in hierarchical assembly \citep{Hopkins:2006,Hopkins:2007,Hopkins:2008b,Hopkins:2008,Jahnke:2011,Treister:2012,Graham:2023,Graham:2023b,Natarajan_2023}.

\section{Experiment with Semi-analytic Models}\label{sec:SAMs}
As an additional benchmark test for the methodology under the SMBH--galaxy context, we practiced the same causal discovery method on data generated by SAMs, where the ground truth causal direction is clearly defined, propagated through a series of coupled partial-differential equations, and is easily customizable. 

SAMs are powerful tools to model galaxy formation using dark matter halo merger trees from $N$-body simulations with some phenomenological descriptions of baryonic physical processes like cosmic reionization, hot gas cooling and cold gas infall, star formation and metal production, supernova feedback, gas stripping and tidal disruption of satellites, galaxy mergers, bulge formation, black hole growth, AGNs feedback, etc.
We adopt the model of \citet{Luo2016}, which is the resolution-independent version of the Munich galaxy formation model: L-Galaxies (mainly based on models of \citealt{Fu2013} and \citealt{Guo2011,Guo2013}).
The dark matter only $N$-body simulation is the JiuTian-1G simulation with $6144^3$ dark matter particles in a 1\,Gpc/$h$ cubic simulation box, based on Planck 2018 \citep{Planck:2020} cosmological parameters.

In the model, there are two processes related to black hole growth and its feedback.
The first is ``quasar mode'' where SMBHs can accrete cold gas directly during galaxy mergers.
The other is ``radio mode'' where SMBHs can accrete hot gas continually from their host galaxies and inject energy into the hot atmosphere.
The quasar mode is the main black hole growth channel, while the radio mode is the main AGNs feedback channel to suppress hot gas cooling.
More details can be found in the supplementary material of \citet{Henriques2015}.
The models used here are in broad agreement with many of the standard studies on AGN feedback \citep[e.g.,][]{Kauffmann:2000,Benson:2003,Granato:2004,DiMatteo:2005,Bower:2006,Croton:2006}.

In SAM galaxies, black hole feedback is actively affecting galaxies and is hard-coded to turn off as soon as a galaxy is quenched.
Therefore, in SAM elliptical galaxies that become quenched, galaxy properties cause the black hole mass via the only remaining mechanism (i.e., mergers/accretion), and in SAM spiral galaxies with black hole feedback still on, black holes primarily cause galaxy properties.
We conducted an additional check where the black hole feedback is manually turned off throughout the entire life of galaxies as ``SAM no feedback'' galaxies.

\begin{figure}
  \centering
\includegraphics[width=\linewidth]{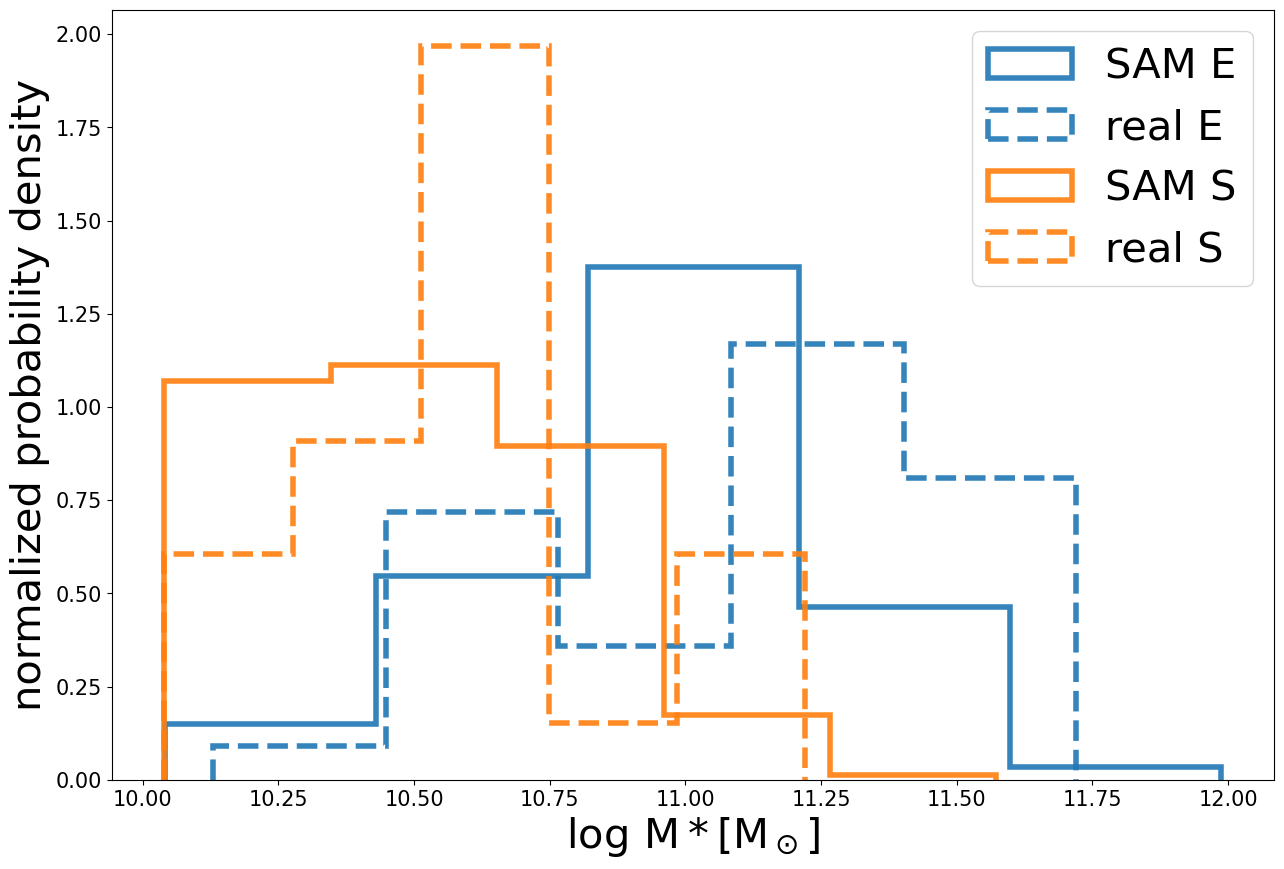}
  \caption{
  The $M^*$ distribution for semi-analytical model (SAM) galaxies compared to the $M^*$ distribution of real galaxies used in this work.
  The $M^*$ distributions of SAM E and SAM S galaxies are similar to that of the real observational data for a fair comparison.
  }
\label{fig:sam_mass}
\end{figure}

This gives us three groups of SAM galaxies: SAM E galaxies, SAM S galaxies, and SAM no-feedback galaxies.
SAM E galaxies are galaxies with $B/T>0.78$ (bulge-to-total ratio, $M^*_\mathrm{bulge}/M^*>0.78$ from \citealt{Graham:2008}), and SAM S galaxies are galaxies with $B/T\leq0.78$.
Additional stellar mass $M^*$ cuts are applied such that the $M^*$ distributions of SAM E and SAM S galaxies are similar to that of the real observational data used in this work for a fair comparison, as shown in Figure~\ref{fig:sam_mass}.
No cuts are applied to SAM no-feedback galaxies since they do not have any realistic counterparts and are generated solely for this test.
This gives us 1189 SAM E galaxies, 1999 SAM S galaxies, and 2663 SAM no-feedback galaxies.

\begin{figure*}
  \centering
  \includegraphics[width=\linewidth]{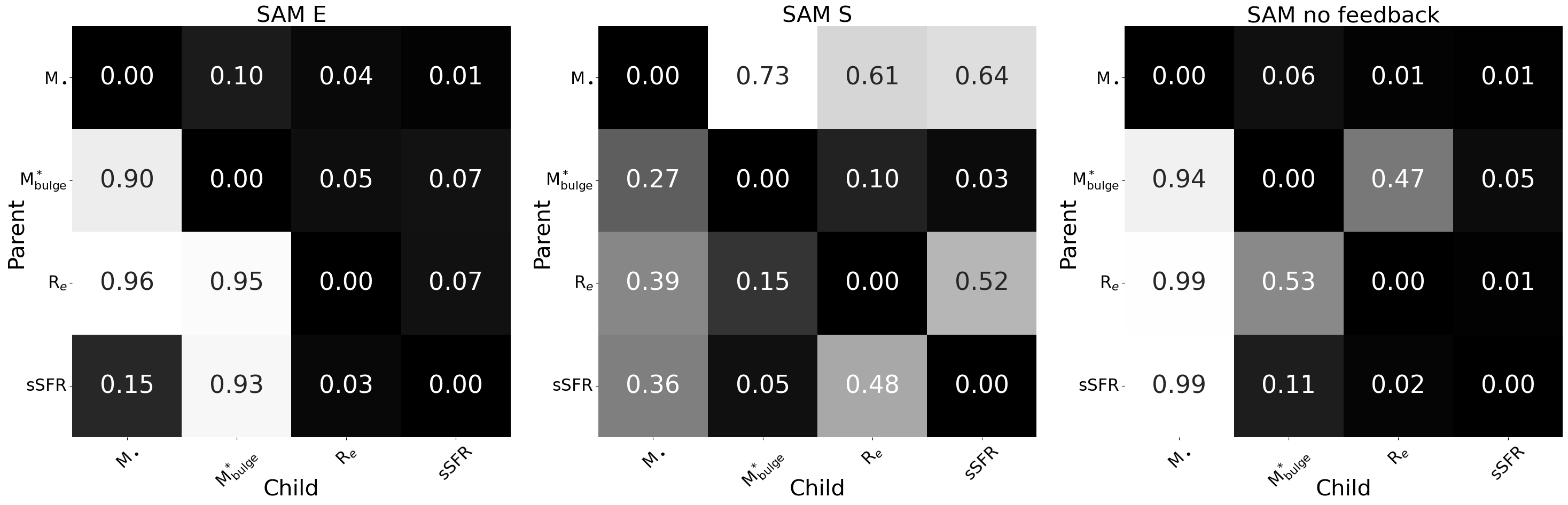}
  \includegraphics[width=\linewidth]{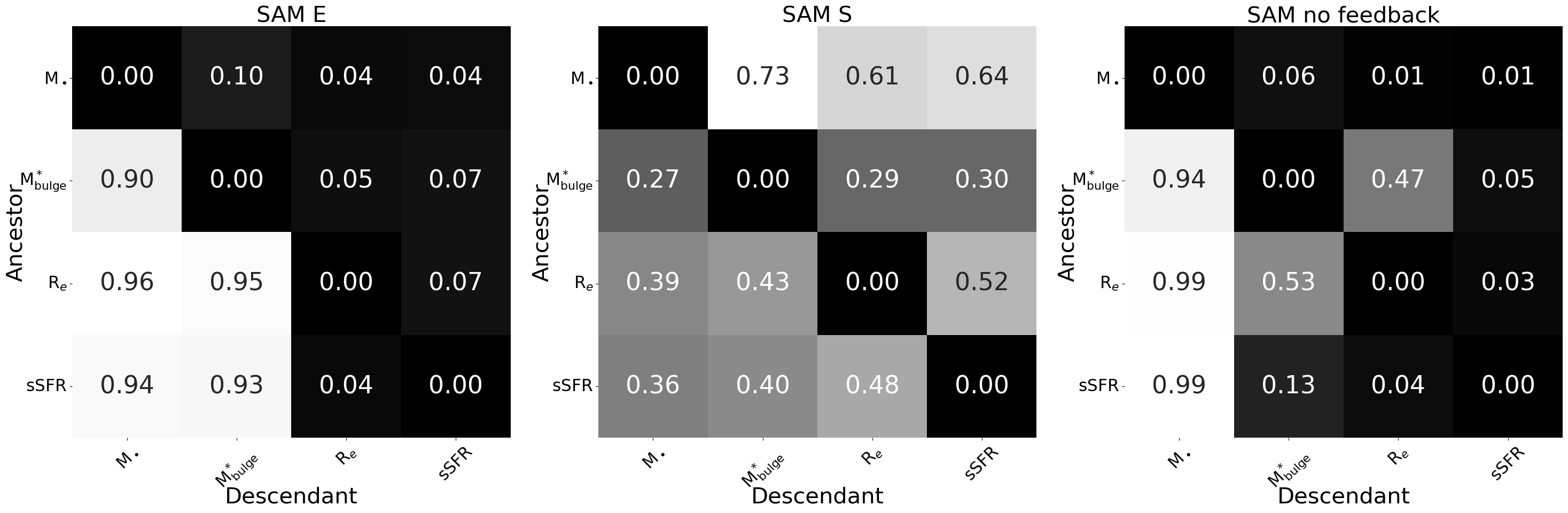}
  \caption{
  \emph{Edge} (\emph{top} matrices) and \emph{path} (\emph{bottom} matrices) marginals for SAM galaxies.
  These matrices are similar to those found in Figure~\ref{fig:exact_matrix}.
  Causal discovery is performed on ellipticals (SAM E), spirals (SAM S), and galaxies with black hole feedback intentionally turned off (SAM no feedback).
  Here, we are restricted to four parameters that are tracked in the SAMs.
  This test shows that the same pattern of causality seen in the real galaxies for elliptical vs.\ spiral (i.e., quenched vs.\ star-forming) galaxies is upheld in these simulated galaxies.
  Moreover, the causal structure identified in SAM no feedback galaxies matches exactly with our design where the black hole $\rightarrow$ galaxy direction is strictly turned off.
  }
\label{fig:SAM}
\end{figure*}

The results we present in Figure~\ref{fig:SAM} indeed confirm the designed causal structure in the SAMs.
The edge marginals clearly show that in SAM E galaxies, galaxy properties cause the black hole mass, in SAM S galaxies, black hole mass causes galaxy properties, and in SAM no feedback galaxies, galaxy properties cause the black hole mass, as the opposite direction is forbidden by construction.

\section{Crosschecking with PC and FCI}\label{sec:PC}

\begin{figure*}
  \centering
  \includegraphics[width=0.35\linewidth]{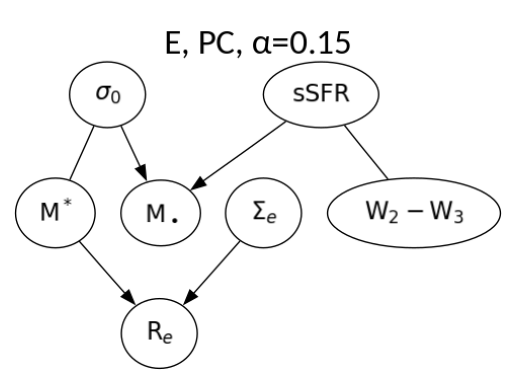}
  \hspace{0.10in}
  \includegraphics[width=0.24\linewidth]{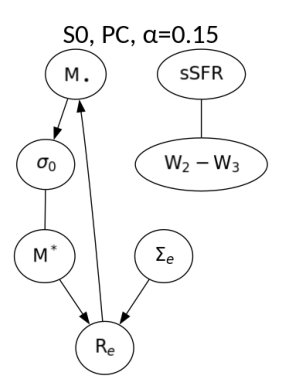}
  \hspace{0.10in}
  \includegraphics[width=0.35\linewidth]{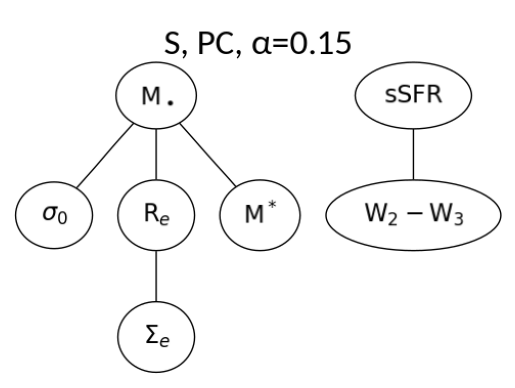}
  \vskip 0.15in
  \includegraphics[width=0.29\linewidth]{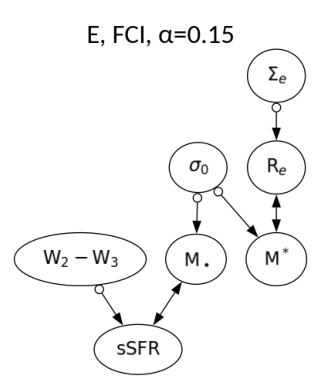}
  \hspace{0.10in}
  \includegraphics[width=0.29\linewidth]{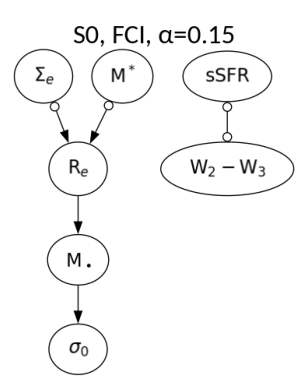}
  \hspace{0.10in}
  \includegraphics[width=0.36\linewidth]{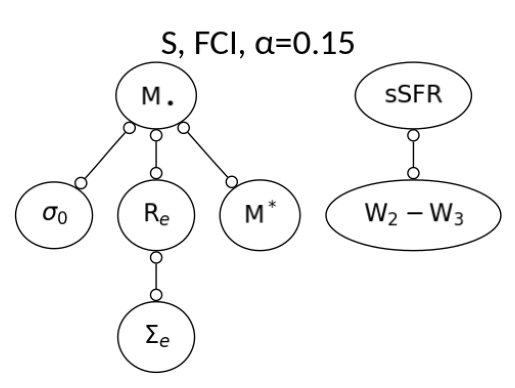}
  \caption{
  Graphs learned by the \emph{PC} algorithm (\emph{upper} row) and by the \emph{FCI} algorithm (\emph{bottom} row).
  The graphs in the \emph{top} row use PDAGs to represent MECs of DAGs by leaving some edges undirected.
  The graphs in the \emph{bottom} row are Partial Ancestral Graphs (PAGs) and introduce additional edge types: $A\longleftrightarrow B$ corresponding to a confounding relation (i.e., a third variable causes both A and B) and empty circles representing uncertainty regarding the ending symbol of the edge (i.e., $A\ \circ\hspace{-1.2mm}\rightarrow B$ may correspond to either $A\rightarrow B$ or to $A\longleftrightarrow B$, but rules out $B\rightarrow A$).
  The significance cutoff for conditional independence tests is set to $\alpha=0.15$ in all graphs.
  This demonstration shows that the same general causal trends are recovered using these \emph{constraint-based methods} when comparing with the primary \emph{score-based method} we utilized in our study.
  }
\label{fig:PCFCI}
\end{figure*}

The PC and FCI algorithms, two \emph{constraint-based methods} (in contrast with the \emph{score-based method} adopted in this work), are also applied to the same observational data to cross-check our results.
The details of these two time-tested algorithms are already presented in \S\ref{sec:constraint-based}.
We adopt the implementation of PC and FCI in the \texttt{Python} package \texttt{causal-learn} \citep{causallearn}, and the results are reported in Figure~\ref{fig:PCFCI}. 
The exact posterior result including edge/path marginals (Figure~\ref{fig:exact_matrix}) and the \emph{top} MECs/DAGs (Figures~\ref{fig:MEC+DAG}, \ref{fig:exact_MECs}, and \ref{fig:exact_DAGs}) are generally consistent with the causal graphs found by PC and FCI.

In ellipticals, the PC algorithm finds $\sigma_0$ and sSFR cause $M_\bullet$.
In our Bayesian approach, $\sigma_0 \rightarrow M_\bullet$ and sSFR$\rightarrow M_\bullet$ indeed have the highest and second highest edge/path marginals among the potential causes of $M_\bullet$ in ellipticals. 
In spirals, the PC algorithm finds $\sigma_0$, $R_\mathrm{e}$, and $M^*$ as effects of $M_\bullet$, and this is also consistent with the edge/path marginals reported in our Bayesian approach. 
The FCI algorithm produces results compatible with those of the PC algorithm, with the difference that, without the assumption of causal sufficiency, it leaves open the possibility that all the relations between SMBH mass and its causal parents are confounded by unobserved variables. 
Particularly, the double arrow between $M_\bullet$ and sSFR in the lower left DAG of Figure~\ref{fig:PCFCI} may indicate an unobserved confounder, possibly the gas fraction, which in the future can be tested through hydrodynamical simulations where the gas fraction is more accessible.

Note that here we adopt a relatively high value of $\alpha=0.15$, the significance cutoff for the $p$-value of conditional independence tests.
Generally, a lower $\alpha$ value gives more false negative errors (i.e., fails to identify causal relations that exist), and a higher $\alpha$ results in more false positive errors (i.e., identifies causal relations that do not exist).
Practically, the choice of $\alpha$ is often empirical and highly depends on the context.
Here in our case, the conditional independence tests, which are the core of PC and FCI, suffer from the limited size of the dataset (35, 38, and 28 for E, S0, and S galaxies, respectively). 
We therefore selected a higher value of $\alpha$ to mitigate this.
These limitations of PC and FCI are one of the main motivations for our adoption of a Bayesian approach by relatively comparing the posterior probabilities across all possible DAGs.

\section{Testing Limitations}\label{sec:limitations}

\subsection{Possible Unobserved Confounders}\label{sec:confounders}

Our posterior calculation approach implicitly adopts the assumption of causal sufficiency, i.e., assuming there are no unobserved confounders\footnote{An unobserved confounder is a variable that is not included in the analysis but is a cause of two or more variables of interest.}. 
With the presence of an unobserved confounder, non-existing causal relations might be falsely identified.
Some potential unobserved confounders, such as the reserve of gas or merger history, are practically difficult to observe but are already integrated into our interpretation.
However, the distance from us to galaxies does not directly play any role in galaxy formation theory nor in our interpretation, but might influence multiple variables we examined, since our ability to measure all these seven variables decreases as distance increases and thus bias our sample towards nearby and more massive BHs/galaxies.
Therefore, we examined the impact of distance by performing causal discovery with distance as one of the seven variables.

\begin{figure*}
  \centering
  \includegraphics[width=\linewidth]{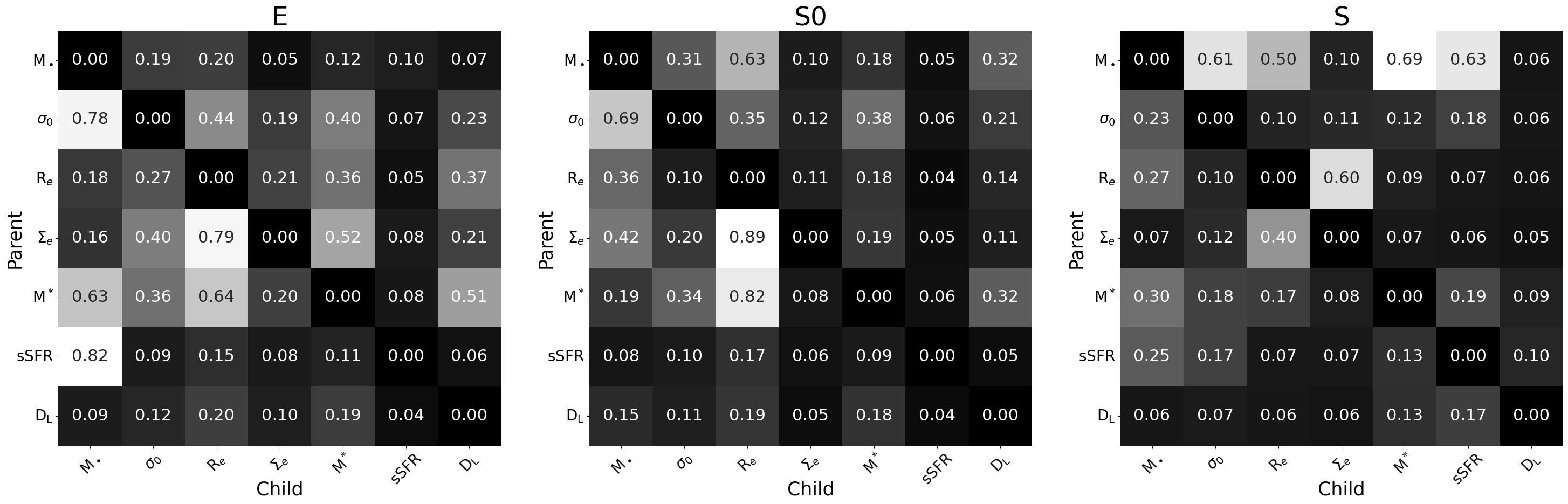}
  \includegraphics[width=\linewidth]{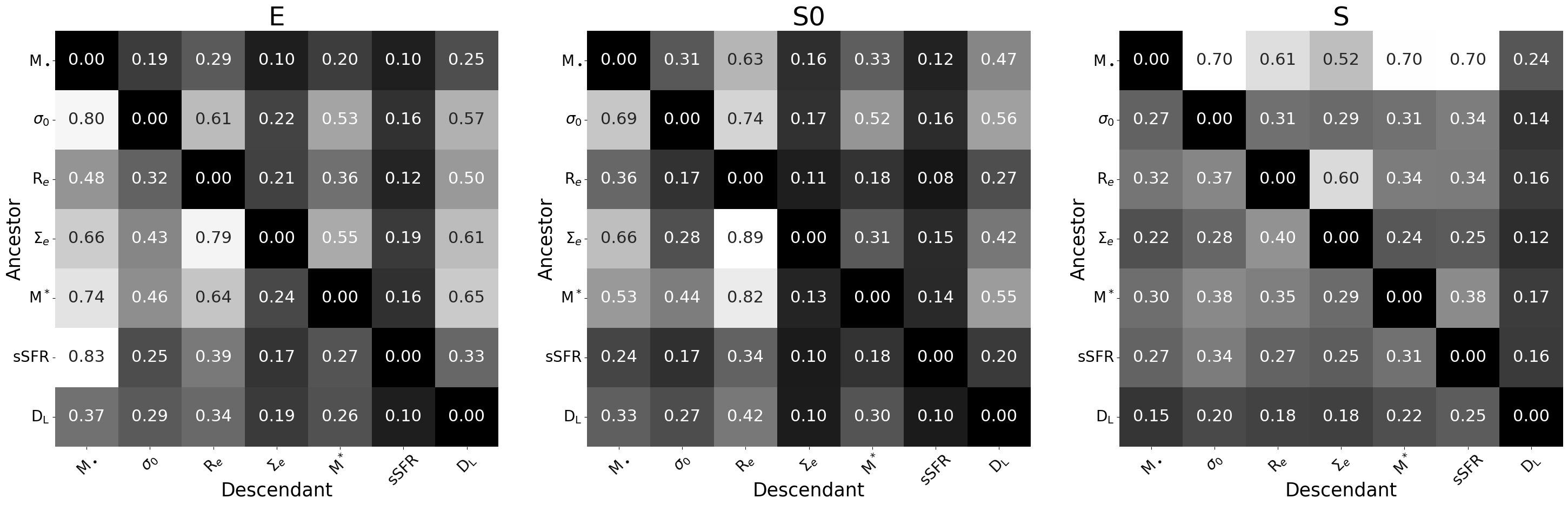}
  \caption{
  \emph{Edge} marginals (\emph{top} matrices) and \emph{path} marginals (\emph{bottom} matrices) with luminosity distance ($D_L$) as one of the variables.
  Qualitatively, we find no change to the causal directions (as compared with Figure~\ref{fig:exact_matrix}) when testing distance as a possible confounding variable.
  }
\label{fig:distance}
\end{figure*}

Since $W2-W3$ and sSFR are highly degenerate with each other, we replaced $W2-W3$ with $D_L$, the luminosity distance to our targets\footnote{
Our luminosity distances are adopted from \citet{Graham:2023}.
Indeed, this sample of dynamically-measured black hole masses comes from galaxies that are all in the local Universe (median $D_L=19.3$\,Mpc; $z=0.00439$ according to \citealt{Planck:2020}).}.
The edge and path marginals with distance included are presented in Figure~\ref{fig:distance}.
Comparing against the original marginals without distance (Figure~\ref{fig:exact_matrix}), the presence of distance barely changes any previously identified causal relations, since the edge and path marginals between galaxy properties and SMBH masses remains unchanged with or without the inclusion of distance.\footnote{
Although we have tested for the biasing effect of distance across the modest range of our nearby galaxies, future work is needed to ascertain the variances encountered when considering galaxies at great distances due to the vast changes in look-back time.
}

\subsection{Stability under Observational Errors}\label{sec:Obs_errors}

\begin{figure*}
  \centering
  \includegraphics[width=\linewidth]{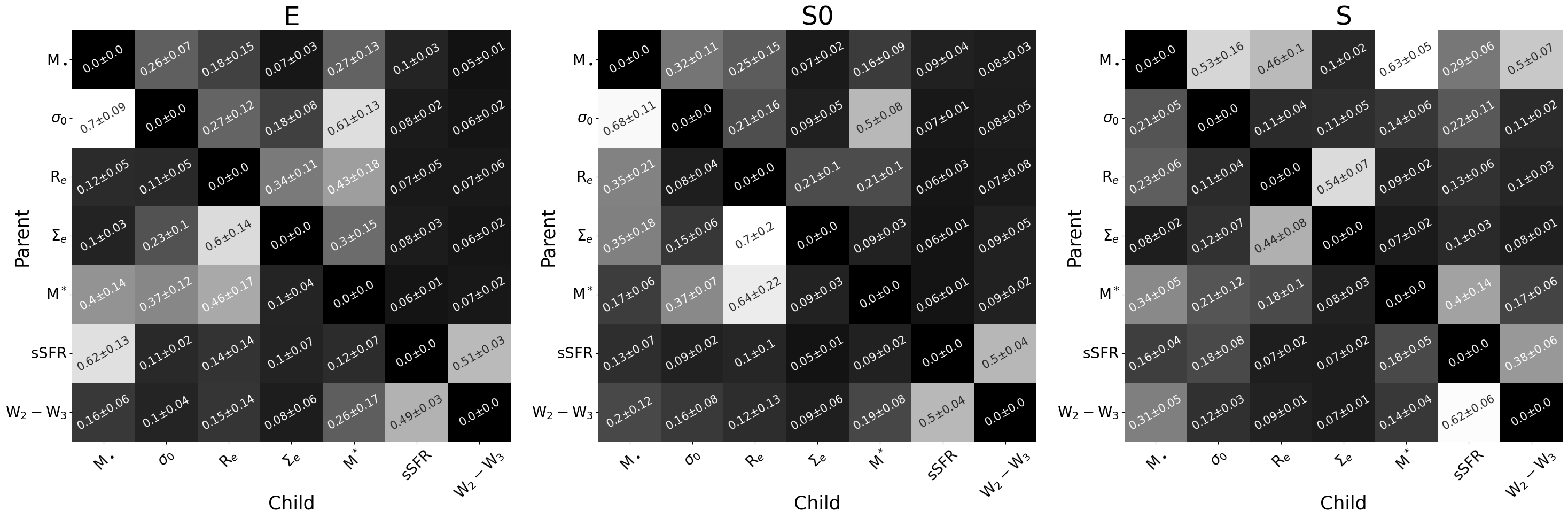}
  \includegraphics[width=\linewidth]{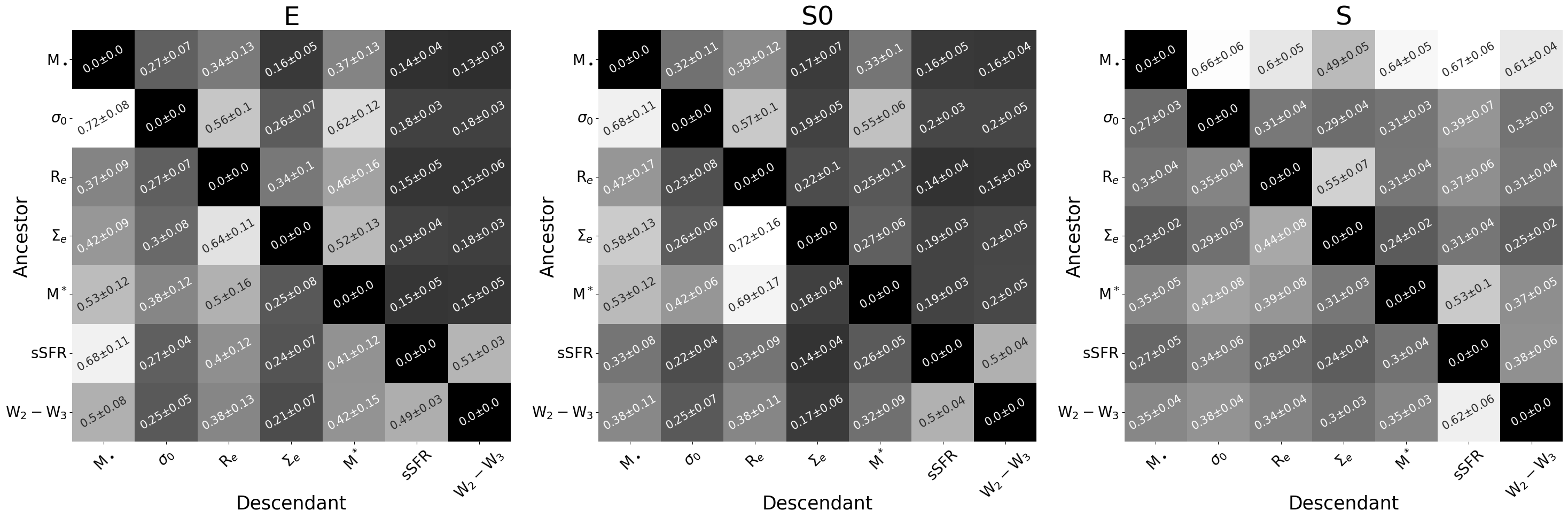}
  \caption{
  The mean and standard deviation of \emph{edge} marginals (\emph{top} matrices) and \emph{path} marginals (\emph{bottom} matrices) over 100 random sampling realizations for each morphological class.
  Qualitatively, we find no change to the causal directions (as compared with Figure~\ref{fig:exact_matrix}) when considering observational errors on all variables, and the uncertainties remain low ($\leq22\%$).
  }
\label{fig:obs_err}
\end{figure*}

The variables used in this work are affected by observational errors and their marginal posterior probability distributions are given in Table~\ref{tab:sample} (assuming Gaussian posteriors). 
However, the causal structures explored so far have been calculated for the mean of these posteriors without considering their uncertainties.
We now quantify the effect of this uncertainty on our inference of the causal structures. 
To do this, we draw samples from the posterior distribution of each variable to produce 100 mock datasets.  
The causal discovery method outlined in this paper is repeated on each of these 100 randomly-sampled datasets to arrive at 100 pairs of different edge marginal and path marginal matrices for each of the three morphological types considered.
The edge marginal and path marginal matrices are summarized in Figure~\ref{fig:obs_err}.

\begin{figure}
  \centering
  \includegraphics[width=\linewidth]{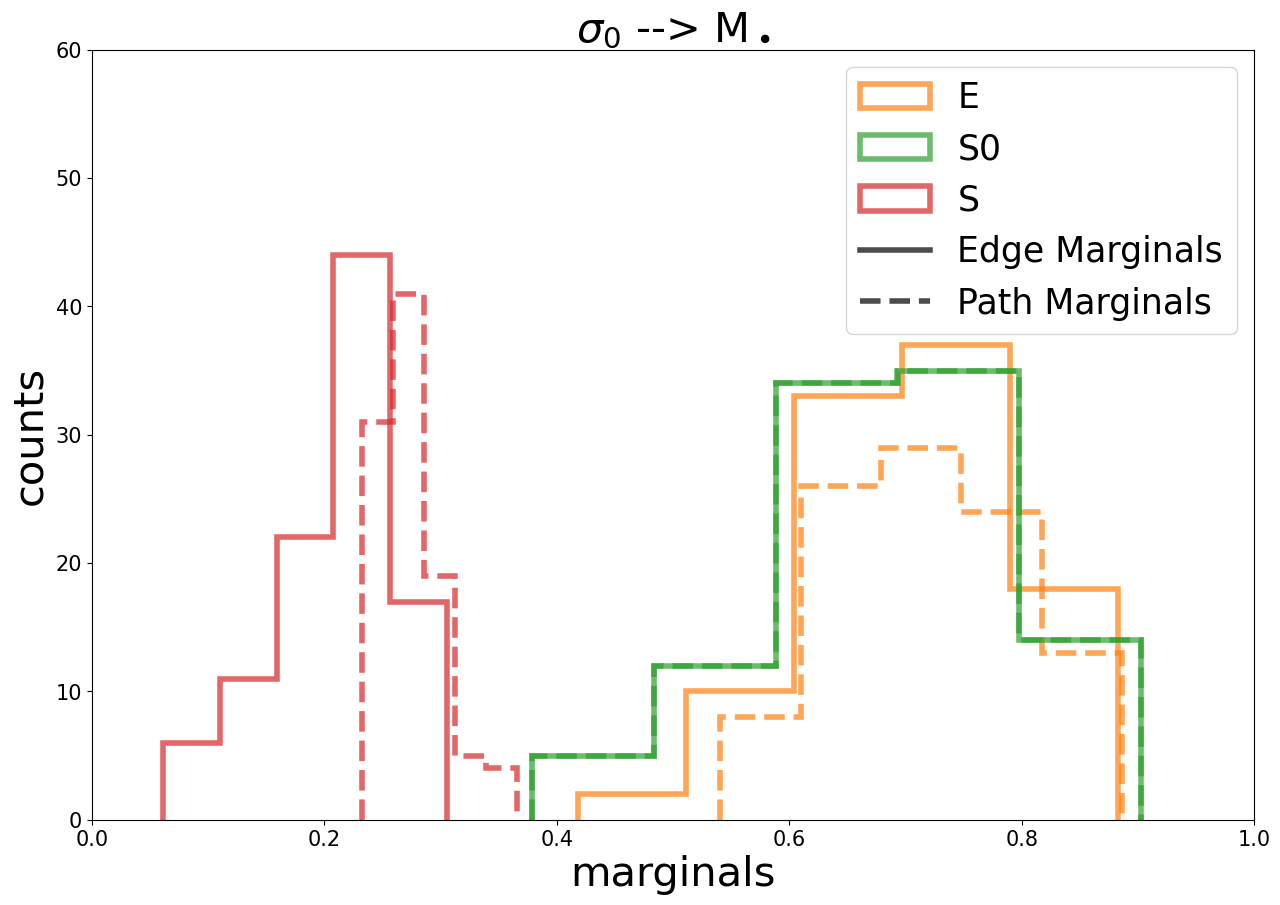}
  \caption{
  \emph{Edge} marginal ({\hdashrule[0.35ex]{8mm}{0.4mm}{}}) and \emph{path} marginal ({\hdashrule[0.35ex]{8mm}{1pt}{1mm}}) distributions of $\sigma_0 \rightarrow M_\bullet$ over 100 random realizations for each morphological class.
  This plot illustrates the distribution of the element at the second row and first column (2, 1) in all matrices from Figure~\ref{fig:obs_err}, showing a clear separation between the causal directions found in late-type vs.\ early-type galaxies.
  }
\label{fig:obs_err_dist}
\end{figure}

We find that, overall, the key findings of this study are robust against these uncertainties.
For example, in ellipticals, the edge marginal between $\sigma_0$ and $M_\bullet$ in both directions across random sampling realizations are $0.70\pm0.09$ and $0.26\pm0.07$, giving a 3.84\,$\sigma$ discrepancy (in other words, the probability that the inferred direction of causality is due to noise and the resulting uncertainties in the variables is about $10^{-4}$).
Figure~\ref{fig:obs_err_dist} shows the distribution of the edge marginals and path marginals for $\sigma_0 \rightarrow M_\bullet$.
The difference between ellipticals and spirals is evident in all realizations.

\subsection{Stability under Possible Outliers}\label{sec:outliers}

\begin{figure*}
  \centering
  \includegraphics[width=\linewidth]{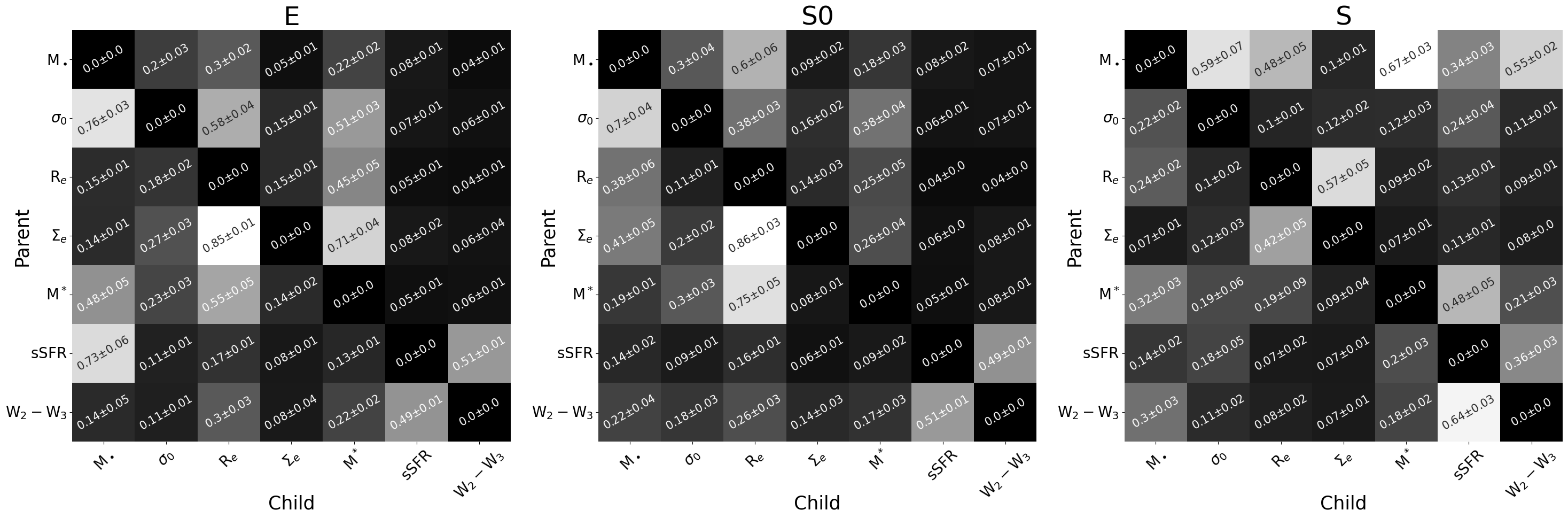}
  \includegraphics[width=\linewidth]{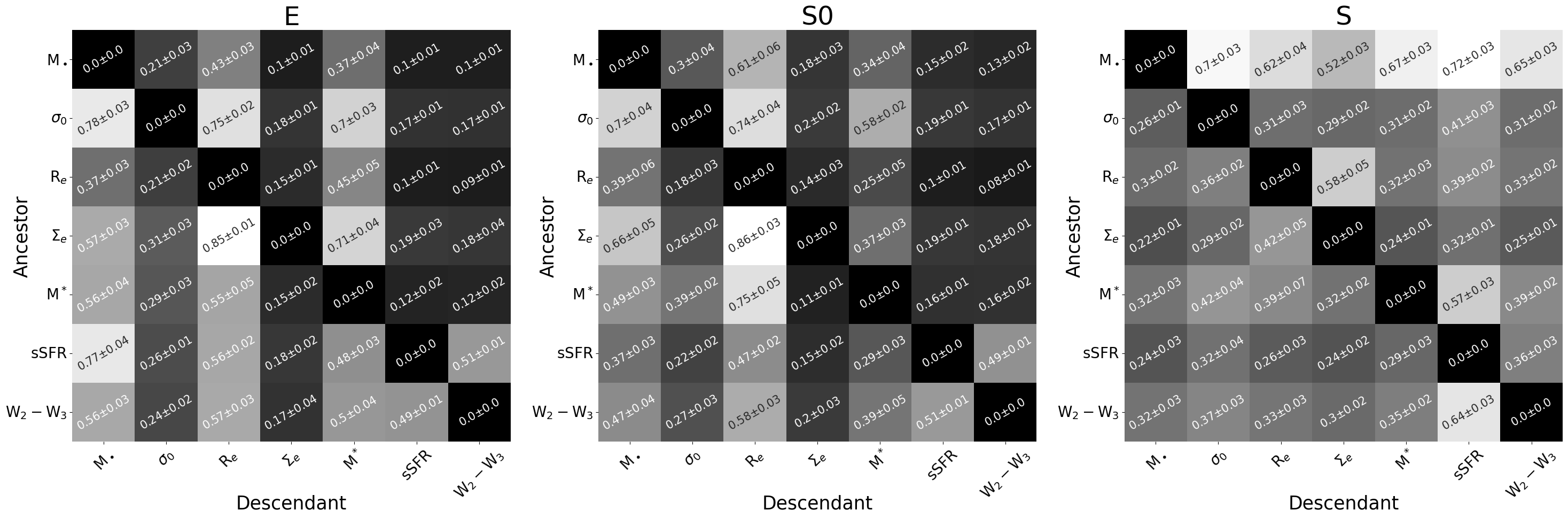}
  \caption{
  The mean and standard deviation of \emph{edge} marginals (\emph{top} matrices) and \emph{path} marginals (\emph{bottom} matrices) over all leave-one-out realizations.
  Qualitatively, we find no change to the causal directions (as compared with Figure~\ref{fig:exact_matrix}) when testing for possible outliers, and the uncertainties remain low ($\leq9\%$).
  }
\label{fig:LOO}
\end{figure*}

\begin{figure}
  \centering
  \includegraphics[width=\linewidth]{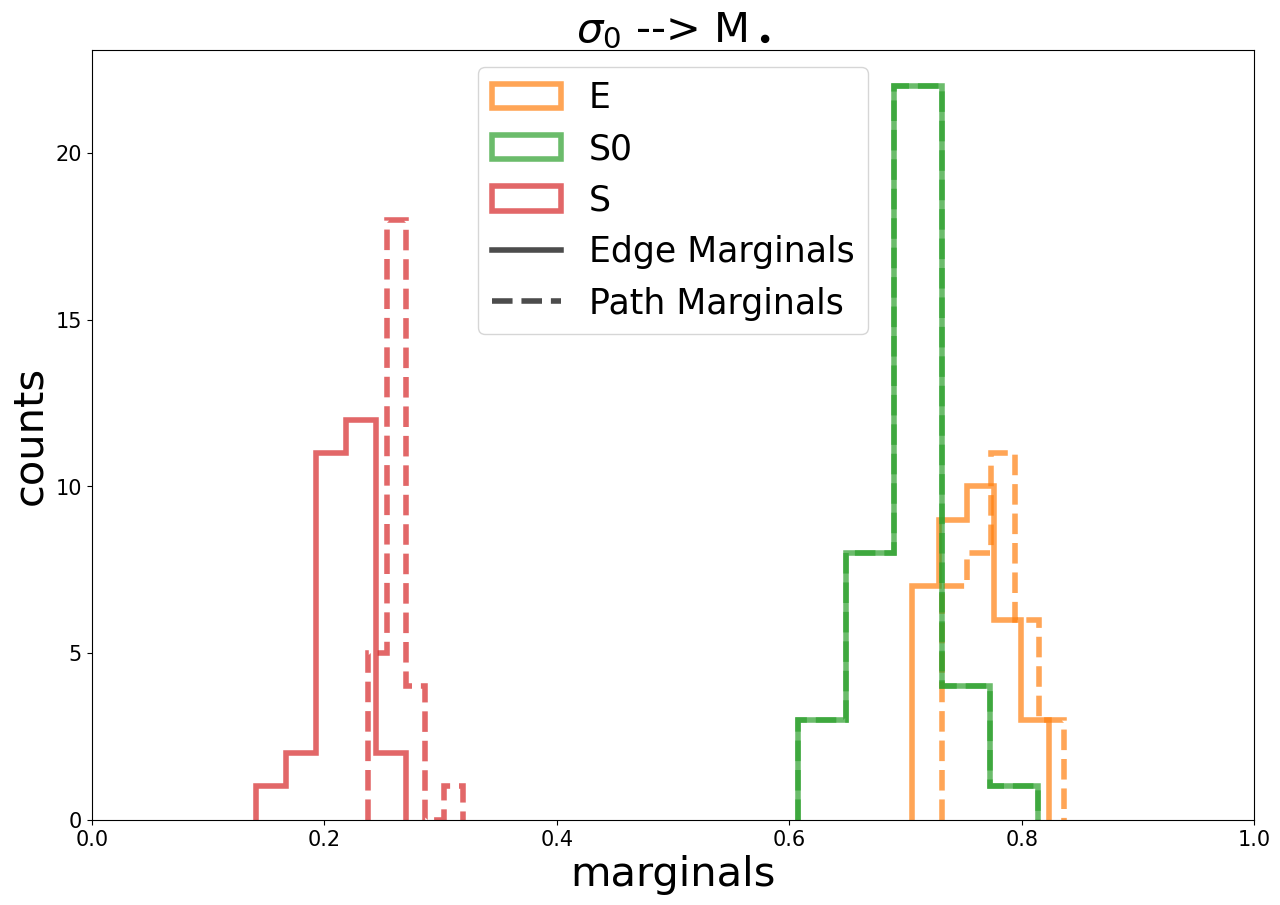}
  \caption{
  \emph{Edge} marginal ({\hdashrule[0.35ex]{8mm}{0.4mm}{}}) and \emph{path} marginal ({\hdashrule[0.35ex]{8mm}{1pt}{1mm}}) distributions of $\sigma_0 \rightarrow M_\bullet$ over all leave-one-out realizations.
  This plot illustrates the distribution of the element at the second row and first column (2, 1) in all matrices from Figure~\ref{fig:LOO}, showing a clear separation between the causal directions found in late-type vs.\ early-type galaxies.
  }
\label{fig:LOO_dist}
\end{figure}

We also explored the possibility of individual outlier galaxies biasing the inferred causal relations.
To do this, we performed leave-one-out cross-validation.
For ellipticals, lenticulars, and spirals, respectively, one galaxy is taken out at a time, and causal discovery is performed repeatedly (e.g., for 35 elliptical galaxies this procedure will be repeated 35 times).
The mean and standard deviation of the resulting marginals are shown in Figure~\ref{fig:LOO}, and the marginals for $\sigma_0 \rightarrow M_\bullet$ are highlighted in Figure~\ref{fig:LOO_dist}.
As can be seen, the fluctuations due to leave-one-out are much smaller than the uncertainties resulting from observational errors (\S\ref{sec:Obs_errors}), suggesting that the results are not driven by any potential individual outlier galaxy.

\subsection{Cyclicity}\label{sec:cyclicity}

By calculating the posterior probabilities of all possible DAGs, we implicitly assumed acyclicity, i.e., no loops in a graph.
In fact, the existence of feedback loops between black hole mass and galaxy properties (i.e., having black hole mass causing the galaxy properties, and then galaxy properties also causing black hole mass at the same time) is trivial in ellipticals and spirals according to galaxy formation theory. 
Black holes affect their host galaxies through black hole feedback, a process that heats the gas and pushes gas out to starve star formation, while galaxies also affect the central black hole through mergers and accretion.
In an ideal spiral galaxy, there have been (at most) only minor mergers, thus killing off the merger path of galaxy $\rightarrow$ black hole.

\begin{figure}
  \centering
\includegraphics[width=\linewidth]{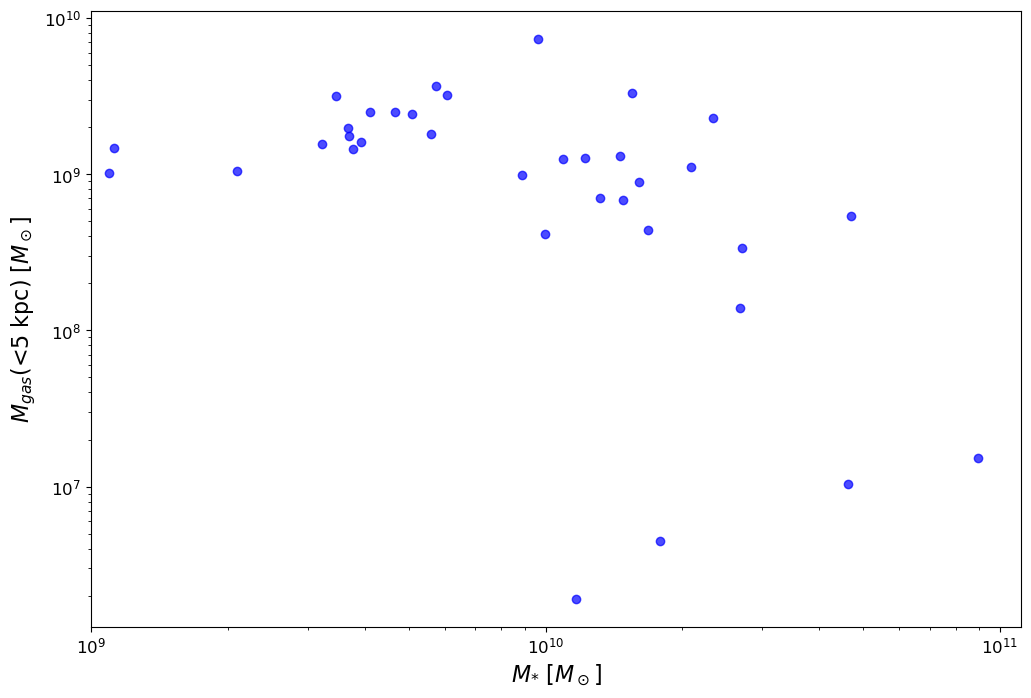}
  \caption{
  Gas mass within 5\,kpc versus total stellar mass in NIHAO simulated galaxies \citep{Wang2015}.
  The central gas mass is fairly constant in gas-rich galaxies, implying that gas accretion onto the black hole, which is mainly regulated by the black hole mass and the local gas density \citep{Bondi1952}, is also quite uniform across galaxies, weakening the galaxy $\rightarrow$ SMBH causal relation in spirals.
  This is shown to demonstrate a weakened argument for the possibility of a cyclic situation where both ``galaxy $\rightarrow$ SMBH'' and ``galaxy $\leftarrow$ SMBH'' are present at the same time in spirals.
  }
\label{fig:NIHAO}
\end{figure}

The accretion onto the black hole is mainly regulated by the black hole mass itself and the gas density in the central region \citep{Bondi1952}.
This latter quantity is found to be relatively constant in gas-rich galaxies, as confirmed by modern numerical simulations, like the NIHAO suite \citep{Wang2015,Blank2019} as shown in Figure~\ref{fig:NIHAO}.
This implies that accretion is fairly constant in all gas-rich galaxies, diminishing the causal relation galaxy $\rightarrow$ black hole.

Therefore, in spiral galaxies, the causal relation of galaxy $\rightarrow$ black hole is expected to be very weak compared to the black hole $\rightarrow$ galaxy direction.
On the other hand, ellipticals are in short supply of gas, therefore the central SMBH lacks the media in which to project its energy to regulate star formation.
As a result, in ellipticals, the black hole $\rightarrow$ galaxy direction is negligible compared to the galaxy $\rightarrow$ black hole path enabled by major mergers.

In all (in spirals and ellipticals), one of the causal directions between SMBHs and galaxies is expected to considerably overwhelm the other, making the causal structure acyclic.
The lenticulars, however, might have both major mergers and black hole feedback simultaneously, thus being more cyclic in their causal structure.
This may be one of the reasons why we see many sub-modes in the posterior distributions of lenticulars, as shown in Figure~\ref{fig:exact_distribution}.
To fully identify cyclic causal structures, time-series data is usually required.
While in our case of SMBH--galaxy co-evolution, which happens on a timescale of billions of years, obtaining time-series data is impossible within the lifetime of humanity\footnote{Except in simulations, which we will investigate in future work.}, studies of samples of galaxies with different ages may provide observational clues about the presence or absence of cyclicity in future studies.

\subsection{Alternative Priors}\label{sec:prior}
As detailed in \S\ref{sec:bge}, the posterior probability of a graph given the data is given by
\begin{equation}
    P(G\mid D) \propto P(D\mid G)P(G).
\end{equation}
$P(D\mid G)$ can be calculated through the BGe score, and $P(G)$ is the prior probability of graphs, i.e., DAGs.
In this work we adopt a uniform prior,
\begin{equation}
    P_{\text{uniform}}(G)=\frac{1}{N_{\text{DAGs}}},
\end{equation}
such that every one of the nearly $1.14\times10^9$ possible DAGs have the same prior probability.
By construction, this uniform prior does not encode any assumptions or biases about the structure of the causal graph.

Although a uniform prior best fits the data-driven purpose of this study, here we explore possible alternative priors to test the robustness of the identified causal structure.
Another possible choice of prior could be a sparsity-based prior, which embraces the idea of Occam's razor and favors simpler graphs with fewer edges.
Here we adopt a sparsity-based prior that penalizes the number of edges exponentially,
\begin{equation}
\label{equ:sparse}
    P_{\text{sparse}}(G)=\frac{e^{-\lambda \cdot E(G)}}{\sum_{\text{DAGs}}e^{-\lambda \cdot E(G)}},
\end{equation}
where $\lambda>0$ is a sparsity penalty parameter and a higher $\lambda$ enforces stronger sparsity.

\begin{figure*}[]
  \centering
  \includegraphics[width=\linewidth]{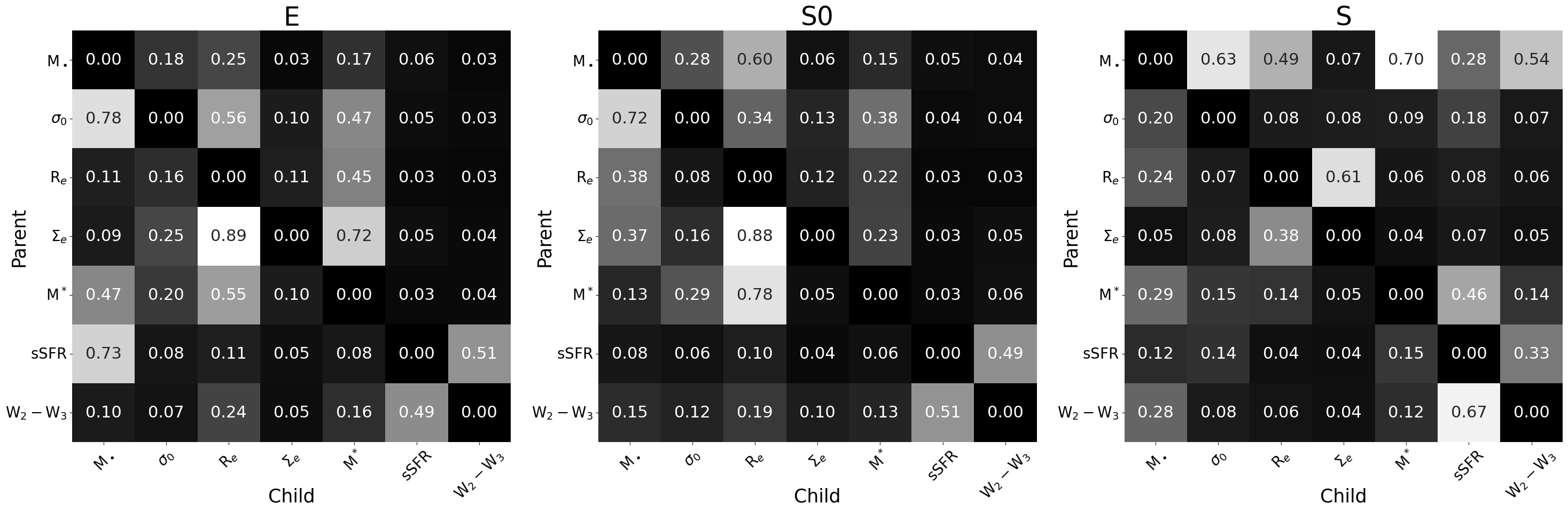}
  \includegraphics[width=\linewidth]{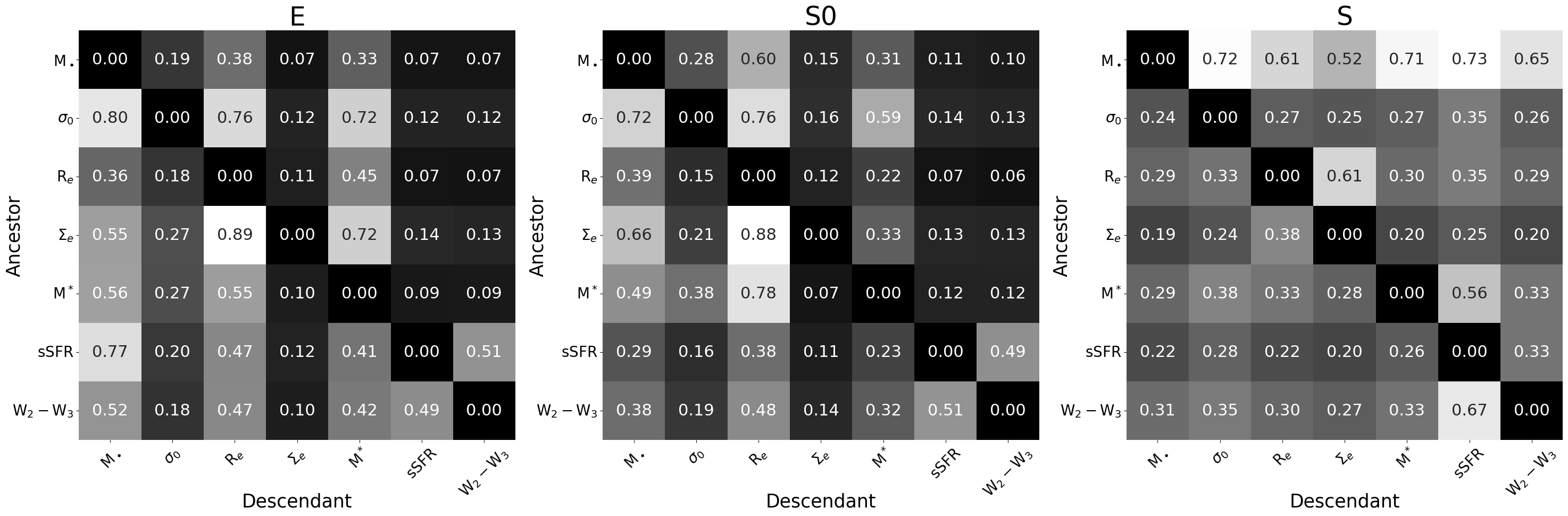}
  \caption{
  Exact posterior \emph{edge} marginals (\emph{top} matrices) and \emph{path} marginals (\emph{bottom} matrices) for \emph{elliptical} (\emph{left} matrices), \emph{lenticular} (\emph{middle} matrices), and \emph{spiral} (\emph{right} matrices) galaxies under a \emph{sparsity-based prior} detailed in Equation~\ref{equ:sparse} and \S\ref{sec:prior}.
  The marginals found are very similar to the ones found under a uniform prior (Figure~\ref{fig:exact_matrix}). 
  Despite the fact that a sparsity-based prior favors simpler DAGs with minimal number of edges, it barely changes any edge or path marginal between galaxy properties and SMBH masses.
  }
\label{fig:sparse}
\end{figure*}

We performed causal discovery under this sparsity-based prior with $\lambda=0.5$, and the edge marginals and path marginals found are presented in Figure~\ref{fig:sparse}.
Compared with the results under a uniform prior (Figure~\ref{fig:exact_matrix}), the result under a sparsity-based prior does show a tendency towards simpler graphs with less edges since the probability of less likely edges becomes even more unlikely under a sparsity-based prior, i.e., a darker cell becomes even darker from a uniform prior (Figure~\ref{fig:exact_matrix}) to a sparsity-based prior (Figure~\ref{fig:sparse}).
For example, in E galaxies, $P(\text{sSFR}\rightarrow\sigma_0)=0.11$ under a uniform prior, but this number drops to $P(\text{sSFR}\rightarrow\sigma_0)=0.08$ under a sparsity-based prior.
On the other hand, the probabilities of causal connections between galaxy properties and SMBH masses, which are the main findings of this work, remain almost the same, and some even increase, i.e., a brighter cell becomes even brighter from a uniform prior (Figure~\ref{fig:exact_matrix}) to a sparsity-based prior (Figure~\ref{fig:sparse}).
For example, in S galaxies, $P(M_\bullet\rightarrow\sigma_0)=0.62$ under a uniform prior, but rises to $P(M_\bullet\rightarrow\sigma_0)=0.63$ under a sparsity-based prior.
Therefore, as darker cells become darker and brighter cells get brighter, the causal connection between galaxy properties and SMBH masses get even more highlighted, further supporting the main claims of this work.

Besides a \emph{sparsity-based} prior and a \emph{uniform} prior, there are many other priors such as a \emph{Erd\H{o}s-R\'{e}nyi} prior \citep{Erdos:1960}, which favors a certain average number of edges per node; a \emph{combinatorial fair} prior, which incorporates combinatorial fairness and treats graphs with different parent configurations equally; and a \emph{biased} prior which favors certain configuration of graphs based on prior domain knowledge, for example, a higher prior probability for $M_\bullet\rightarrow\sigma_0$ over $\sigma_0\rightarrow M_\bullet$.
Given the aim of this paper is to make the result free of presumed models and possible biases, and then use this data-driven result to compare against theoretical galaxy formation models, we stick to the uniform prior.

\section{Possible Extension to More Variables with \texttt{DAG-GFN}}\label{sec:DAG-GFN}
The general timescale to perform the exact posterior search for seven variables outlined in this work, including generating all possible DAGs, transitive closures, calculating posterior probabilities, and getting edge/path marginals, is approximately a few hours.
However, as the number of possible DAGs grows by a factor of $\sim$$10^2$ when the number of variables increases from seven to eight, and by a factor of $\sim$$10^6$ when the number of variables increases from seven to nine, an exact search becomes impractical.
Here, we explore \texttt{DAG-GFN} as a feasible way to approximate the posteriors as the number of variables increases.

The \texttt{DAG-GFN} method \citep{deleu2022daggflownet} uses the framework of Generative Flow Networks \citep{bengio2021gflownet,bengio2023gflownetfoundations}, GFlowNets, to (approximately) sample from the posterior distribution.
GFlowNets treat the problem of sampling from an unnormalized distribution over discrete and compositional objects as a sequential decision-making problem, where actions are taken by sampling from a learned policy at each step of generation.
In the context of (Bayesian) causal discovery, DAGs are constructed one edge at a time, starting from the empty graph.
The objective is to find a policy $\pi(G'\mid G)$ giving the probability of adding an edge to the DAG $G$ to transform it into a new graph $G'$, such that sampling sequentially from it would yield samples from a distribution proportional to $R(G)$ (i.e., an unnormalized distribution).
\citet{deleu2022daggflownet} showed that such a policy satisfies
\begin{equation}
    \frac{1}{|G| + 1}R(G')\pi(\mathrm{stop}\mid G) = R(G)\pi(G'\mid G)\pi(\mathrm{stop}\mid G'),
\end{equation}
where $|G|$ is the number of edges in $G$, and $\pi(\mathrm{stop}\mid G)$ is the probability of stopping the sequential process, effectively returning $G$ as a sample of the posterior.
To sample the posterior $P(G\mid D) \propto P(D\mid G)P(G)$ (by Bayes' rule), we can then use $R(G) = P(D\mid G)P(G)$.

\begin{figure*}
  \centering
  \includegraphics[width=0.325\linewidth]{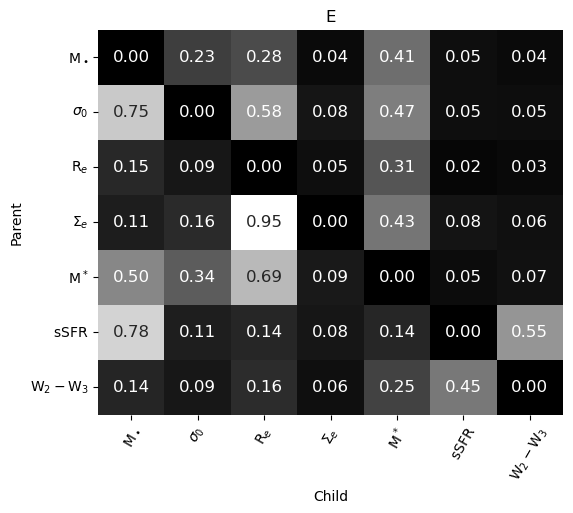}
  \includegraphics[width=0.325\linewidth]{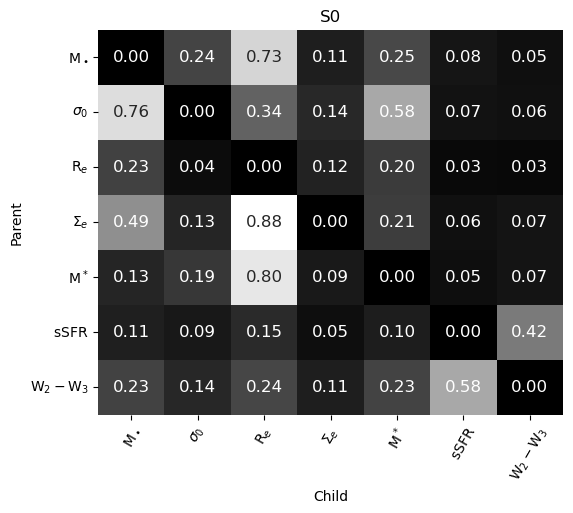}
  \includegraphics[width=0.325\linewidth]{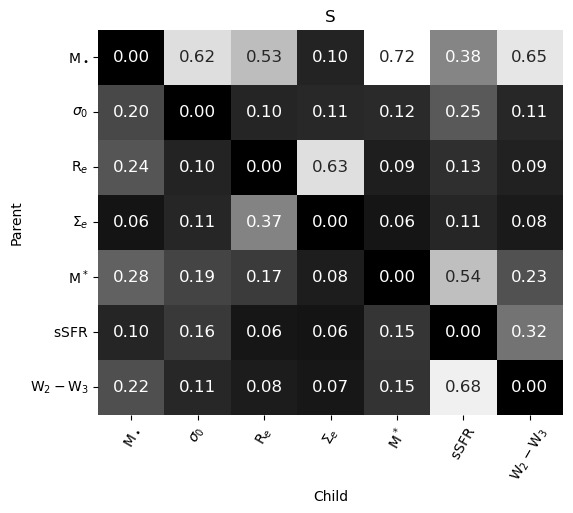}
  \includegraphics[width=0.325\linewidth]{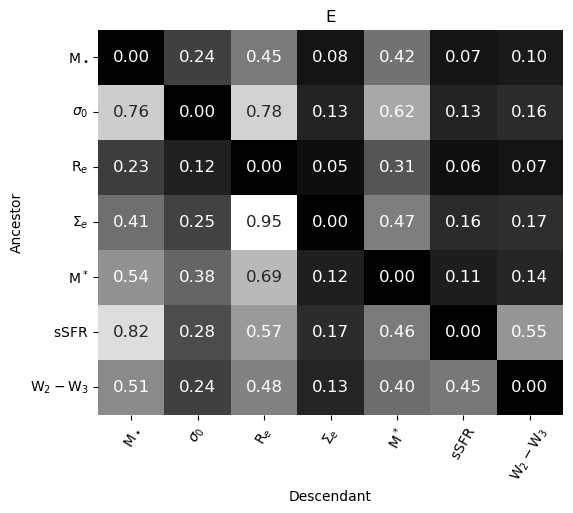}
  \includegraphics[width=0.325\linewidth]{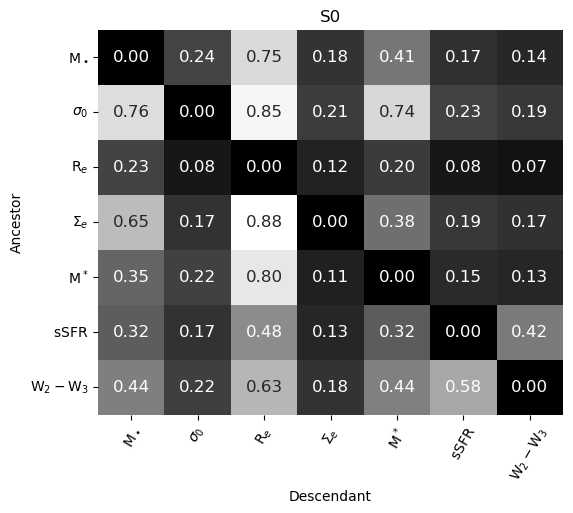}
  \includegraphics[width=0.325\linewidth]{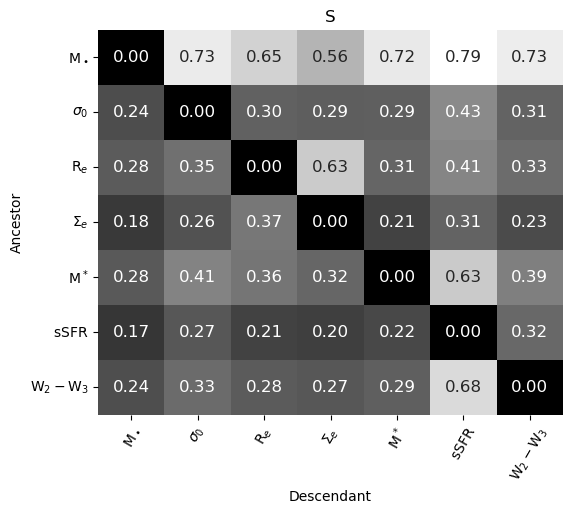}
  \caption{
  \emph{Edge} marginals (\emph{top} matrices) and \emph{path} marginals (\emph{bottom} matrices) approximated by \texttt{DAG-GFN}.
  Qualitatively, we find no change to the causal directions (as compared with Figure~\ref{fig:exact_matrix}) when using \texttt{DAG-GFN} to draw representative samples from the full Bayesian posterior.
  }
\label{fig:GFN_matrix}
\end{figure*}

\begin{figure*}
  \centering
  \includegraphics[width=0.22\linewidth]{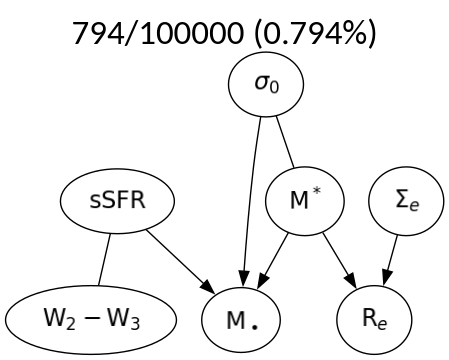}
  \hspace{0.02\linewidth}
  \includegraphics[width=0.22\linewidth]{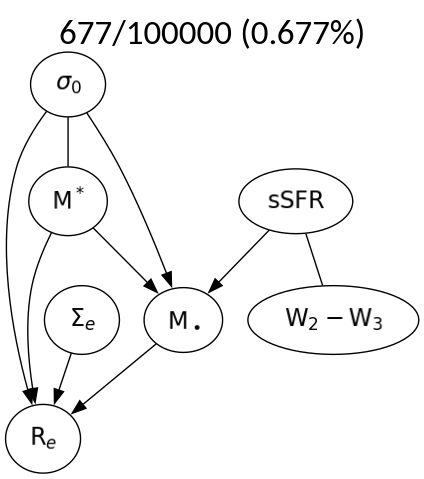}
  \hspace{0.02\linewidth}
  \includegraphics[width=0.22\linewidth]{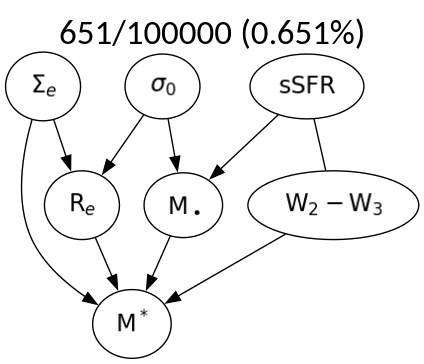}
  \hspace{0.02\linewidth}
  \includegraphics[width=0.22\linewidth]{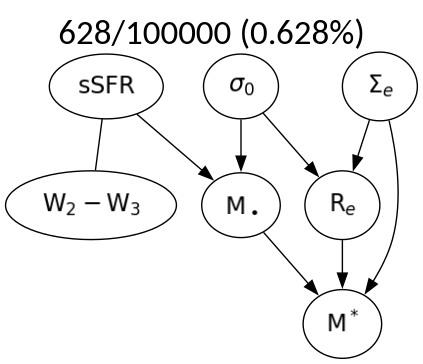}
  \noindent\rule{\textwidth}{0.5pt}
  \vskip 0.1in
  \includegraphics[width=0.22\linewidth]{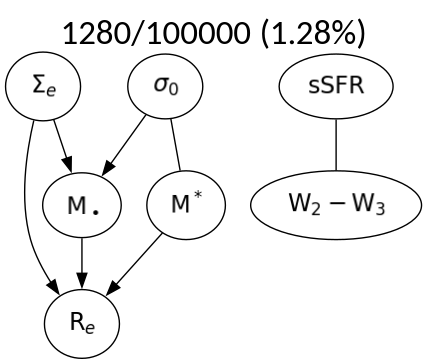}
  \hspace{0.02\linewidth}
  \includegraphics[width=0.22\linewidth]{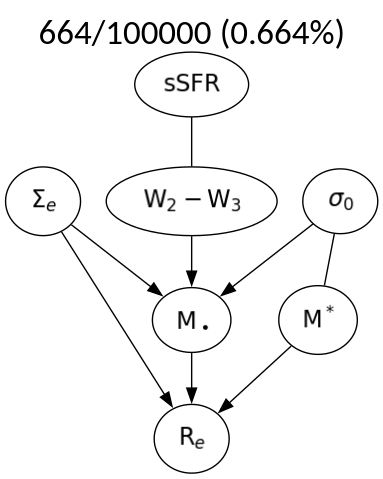}
  \hspace{0.02\linewidth}
  \includegraphics[width=0.22\linewidth]{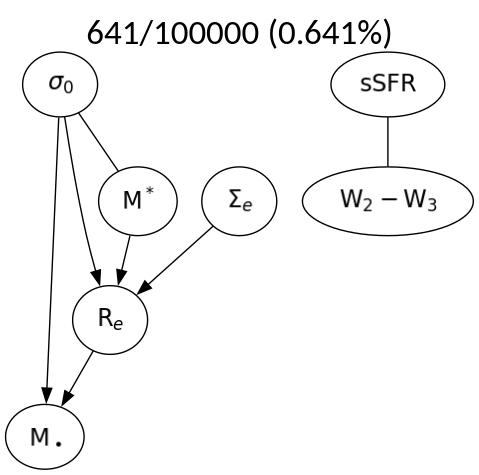}
  \hspace{0.02\linewidth}
  \includegraphics[width=0.22\linewidth]{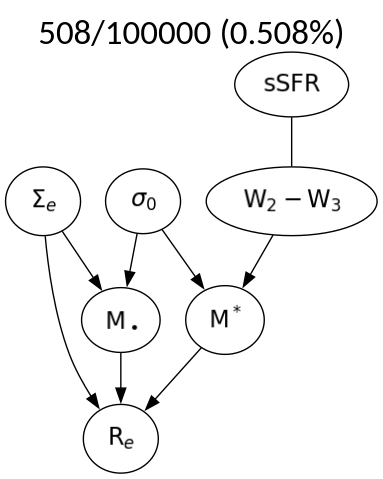}
  \noindent\rule{\textwidth}{0.5pt}
  \vskip 0.1in
  \includegraphics[width=0.22\linewidth]{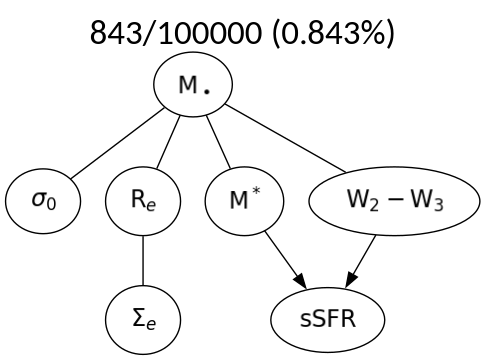}
  \hspace{0.02\linewidth}
  \includegraphics[width=0.22\linewidth]{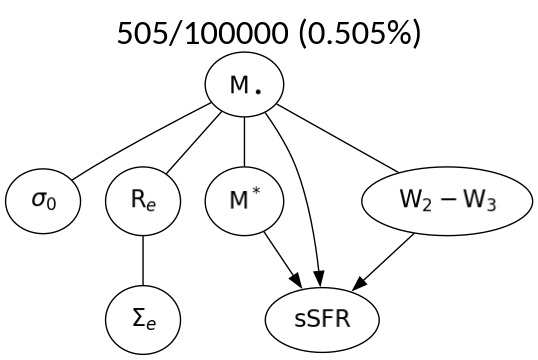}
  \hspace{0.02\linewidth}
  \includegraphics[width=0.22\linewidth]{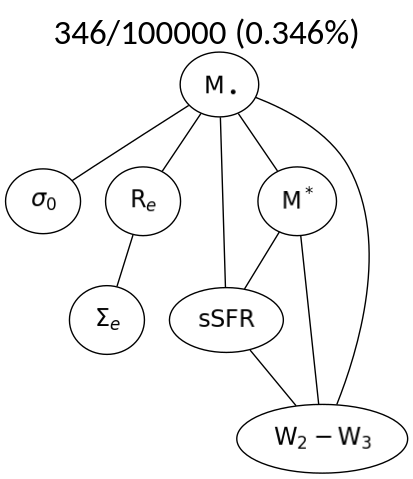}
  \hspace{0.02\linewidth}
  \includegraphics[width=0.22\linewidth]{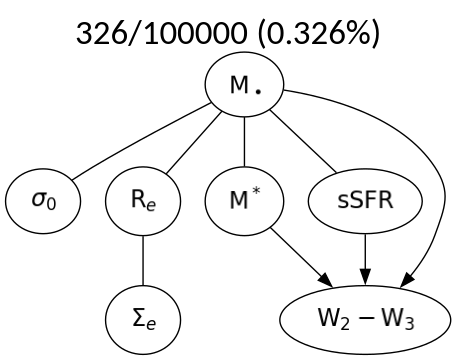}
  \caption{
  Top four MECs sampled by \texttt{DAG-GFN} for \emph{elliptical} (\emph{top} panel), \emph{lenticular} (\emph{middle} panel), and \emph{spiral} (\emph{bottom} panel) galaxies.
  Qualitatively, we find no change to the causal structures (as compared with Figure~\ref{fig:exact_MECs}) when using \texttt{DAG-GFN} to draw representative samples from the full Bayesian posterior.
  }
\label{fig:GFN_MECs}
\end{figure*}

\begin{figure*}
  \centering
  \includegraphics[width=0.135\linewidth]{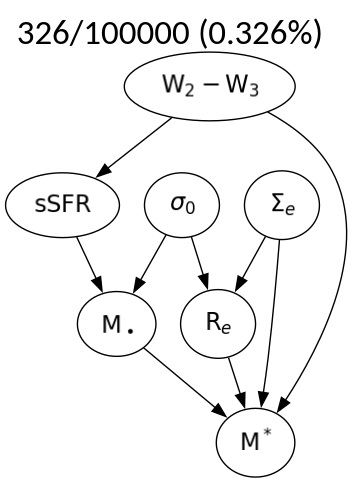}
  \hspace{0.035\linewidth}
  \includegraphics[width=0.135\linewidth]{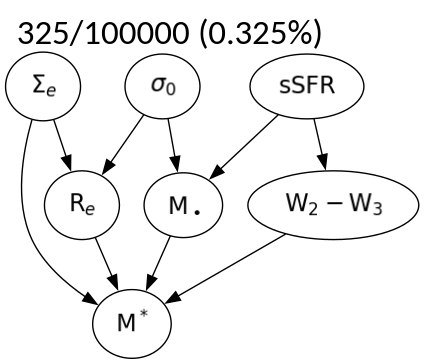}
  \hspace{0.035\linewidth}
  \includegraphics[width=0.135\linewidth]{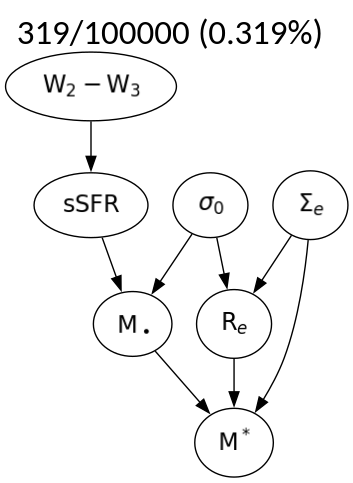}
  \hspace{0.035\linewidth}
  \includegraphics[width=0.135\linewidth]{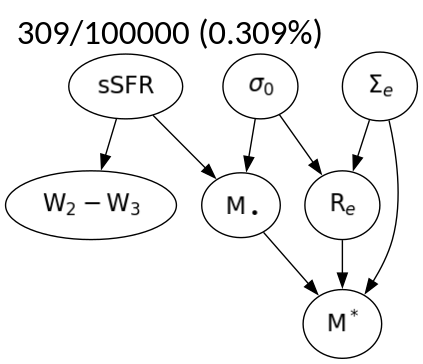}
  \hspace{0.035\linewidth}
  \includegraphics[width=0.135\linewidth]{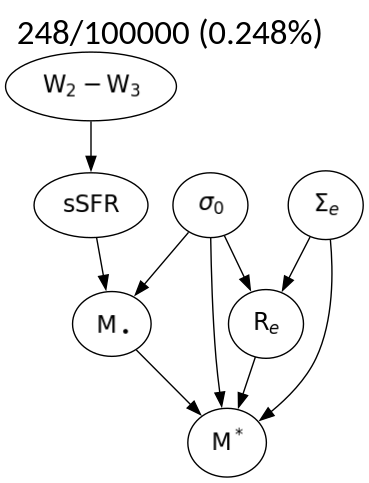}
  \vskip 0.1in
  \includegraphics[width=0.135\linewidth]{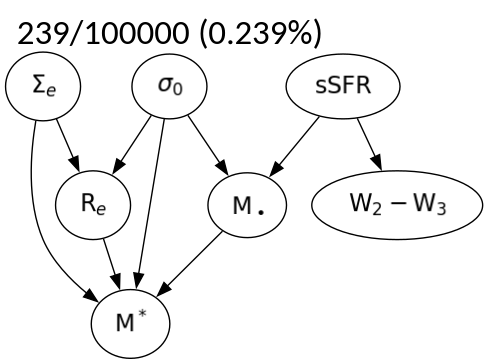}
  \hspace{0.035\linewidth}
  \includegraphics[width=0.135\linewidth]{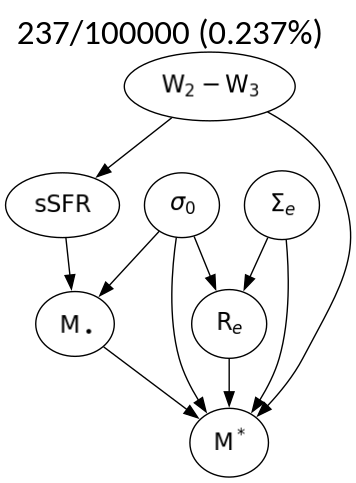}
  \hspace{0.035\linewidth}
  \includegraphics[width=0.135\linewidth]{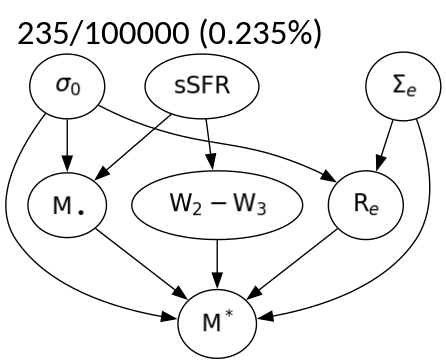}
  \hspace{0.035\linewidth}
  \includegraphics[width=0.135\linewidth]{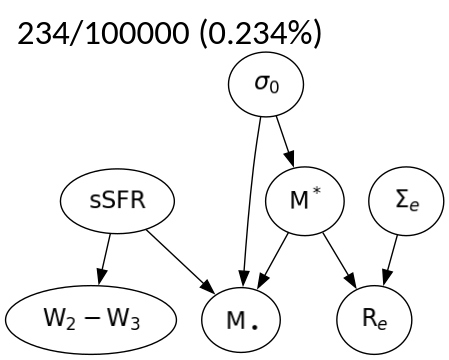}
  \hspace{0.035\linewidth}
  \includegraphics[width=0.135\linewidth]{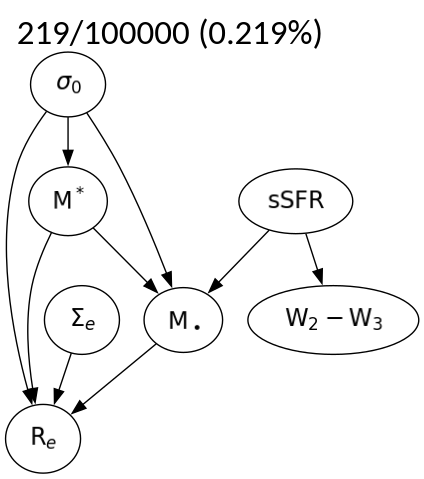}
  \noindent\rule{\textwidth}{0.5pt}
  \vskip 0.1in
  \includegraphics[width=0.135\linewidth]{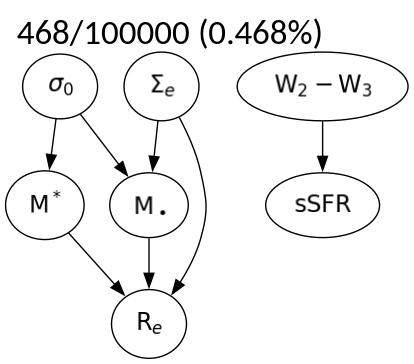}
  \hspace{0.035\linewidth}
  \includegraphics[width=0.135\linewidth]{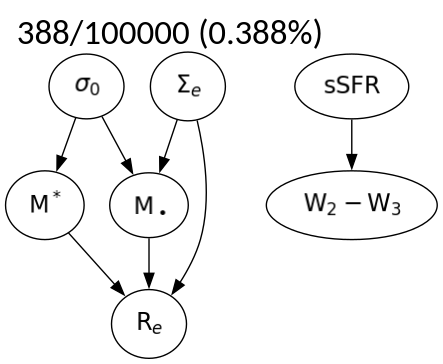}
  \hspace{0.035\linewidth}
  \includegraphics[width=0.135\linewidth]{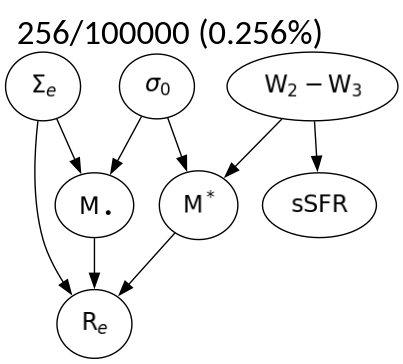}
  \hspace{0.035\linewidth}
  \includegraphics[width=0.135\linewidth]{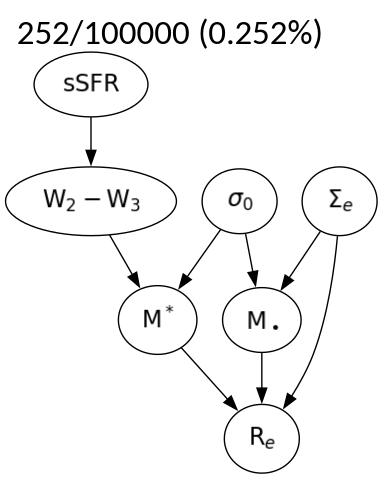}
  \hspace{0.035\linewidth}
  \includegraphics[width=0.135\linewidth]{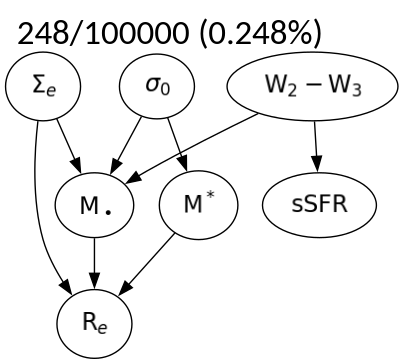}
  \vskip 0.1in
  \includegraphics[width=0.135\linewidth]{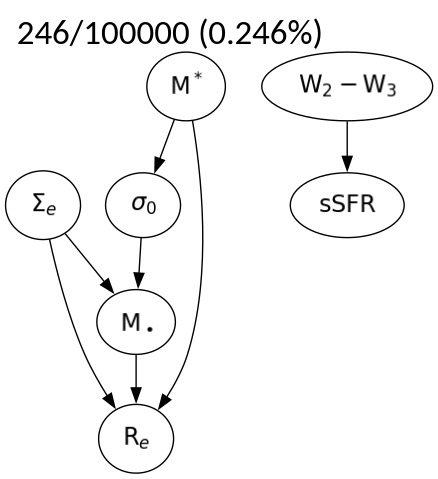}
  \hspace{0.035\linewidth}
  \includegraphics[width=0.135\linewidth]{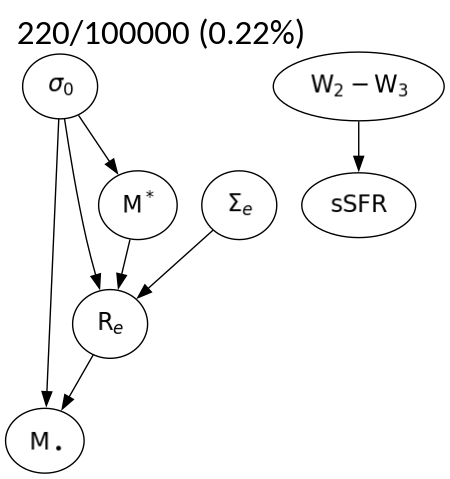}
  \hspace{0.035\linewidth}
  \includegraphics[width=0.135\linewidth]{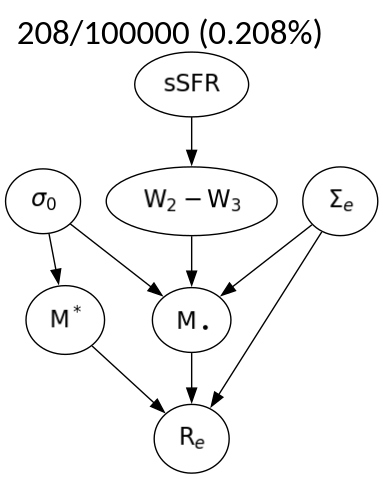}
  \hspace{0.035\linewidth}
  \includegraphics[width=0.135\linewidth]{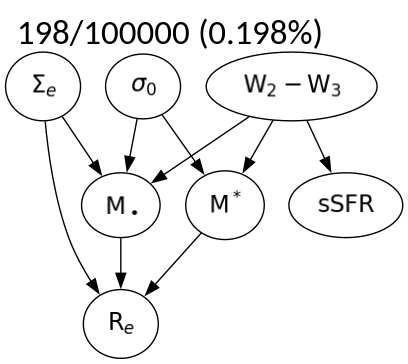}
  \hspace{0.035\linewidth}
  \includegraphics[width=0.135\linewidth]{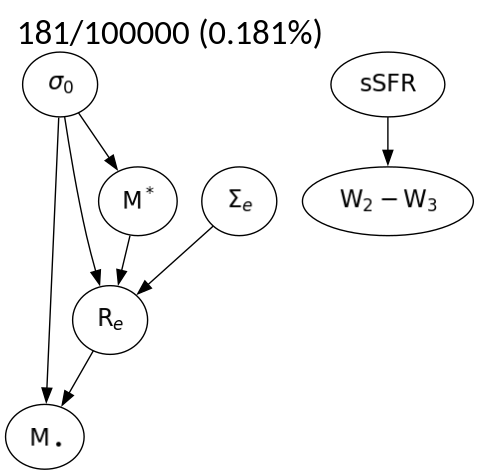}
  \noindent\rule{\textwidth}{0.5pt}
  \vskip 0.1in
  \includegraphics[width=0.135\linewidth]{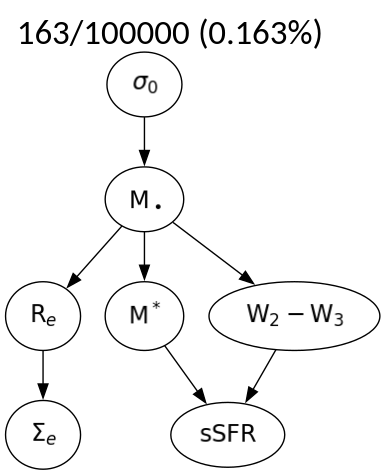}
  \hspace{0.035\linewidth}
  \includegraphics[width=0.135\linewidth]{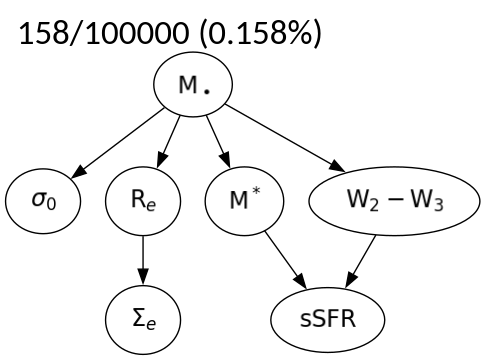}
  \hspace{0.035\linewidth}
  \includegraphics[width=0.135\linewidth]{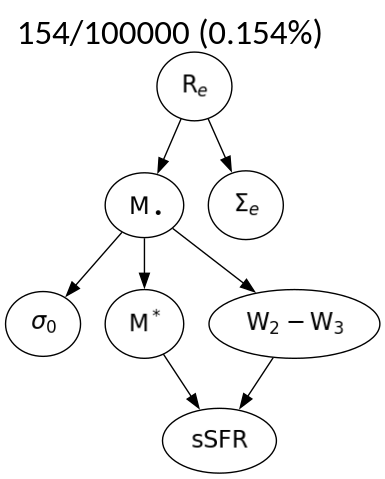}
  \hspace{0.035\linewidth}
  \includegraphics[width=0.135\linewidth]{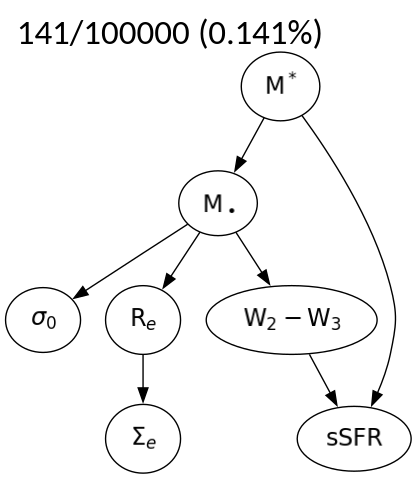}
 \hspace{0.035\linewidth}
  \includegraphics[width=0.135\linewidth]{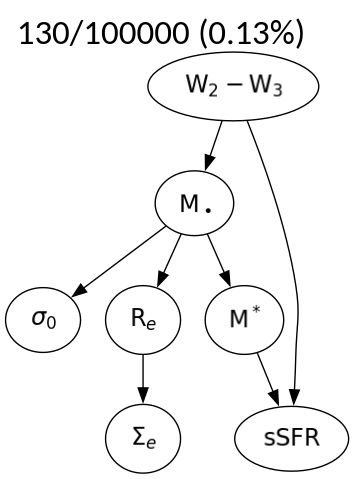}
  \vskip 0.1in
  \includegraphics[width=0.135\linewidth]{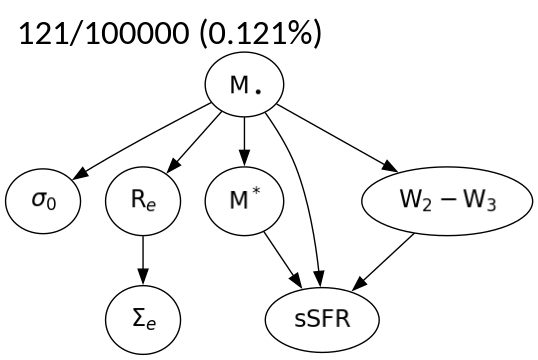}
  \hspace{0.035\linewidth}
  \includegraphics[width=0.135\linewidth]{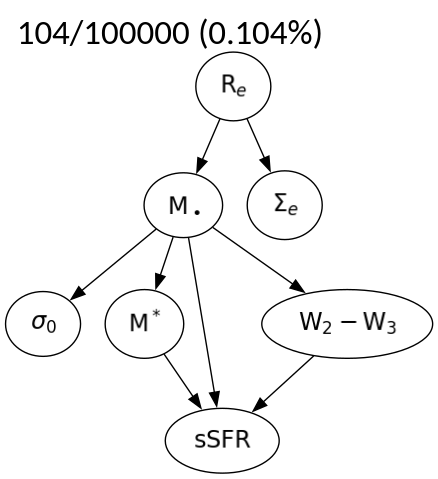}
  \hspace{0.035\linewidth}
  \includegraphics[width=0.135\linewidth]{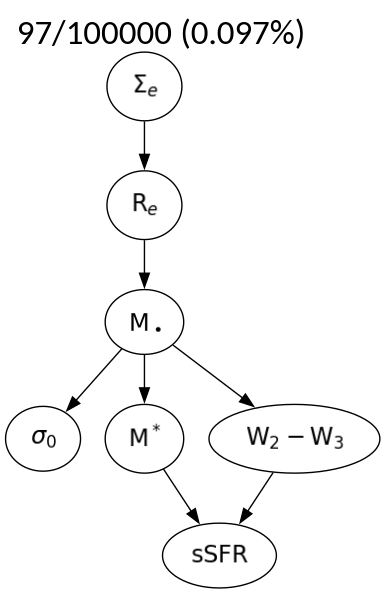}
  \hspace{0.035\linewidth}
  \includegraphics[width=0.135\linewidth]{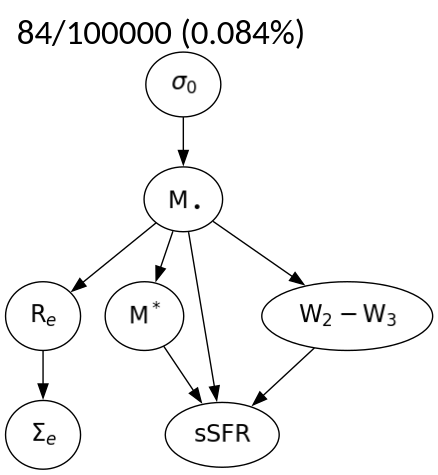}
  \hspace{0.035\linewidth}
  \includegraphics[width=0.135\linewidth]{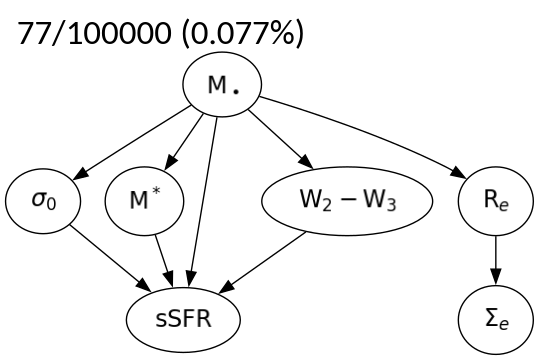}
  \caption{
  Top 10 DAGs sampled by \texttt{DAG-GFN} for \emph{elliptical} (\emph{top} panel), \emph{lenticular} (\emph{middle} panel), and \emph{spiral} (\emph{bottom} panel) galaxies.
  Qualitatively, we find no change to the causal structures (as compared with Figure~\ref{fig:exact_DAGs}) when using \texttt{DAG-GFN} to draw representative samples from the full Bayesian posterior.
  }
\label{fig:GFN_DAGs}
\end{figure*}

\texttt{DAG-GFN} was trained on our data, and $10^5$ DAGs were sampled from the trained network.
The frequency of each sampled unique DAG gives the approximated posterior probability of that DAG. 
The marginals, as well as the top MECs and DAGs are presented in Figures~\ref{fig:GFN_matrix}, \ref{fig:GFN_MECs}, and \ref{fig:GFN_DAGs}, respectively.
The approximated posteriors by \texttt{DAG-GFN} are highly consistent with the exact posteriors from our primary analysis.
Visual inspection reveals that Figure~\ref{fig:exact_matrix} and Figure~\ref{fig:GFN_matrix} present noticeable similarities.

\section{Conclusions}\label{sec:conclusions}

We present the first data-driven evidence on the direction of the causal relationship between supermassive black holes and their host galaxies.
Our findings suggest that in elliptical galaxies, bulge properties influence SMBH growth, whereas in spiral galaxies, SMBHs shape galaxy characteristics.
The process of quenching can be causally explained as follows:
\begin{enumerate}
    \item quenching starts in gas-rich (i.e., spiral) galaxies, and hence there is a causal connection; and
    \item the quenching is over in elliptical galaxies, where we only see the end product of such quenching, and the causal connection is now reversed.
\end{enumerate}
These findings support theoretical models of galactic evolution driven by feedback processes and mergers.
Withal, our causal mechanisms are defined from a relatively modest number of galaxies in the local Universe.
Further insights can be gained in the future with wider and deeper surveys for more SMBH mass measurements, or by using time-series data and control variables in galaxy simulations \citep{Waterval:2024} across a wide range of redshifts to test the causal findings and explanations presented here.
Besides that, continual advances in the nascent field of causal discovery will verily help alleviate potential biases imposed by unobserved confounders \citep{Bernhard:2021,Zhang:2024,Jin:2024} or cyclicity \citep{Ghassami:2019,Dai:2024}.

With the knowledge we gain from learning the underlying causal structures and mechanisms behind galaxy--SMBH co-evolution, it should ultimately be possible to create physically-motivated black 
mass scaling relations that faithfully model the reality of action/interaction.
The successful application of causal discovery to this astrophysical dataset paves the way for a deeper understanding of the fundamental physical processes driving galaxy evolution and establishes causal discovery as a promising tool for data-driven insights across various scientific disciplines.

\begin{acknowledgments}
This research was carried out on the high-performance computing resources at New York University Abu Dhabi.
We acknowledge the usage of the HyperLeda database (\url{http://leda.univ-lyon1.fr}). 
Z.J.\ and M.P.\ wish to extend their heartfelt thanks to Jithendaraa Subramanian for providing in-depth support and clarifications regarding \texttt{DAG-GFN}, and to Michelle Liu for comments and discussion.
Y.H. thanks Andrew Benson and Dhanya Sridhar for helpful discussions.
Z.J.\ thanks Michael Blanton and Joseph Gelfand for useful suggestions.
Z.J.\ genuinely thanks Mohamad Ali-Dib for his very timely help with HPC technical issues.

This material is based upon work supported by Tamkeen under the NYU Abu Dhabi Research Institute grant CASS.
This work is partially supported by Schmidt Futures, a philanthropic initiative founded by Eric and Wendy Schmidt as part of the Virtual Institute for Astrophysics (VIA).
M.P.\ acknowledges financial support from the European Union's Horizon 2020 research and innovation program under the Marie Sk\l{}odowska-Curie grant agreement No.\ 896248.

The data and code used for this work are available for download from the following GitHub repository: \url{https://github.com/ZehaoJin/causalbh}.
\end{acknowledgments}

\software{
\\
\href{https://causal-learn.readthedocs.io/en/latest/}{\textcolor{linkcolor}{\texttt{causal-learn}}} \citep{causallearn}\\
\href{https://github.com/tristandeleu/jax-dag-gflownet}{\textcolor{linkcolor}{\texttt{DAG-GFN}}} \citep{deleu2022daggflownet}\\
\href{https://www.gymlibrary.dev/index.html}{\textcolor{linkcolor}{\texttt{Gym}}} \citep{gym}\\
\href{https://jax.readthedocs.io/en/latest/}{\textcolor{linkcolor}{\texttt{JAX}}} \citep{jax2018github}\\
\href{https://github.com/matplotlib/matplotlib}{\textcolor{linkcolor}{\texttt{Matplotlib}}} \citep{Hunter:2007}\\
\href{https://networkx.org/}{\textcolor{linkcolor}
{\texttt{NetworkX}}} \citep{networkx}\\
\href{https://github.com/numpy/numpy}{\textcolor{linkcolor}{\texttt{NumPy}}} \citep{harris2020array}\\
\href{https://pandas.pydata.org/}{\textcolor{linkcolor}{\texttt{Pandas}}} \citep{McKinney_2010}\\
\href{https://pgmpy.org/}{\textcolor{linkcolor}{\texttt{pgmpy}}} \citep{ankan2015pgmpy}\\
\href{https://pygraphviz.github.io/}{\textcolor{linkcolor}{\texttt{PyGraphviz}}}\\
\href{https://www.python.org/}{\textcolor{linkcolor}{\texttt{Python}}} \citep{Python}\\
\href{https://github.com/scipy/scipy}{\textcolor{linkcolor}{\texttt{SciPy}}} \citep{Virtanen_2020}\\
\href{https://seaborn.pydata.org/}{\textcolor{linkcolor}{\texttt{seaborn}}} \citep{Waskom2021}
          }

\clearpage
\section*{ORCID iDs}

\begin{CJK*}{UTF8}{gbsn}
\begin{flushleft}
Zehao Jin (金泽灏) \scalerel*{\includegraphics{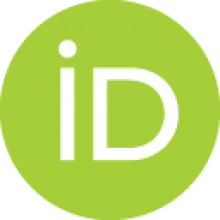}}{B} \url{https://orcid.org/0009-0000-2506-6645}\\
Mario Pasquato \scalerel*{\includegraphics{orcid-ID.png}}{B} \url{https://orcid.org/0000-0003-3784-5245}\\
Benjamin L.\ Davis \scalerel*{\includegraphics{orcid-ID.png}}{B} \url{https://orcid.org/0000-0002-4306-5950}\\
Tristan Deleu \scalerel*{\includegraphics{orcid-ID.png}}{B} \url{https://orcid.org/0009-0005-1943-3484}\\
Yu Luo (罗煜) \scalerel*{\includegraphics{orcid-ID.png}}{B} \url{https://orcid.org/0000-0003-2341-9755}\\
Changhyun Cho \scalerel*{\includegraphics{orcid-ID.png}}{B} \url{https://orcid.org/0000-0002-9879-1749}\\
Pablo Lemos \scalerel*{\includegraphics{orcid-ID.png}}{B} \url{https://orcid.org/0000-0002-4728-8473}\\
Laurence Perreault-Levasseur \scalerel*{\includegraphics{orcid-ID.png}}{B} \url{https://orcid.org/0000-0003-3544-3939}\\
Yoshua Bengio \scalerel*{\includegraphics{orcid-ID.png}}{B} \url{https://orcid.org/0000-0002-9322-3515}\\
Xi Kang (康熙) \scalerel*{\includegraphics{orcid-ID.png}}{B} \url{https://orcid.org/0000-0002-5458-4254}\\
Andrea Valerio Macci\`{o} \scalerel*{\includegraphics{orcid-ID.png}}{B} \url{https://orcid.org/0000-0002-8171-6507}\\
Yashar Hezaveh \scalerel*{\includegraphics{orcid-ID.png}}{B} \url{https://orcid.org/0000-0002-8669-5733}
\end{flushleft}
\end{CJK*}

\bibliography{bibliography}{}

\end{document}